\documentclass[12pt]{iopart} 
\usepackage{iopams} \usepackage{natbib} \usepackage{graphicx} 
\newcommand{\PP}{{\cal{P}}} \newcommand{\vers}[1]{\vec u_#1}
\def\uu{4U\,0142+614\,}  
 
\def\ea{1E\,2259+586\,}

\usepackage{deluxetable}
\usepackage{xcolor}
\usepackage{footnote}

\begin{document}

\maketitle
\review[Magnetars: the physics behind observations] {Magnetars: the physics
behind observations}

\author{R Turolla$^{1,2}$, S Zane$^2$, A L Watts$^3$}

\address{$^1$ Department of Physics and Astronomy, University of Padova, via Marzolo 8, 35131 Padova, Italy}
\address{$^2$ Mullard Space Science Laboratory, University College London, Holbury St. Mary, Surrey, RH5 6NT, UK}
\address{$^3$ Anton Pannekoek Institute for Astronomy, University of Amsterdam,
Postbus 94249, 1090 GE Amsterdam, The Netherlands}
\ead{turolla@pd.infn.it}
\begin{abstract}
Magnetars are the strongest magnets in the present universe and the
combination of extreme magnetic field, gravity and density makes them
unique laboratories to probe current physical theories (from quantum
electrodynamics to general relativity) in the strong field limit.
Magnetars are observed as peculiar, burst--active X-ray pulsars, the
Anomalous X-ray Pulsars (AXPs) and the Soft Gamma Repeaters (SGRs); the
latter emitted also three ``giant flares'', extremely powerful events
during which luminosities can reach up to $10^{47}$~erg/s for about one
second. The last five years have witnessed an explosion in magnetar
research which has led, among other things, to the discovery of transient,
or ``outbursting'', and ``low-field'' magnetars. Substantial progress has
been made also on the theoretical side. Quite detailed models for
explaining the magnetars' persistent X-ray emission, the properties of the
bursts, the flux evolution in transient sources have been developed and
confronted with observations. New insight on neutron star asteroseismology
has been gained through improved models of magnetar oscillations. The
long-debated
issue of magnetic field
decay in neutron stars has been addressed, and its importance recognized in
relation to the evolution of magnetars and to the links among magnetars
and other families of isolated neutron stars. The aim of this paper is to
present a comprehensive overview in which the observational results are
discussed in the light of the most up-to-date theoretical models and their
implications. This addresses not only the particular case of magnetar
sources, but the more fundamental issue of how physics in strong magnetic
fields can be constrained by the observations of these unique sources.
\end{abstract}

\maketitle
\section{Introduction}
\label{intro}

Neutron stars (NSs), the endpoint of the evolution of massive stars with
$10\lesssim M/M_\odot\lesssim 25$, are extremely compact remnants endowed
with strong magnetic fields. Isolated (i.e. not belonging to a binary
system) neutron stars were for a long time identified with radio-pulsars,
and only in the last two decades, mainly thanks to high-energy
observations, was the existence of other manifestations of isolated neutron
stars recognized. Among them, there are two groups of X-ray pulsars
with remarkably peculiar properties, the Soft Gamma Repeaters (SGRs) and
the Anomalous X-ray Pulsars \cite[AXPs, see e.g.][for a review]{mere08}.
The separation into two classes reflects the way
in which these sources were originally discovered. SGRs were revealed
through the detection of short, intense bursts in the hard X-/soft
gamma-ray range \cite[][]{maz79a,maz79b}, and because of this were
initially associated with gamma ray bursts (GRBs); however, burst emission
from SGRs was soon recognized to repeat, at variance to what was observed in
GRBs, setting the two phenomena apart. On the other hand, AXPs were
identified as X-ray pulsar in the soft X-ray range
\cite[$<10$~keV,][]{merst95}. They were dubbed ``anomalous'' because their
high X-ray luminosity ($\sim 10^{34}-10^{36}$~erg/s) cannot be easily
explained in terms of the conventional processes which apply to other
classes of pulsars, i.e. accretion from a binary companion or injection of
rotational energy in the pulsar wind/magnetosphere. Over the last few
years, observations have revealed many similarities between these two
classes of objects \cite[see e.g.][]{woodth06}, including the discovery
that AXPs too emit short, SGR-like bursts \cite[][]{kas00,kas03} and
nowadays the idea that SGRs and AXPs belong to a single, unified class is
widely accepted.

The main observational characteristics of SGRs and AXPs are: a) lack of
evidence of binary companions; b) persistent (i.e. non-bursting), often
variable X-ray luminosity in the range $\sim 10^{33}$--$10^{36}$~erg/s,
emitted in the soft (0.5--10 keV) and hard (20--100 keV) X-ray range; c)
pulsations at relatively long spin periods, clustered in the range $\sim
2$--12~s; d) large secular spin-down rate, $\dot P \sim
10^{-13}-10^{-11}$~s/s, which, if interpreted in terms of electromagnetic
losses from a rotating dipole in vacuo, leads to huge magnetic fields,
$\sim 10^{14}$--$10^{15}$~G. SGRs (and AXPs, to a somewhat lesser extent)
exhibit spectacular and frequent bursting activity, which is observed in
the X-/gamma-rays on several timescales, ranging from sub-s to several
tens of seconds. In particular, three different kinds of bursting events
have been observed (see Sec. \ref{bursts}):

\begin{itemize} \item{} short bursts: these are the most common, with
typical duration of $\sim 0.1$--1~s, peak luminosity of
$\sim10^{39}$--$10^{41}$~erg/s, and soft ($\sim 10$~keV), thermal spectra;
they are detected from both SGRs and AXPs;

\item{} intermediate bursts, which last $\sim 1$--40~s and have a peak
luminosity of $\sim 10^{41}-10^{43}$~erg/s. These are characterized by an
abrupt onset and usually also show thermal spectra;  again, they were seen
in both SGRs and AXPs;

\item{} giant flares. These are exceptional, rare events, with an energy
output of $\sim 10^{44}-10^{47}$~erg/s, only exceeded by blazars and GRBs.
They have been observed only in SGRs and only three times since SGRs were
discovered: from SGR 0526-66 in 1979 \cite[][]{maz79b}, from SGR 1900+14
in 1998 \cite[][]{hur99}, and from SGR 1806-20 in 2004
\cite[e.g.][]{hur05,pal05}. All three events started with an initial spike
of $\sim 0.1$--0.2~s duration, followed by a long pulsating tail (lasting
a few hundred seconds) modulated at the neutron star spin period.
\end{itemize}

All together, these properties find an explanation in the so-called
``magnetar'' scenario \cite[][]{dt92,thdun93,thdun95}, according to which the
relatively high X-ray luminosity and the bursting/outbursting activity are
powered by the dissipation and decay of a superstrong magnetic field,
$\approx 10^{14}$--$10^{15}$ G on the surface and possibly higher in the
star's interior. Despite no indisputable measure of an ultra-high magnetic field has been obtained as yet,
a number of indipendent arguments strongly support the idea that SGRs/AXPs are indeed magnetically-powered, as first discussed by \cite{thdun95}. A few of them are
\begin{itemize}
\item{the rotational energy loss rate $\dot E$ (which is believed to fuel standard pulsars) is well below the persistent X-ray luminosity, $\dot E \ll L_X$;}
\item{long spin periods ($\approx 10$ s}) can be attained in $\approx 10^3-10^4$ yrs (the source age as inferred from that of the associated SNR) via magneto-dipolar braking only 
for fields $\gtrsim 10^{14}$ G;
\item{huge spin-down rates have been indeed measured in these sources, implying dipole magnetic fields in the range $\approx 10^{14}-10^{15}$ G;}
\item{the decrease of the scattering opacity in a superstrong magnetic field ($B\gtrsim 10^{14}$ G) pushes upwards the Eddington limit and allows a much larger luminosity
to escape from a (magnetically) confined plasma: this can explain the apparently super-Eddington luminosity of a number of bursts;}
\item{no stellar companions have been discovered in SGRs/AXPs, ruling out accretion as a possible source of energy;}
\item{if no more than a fraction of the magnetic energy was released in a giant flare, this requires $B\gtrsim 10^{14}$ G; in order to power $\approx 100$ giant flares like that emitted
in 2004 by SGR 1806-20 over the source lifetime an internal field $\approx 10^{16}$ G is needed \cite[][]{stella05}. }
\end{itemize}
Although alternative interpretations have been proposed
\cite[see e.g.][for a review and references therein]{turesp13}, the
magnetar model more
naturally explains the properties of SGRs and AXPs, including the bursting
activity and the hyper-energetic giant flares, and will be the focus of
this review.

Even the ``persistent'' emission of these sources is far from being
steady. Magnetars' spin-down is quite irregular, and often accompanied by
glitches and timing noise. Long term variations in magnetars' emission can
occur either as gradual and moderate changes in the flux, accompanied by
variations in the spectrum, pulse profile, and spin-down rate, or as
sudden outbursts, i.e. events during which the flux raises up to a factor
$\sim 1000$
and then decays back to a level compatible with the quiescent state over a
time scale of months/years (see Sec.~\ref{transient}). Within the magnetar
scenario, the first kind of variability is thought to be driven by plastic
deformations in the crust which, in turn, induce changes in the magnetic
current configurations. The more violent outbursts, as well as the
glitches, the bursting activity and even the hyper-energetic giant flares
could instead be due to sudden reconfigurations of the magnetosphere, when
unstable conditions are reached. This may lead to crustal fractures
(starquakes) and/or instabilities in the outer magnetosphere (possibly
involving magnetic reconnection).

Although originally discovered in the X-/soft gamma-rays, magnetars have
been detected at different wavelengths, revealing a rich phenomenology
across the electromagnetic spectrum. AXPs and SGRs have been discovered to
emit in the optical and/or near-infrared (NIR) bands \cite[e.g.][]{hul00,
isr04, dur06}. The optical/NIR counterparts are faint ($K\sim 20$) and the
flux is only a small fraction of the bolometric flux, but still its
detection can place important constraints on models. Several AXPs have
exhibited long-term variability both in their optical/infrared emission
and in X-rays \cite[][]{isr02, hul04, rea05}. Unavoidably this introduces
additional uncertainties in the modelling of broad band spectra, based on
observations at different wavelengths taken at different times.

Magnetars were traditionally considered to be radio-silent, but this
picture was challenged by the (unexpected) discovery of a pulsed radio
counterpart in some sources, a property that seems to be peculiar to
transient magnetars \cite[][see also Sec.~\ref{radiomag} and references
therein]{gelgen07}. When detected, the radio emission of magnetars appears
to be different from that of standard radio-pulsars: the spectrum is
flatter and the flux and pulse profile show strong variations with time,
indicating that the mechanisms causing the emission (or the topology of
the emission region) may differ in the two kinds of sources.

Association with supernova remnants or, possibly, young stellar clusters
has been proposed for a number of sources \cite[see e.g. Table~1
in][]{mere08, muno06, vr00, eik01,fig05,kl04}, which, if confirmed, leads
in some cases to a progenitor with high mass ($>20~M_\odot$), high
metallicity, and to a relatively young age $\sim 10^4$~yr for the neutron
star (see Sec.~\ref{formation}).

The magnetar paradigm that bursting activity is necessarily associated
with a high dipolar field has been revolutionized by the recent discovery
of a few full-fledged magnetars (i.e. neutron stars that displayed
bursting, SGR-like activity) with a dipolar magnetic field comparable with
that of standard radio pulsars (see Sec. \ref{transient} and references
therein). The properties of these sources are compatible with those
expected from aged magnetars, which may still retain a large toroidal
field in the interior, occasionally capable of cracking the star's crust.
This discovery suggests that magnetars could be far more numerous than
previously expected \cite[][]{rea10, tiengo13}, and has had a number of
profound implications, e.g. for star formation, supernovae, gamma ray
bursts \cite[see][for a complete discussion]{rea14b}.

Despite the wide interest in the astrophysical community, review
papers on magnetars were comparatively few, and mainly devoted to the
diverse aspects of their phenomenology. Theoretical results are often
scattered across many specialized papers, the comparison and
interpretation of which are quite a challenge even for an informed reader.
It is outside the scope of this paper to provide a detailed summary of
magnetars' observational properties, about which excellent review papers
have been already published \cite[][]{woodth06, kas07, mere08, hur11b,
reaesp11}; an updated list of sources, containing all the essential data,
is available in the online McGill magnetar catalogue\footnote{The on line
McGill catalogue can be found at \\
http://www.physics.mcgill.ca/\~~pulsar/magnetar/main.html.}
\cite[][]{ol14}, and while preparing this review we also created a
Magnetar Burst Library which is now maintained by the Univ. of Amsterdam
\footnote{See the Amsterdam Magnetar
Burst Library, http://staff.fnwi.uva.nl/a.l.watts/magnetar/mb.html}. Here
we will focus on the theoretical
interpretation of
the emission properties of magnetars and on a cross comparison of the
models presented so far to describe them. A brief summary of the
observational properties, which is not necessarily complete but sets the
context for the subsequent discussion, is placed at the beginning of each
section, when needed. Our main aims are to review the state of the art in
the theoretical modelling, to outline which observational facts are
robustly explained by current models and to discuss the open issues which
still remain to be addressed.

The paper is organized as follows. We begin with a summary of the
mechanisms that can lead to the birth of a highly magnetized neutron star,
discuss how magnetars evolve and briefly touch the link between magnetars
and other classes of Galactic, isolated neutron stars
(Sec.~\ref{magevol}). Sec.~\ref{pers} is dedicated to the twisted
magnetosphere model and its
ability to explain the observed persistent emission in different
wavebands. Transient magnetars and their observations in the radio band
are reviewed in Sec.~\ref{transient}, while Sec.~\ref{bursts} contains a
thorough discussion of burst emission and magnetar seismology. Conclusions
follow.

\section{Birth and evolution of a magnetar}
\label{magevol}

\subsection{Magnetars formation}
\label{formation}

According to the original picture by Duncan and Thompson
\cite[][]{dt92,thdun93}, ultra-magnetized neutron stars form
through magnetic field amplification by a vigorous dynamo action
in the early, highly convective stages. Rotation and convection
produce two types of dynamo effects in an astrophysical plasma:
the $\alpha$ dynamo, arising from the coupling of convective
motions and rotation, and the $\omega$ dynamo, driven by
differential rotation. In proto NSs both effects are present and
since the $\alpha$-$\omega$ dynamo operates at low Rossby numbers,
the initial spin period must be short, $\lesssim 3$ ms, to ensure
efficient convective mixing \cite[][]{dt92}. Magnetars would be,
then, the endpoint of the evolution of massive stars with rapidly
rotating cores.
Rapidly spinning, collapsing stellar cores are
expected to produce highly energetic supernovae
\cite[][]{dt92,th04,buc07}, because a significant fraction of the
rotational energy, $E_{rot}\sim 3\times 10^{52} (P/\rm{1\,
ms})^{-2}\ \rm{erg}$, is transferred to the ejecta via the strong
magnetic coupling with the proto-neutron star. Any observational
signatures that magnetars were born in (above-average) energetic
events were searched for in a number of supernova remnants
positively associated with SGRs/AXPs, but no evidence has yet been
found \cite[][]{vk06,v08}. If the internal magnetic field is $\sim
10^{16}\ \mathrm G$, however, rotational energy can be efficiently
carried away by gravitational waves, which do not interact with
the ejecta \cite[][]{dallosso09}.

Alternatively, it has been suggested that ultra-strong fields in
neutron stars result from magnetic flux conservation \cite[the
fossil field scenario;][]{fw06,fw08}. \cite{fw06}, starting from a
parameterized model of the distribution of magnetic flux on the
main sequence and of the spin period of neutron stars at birth,
derived the expected properties of isolated radio pulsars in the
Galaxy, given the spatial distribution of the initial mass
function and star formation rate. Comparison with the data in the
1374-MHz Parkes Multi-Beam Survey was then used to constrain the
model parameters. They find that the distribution of the magnetic
field in the core of the OB progenitors comprises $\sim 8\%$ of
stars with a magnetic field in excess of $\sim 1000$ G. The
core-collapse supernovae of these high-field stars can produce
$\sim 25$ magnetars with properties (surface magnetic field, spin
period, age) in agreement with those observed in SGRs/AXPs. As
first noted by \cite{spruit08}, the number of Galactic magnetars
predicted by the fossil field model may be too low, and this is a
more and more serious issue, given the steady increase of the
magnetar population and the possibility that many ``dormant''
SGRs/AXPs lurk among ``standard'' radio pulsar (see
Sec.~\ref{transient}). A possibility is that magnetars are formed
through different channels: for instance it has been suggested
that at least part of the magnetars may be born as rapidly
rotating neutron stars in systems in which the core of the
collapsing star was accelerated by tidal synchronization in a very
close binary \cite[][]{popro06,bog09}.

Interestingly, the high-field progenitors of magnetars
should be in the far end of the mass distribution of OB stars,
with masses $\sim 20$--$45M_\odot$, which, in standard
evolutionary models, should mostly have given rise to black holes
\cite[][see also \citealt{clark05}]{fw08}. The notion that
magnetars descend from massive stars (typically above the
canonical neutron star-black hole divide) received further support
from the observational evidence that (some) SGRs/AXPs are
associated with young clusters of massive stars.  The progenitor
mass of SGR 1806-20 and the AXP 1E 1048.1-5937 has been estimated
to be in excess of $\sim 30M_\odot$ \cite[][]{bibby08,gaens05a};
the progenitor mass of SGR 1900+14 appears, however, to be $\sim
17 M_\odot$ \cite[][]{clark08,davies09}. One of the
strongest evidence in favour of high-mass progenitors of SGRs/AXPs
came from the robust association of the AXP  CXO J164710.2-455216
with the young cluster Westerlund 1 \cite[][]{muno06}. Since the
cluster is $\sim 4$ Myr old \cite[][]{clark05}, the minimum mass
of a star that could have reached the supernova stage is $\sim 40
M_\odot$. Hence the claim that CXO J164710.2-455216 originated
from a star with $M\gtrsim 40 M_\odot$. Very recently
\cite{clark14} proposed that CXO J164710.2-455216 was born in a
massive binary and found evidence for the former companion, the
runaway star Wd1-5, ejected from the system when the magnetar
progenitor exploded. If Wd1-5 and CXO J164710.2-455216 were indeed
related, evolution in the binary would lead to a decrease of the
progenitor mass through strong mass loss when it entered a
Wolf-Rayet phase, and common envelope evolution would prevent
spin-down of its core. Magnetar birth in a binary may then be a
key ingredient to bring the progenitor mass within the neutron
star formation range, and to provide the high core rotational speed
required for the onset of the convective dynamo.

Magnetars are also increasingly popular as the central engine
powering GRBs, following the original
suggestion by \citet[][see also \citealt{zhan01,metz11}]{usov92}.
Ultra-magnetized neutron stars have been invoked to explain the
properties of both short and long GRBs. The proto-magnetar would
result from coalescence in a double-degenerate binary (or
accretion-induced collapse of a white dwarf) in this first case
and in a core-collapse supernova in the second
\cite[e.g.][]{pac86,ross03,giaco13,metz08,woo93,macfad99}. Indeed,
a significant fraction of the {\it Swift} long GRBs exhibit late
flares and plateau phases in the lightcurve that provide evidence
for longevity and on-going activity of the central engine
\cite[see e.g.][]{cur08, mar10, ber11, nou06, zhan06}. The
plateau, which occurs around $10^2-10^4$~s after the trigger, has
a fluence that can be as high as the fluence of the prompt
emission. According to the magnetar model the plateau phase is
powered by the initial  spin-down of a newly born magnetar,
powering a relativistic wind \cite[][]{fa06,tro07, ly10, dal11, ber12,
metz11}. Moreover, \cite{row13} have recently shown that 18 of the
{\it Swift} short GRBs (i.e. $64\%$ of the entire sample) can be
clearly fitted with a magnetar plateau phase, while for the rest
the quality of the data is  insufficient to prove or exclude the
presence of the plateau. Out of 18 robust candidates, 10 are
thought to collapse later to a black hole, while the others may
have left behind a rapidly rotating new magnetar. Although these
studies are not a direct, conclusive proof of the magnetar
paradigm, they certainly indicate the frequent occurrence of late
central activity, which has crucial implications for the origin of
the central
engine.
A smoking gun that may allow to differentiate between
models would be the detection of gravitational waves associated to
the event \cite[][and references therein]{row13}.

Another link between magnetars and GRBs has been proposed
following the observations of giant flares. Since all these events
started with an initial, very energetic sub--s spike, it has been
proposed that giant flares, if emitted by extragalactic SGRs, may
appear at Earth as short gamma-ray bursts \cite[][]{pal05,
hur05,hur11a}. The main causes of uncertainty for proving this
idea are in maximum energy released in the flare and in the
spectral properties of the narrow peak. By considering the flare
emitted by SGR 1806-20 and by varying the assumptions about the
peak spectral shape, \cite{po06} computed the possibility of
detection by BATSE of giant flares with an energy of $10^{44}$ or
$10^{46}$~erg, as a function of the distance. They found that the
first kind of event can be seen up to a few Mpc (therefore in M82,
M83, NGC253 and NGC4945), while the second can in principle be
visible up to the Virgo cluster. However, as already noted by
\cite{po06}, this prediction may be too optimistic, since no
evidence has been found for an excess of BATSE short GRBs
from the direction of M82, M83, NGC253 and NGC4945, nor from the
Virgo cluster \citep{pal05}. Similarly, negative results have been
reported by \cite{la05, tan05}, and overall these studies suggest
that no more than a few percent, maybe up to $\sim 8\%$ of the
short GRBs seen by BATSE could be giant flares from extragalactic
SGRs \cite[see also][the latter for a recent update on the detection
upper limits]{hur05, cro11, svin15}.

\subsection{Magneto-thermal evolution}
\label{magtherm}

A major issue in establishing the magnetic evolution of NSs (and
of magnetars in particular) is that observations place very
little, if any, constraint on the structure and strength of the
internal magnetic field. Clearly, in a magnetar the internal field must be strong enough to sustain
the source activity and its geometry must allow magnetic energy to be released.  While there are several indications that
the large-scale, external field can be reasonably assumed to be
(nearly)  dipolar, the internal field most likely contains both
toroidal and poloidal components \cite[e.g.][and Sec.
\ref{twist}]{gep04,gep06}. 
A further complication comes from the at
present poor knowledge of where the internal field resides. The
field can either permeate the entire star (``core'' fields), or be
mostly confined in the crust (``crustal'' fields), depending on
where its supporting (super)currents are located.

The more general configuration for the internal field in a NS will
be, then, that produced by the superposition of current systems in
the core and the crust. As stressed by \cite{geppons07}, the
relative contribution of the core/crustal fields is likely
different in different types of NSs. In old isolated radio pulsars, where
no field decay is observed, the long-lived core component may
dominate, while a sizeable, more volatile crustal field is
probably present in magnetars, for which substantial field decay
over a timescale $\approx 10^3$--$10^5$ yr is expected
\cite[e.g.][]{goldreis92}. As pointed out by \cite{Glam11} ambipolar diffusion plays
little role in magnetar cores during
their active lifetime (after crystallization, the absence of
convective motions already quenches ambipolar diffusion in the
crust). Therefore, if the decay/evolution of the  magnetic field
is indeed the cause of magnetar activity, it is likely to take
place outside the core and be governed by Hall/Ohmic
diffusion in the stellar crust. Other mechanisms, e.g. flux
expulsion from the superconducting core, due to the interaction
between neutron vortices and magnetic flux tubes, are highly
uncertain and very difficult to model. For these reasons, recent
investigations of the magnetic field evolution in magnetars
has focused only on the crustal component of the field.

The relative importance of the Ohmnic decay and Hall drift is
strongly density-
and temperature-dependent. Thus, any self-consistent study of the
magnetic field evolution must be coupled to a detailed modelling
of the neutron star thermal evolution, and vice versa. This
basically means that the induction equation for $\vec B$ must
be solved together with the cooling, a quite challenging numerical
task. Early efforts in this direction used a split approach.
\cite{geppons07} studied the evolution of the field by solving the
complete induction equation in an isothermal crust, but assuming a
prescribed time dependence for the temperature. They found that
crustal  magnetic fields in NSs suffer significant decay during
the first $\approx 10^6$~yr and that the Hall drift, although
inherently conservative (i.e. alone it cannot dissipate magnetic
energy), plays an important role since it may reorganize the field
from the larger to the smaller (spatial) scales where Ohmic
dissipation proceeds faster.

\begin{figure*}
\centering
\includegraphics[width=3in,angle=0]{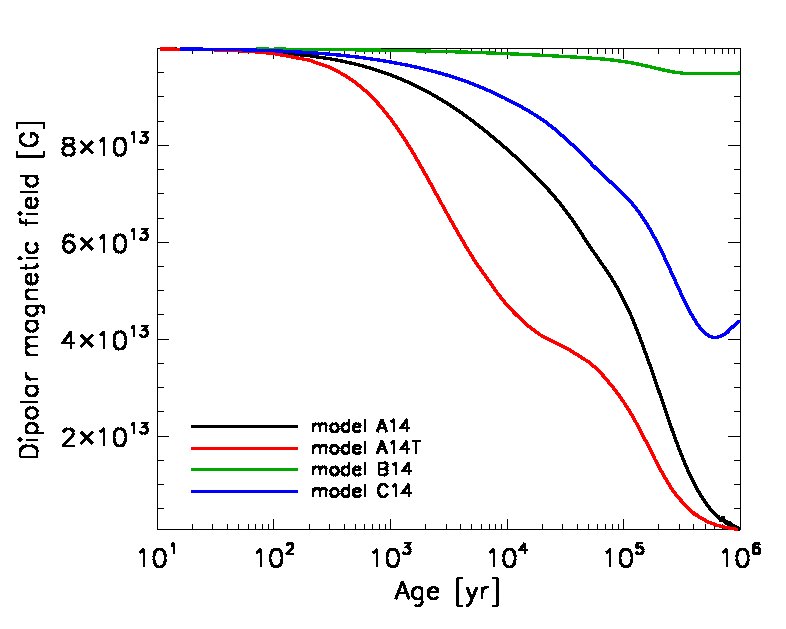}
\includegraphics[width=3in,angle=0]{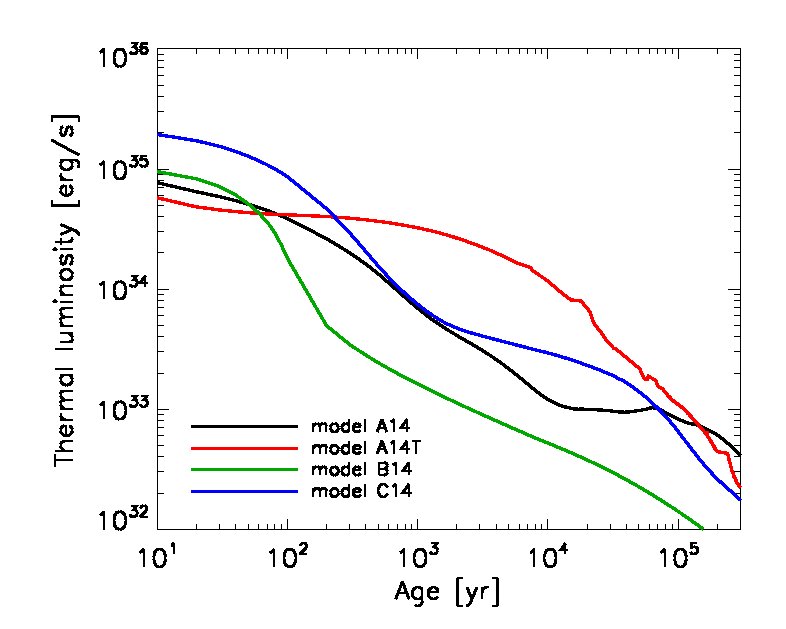}
\caption{\label{magevolfig} Left panel: evolution of the dipole
field polar strength. Right panel: same for the thermal
luminosity. In all models the initial poloidal field is $10^{14}\
\rm G$; the initial toroidal field is zero apart from model A14T,
where it is $5\times 10^{15}\ \rm G$. The star mass is
$1.4M_\odot$
\cite[from][with OUP permission]{vigano13}.} \end{figure*}

The cooling of magnetized NSs with field decay was investigated by
\citet[][see also \citealt{aguil09,kam06,kam07,kam09}]{aguil08} by
adopting a simple, analytical law for the time variation of the
field which incorporates the main features of the Ohmic and Hall
processes. The fully coupled magneto-thermal evolution of a NS was
addressed by \cite{pons09}, including all realistic microphysics.
However, owing to numerical difficulties in treating the Hall
term, their models account only for Ohmic diffusion. A complete
treatment of magneto-thermal evolution, properly including the
Hall term, was recently presented by \citet[][see also
\citealt{dan12a}]{vigano13}. Their calculations confirm the basic
picture outlined in \cite{pons09}, although the presence of the
Hall drift introduces some remarkable differences. Contrary to the
purely dissipative case, evolution is not very sensitive to the
initial relative strength of the toroidal component with respect
to the poloidal one, unless the former is much higher than the
latter. This is because a toroidal component builds up anyway due to the
Hall term, even starting with a purely poloidal
field configuration. Models are not strongly dependent on other
parameters (notably the star mass) either, so that the evolution
is mostly controlled by the initial value of the dipolar field.
Fig.~\ref{magevolfig} shows the evolution of the dipolar field
and of the thermal luminosity for different initial magnetic
geometries: core field (model B14), core+crustal field (C14) and
purely crustal field (A14, AT14). The different decay pattern of
crustal vs. core fields is evident.

\subsection{Magnetars and other neutron star classes}\label{nszoo}

\begin{figure*}
\centering
\includegraphics[width=6in,angle=0]{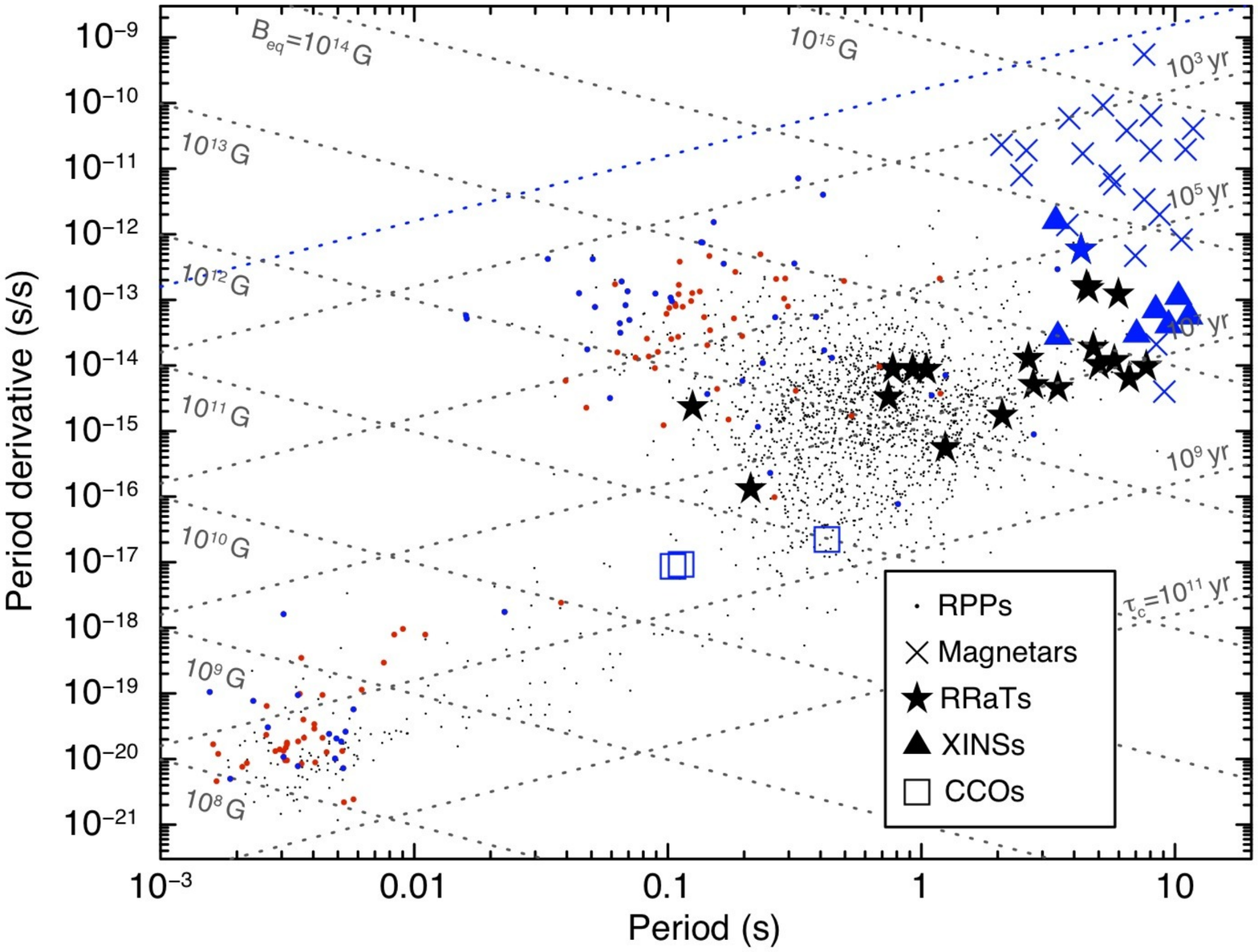}
\caption{\label{ppdot}
The $P$--$\dot P$ diagram illustrating the placement of
the different isolated neutron star classes. The blue dots mark
pulsars detected both
in the radio and X-ray bands,
the red ones those observed  only at X-ray energies. The
lines of constant age and
magnetic field are also shown (courtesy R.P. Mignani).}
\end{figure*}

Over the last two decades our picture of the Galactic neutron star
population has changed drastically, mainly thanks to high-energy
observations. Besides SGRs/AXPs, the existence of several new
classes of isolated neutron stars (INSs), with properties quite at
variance with those of ordinary radio-pulsars (PSRs), has emerged: the
central compact objects in supernova remnants (CCOs in SNRs), the
rotating radio transients (RRaTs) and the X-ray dim INSs (XDINSs
or M7) \cite[see e.g.][for
reviews]{kaspi10,hard13,delu08,ho12,burspo12,tur09}. All these
sources are radio-silent or, in the case of RRaTs (and SGRs/AXPs),
show only sporadic (transient) radio emission (see Sec.
\ref{radiomag}). They were discovered as X-ray pulsators, with the
exception of the RRaTs \cite[only one is currently known as an
X-ray source,][]{mcl07}, and their spectrum is mostly thermal.
While the periods are quite long (from $\approx 0.1$--0.4 s for
the CCOs to $\approx 1$--10 s for the XDINSs and RRaTs), their
period derivatives span a large interval, with implied magnetic
fields ranging from as low as $\sim 3\times 10^{10}\ \rm G$ in
some of the CCOs (which are sometimes referred to as the
``anti-magnetars''), to $\sim10^{12}-10^{13}$~G in RRaTs and XDINSs
\cite[see][]{keane11, tur09}. Ages are also very different, CCOs being quite young (the
associated SNR age is $\lesssim 10^4\ \rm{yr}$) and XDINSs much
older \cite[the estimated dynamical ages are $\approx 10^5 \
\rm{yr}$, e.g.][and references therein]{mign13}; in both cases the
``true'' ages turn out to be shorter than the spin-down ages. Like
PSRs, RRaTs appear to be rotationally-powered, while the (thermal)
X-ray emission from XDINSs and CCOs is powered by the release of
residual heat. The position of the various sources in the
$P$--$\dot P$ diagram is shown in Fig.~\ref{ppdot}.

Although the number of detected sources in each class is fairly
limited in comparison to that of PSRs (7 XDINSs, $\gtrsim 70$
RRaTs, 8 CCOs, about 20 SGRs/AXPs, and few candidates in each class,
vs. $> 2000$ PSRs\footnote{ATNF
pulsar catalogue,
\tt{http://www.atnf.csiro.au/research/pulsar/psrcat/}}), the
estimated birthrate of XIDNSs and RRaTs is comparable to and possibly
higher than that of PSRs, $\beta_{\rm{PSR}}\sim 0.015$--$0.03\
\rm{yr}^{-1}$ \cite[][and references therein]{pop06,kk08}. The
magnetar birthrate is lower than those of other classes,
$\beta_{\rm{mag}}\sim 0.003\ \rm{yr}^{-1}$, although this is
likely a lower limit given the increasing number of SGRs/AXPs
discovered recently.

This clearly is an issue, since the sum of the birthrates of the various INS types
cannot exceed
the core-collapse supernova rate in the Galaxy, $\beta_{\rm{SN}}\sim 0.02\pm 0.01\ \rm{yr}^{-1}$. Unless
the current figures for the INS birthrates are grossly overestimated (and/or INSs can form through other
channels), this implies that some evolutionary links exist among the different classes \cite[][]{kk08}.
That XDINSs might be aged, worn-out magnetars has been suggested
repeatedly, on the basis of the similarity
of the periods and the (relatively) high magnetic fields of the former \cite[e.g.][]{tur09}.
Besides the need to find evolutionary links among the groups, the variety of INS manifestations brings in an
even more fundamental question: which initial parameters determine whether
a proto NS will become, say, a magnetar or a PSR ?
The idea that the properties (and the evolution) of an INS are governed by a limited number of macrophysical
quantities at birth (e.g. mass, magnetic field, period) may indeed open the way to what has been called the ``grand unification of
neutron stars'', or GUNS for short \cite[e.g.][]{kaspi10,igos14}.

\begin{figure*}
\centering
\includegraphics[width=5in,angle=0]{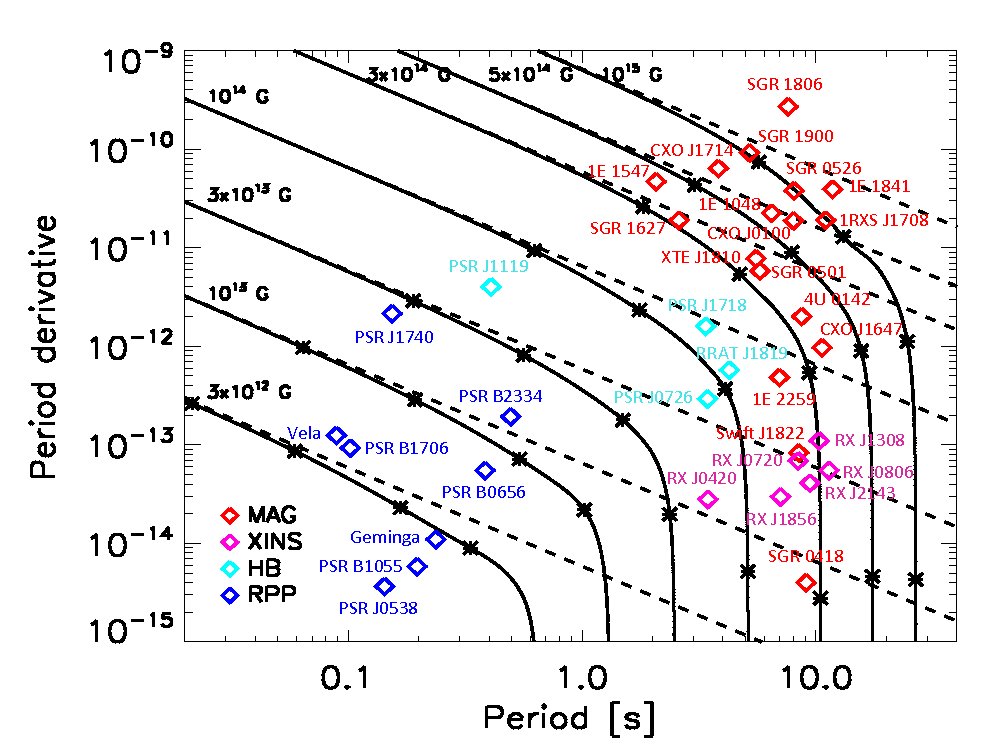}
\caption{\label{ppdot-evol} Evolutionary tracks in the $P$--$\dot P$
plane of INSs with different initial magnetic fields. Asterisks
mark the real age of the source along the track ($10^3,\,
10^4,\,10^5,\,5\times 10^5\ \rm{yr}$) and the dashed lines give
the tracks with constant $B$. MAG = SGRs/AXPs, XIN = XDINSs, HB =
high-B PSRs, RPP = PSRs \cite[from][with OUP permission]{vigano13}.}
\end{figure*}

Magnetic field decay is bound to play a central role in any
attempt to build a GUNS. \cite{pop10} were the first to perform
INS population synthesis calculations including magneto-thermal
evolution, adopting the treatment of \cite{pons09}. Their model
satisfactorily reproduces all INS populations if the initial
magnetic field follows a log-normal distribution with a mean value
$B_0=1.8\times 10^{13}\ \rm G$. Their picture confirms that the
magnetic field decays substantially (by a factor $\gtrsim 10$) in
the most magnetic stars, but provides no clear indications for
evolutionary links among the different INS groups. New population
synthesis calculations, including more updated magneto-thermal
evolutionary models, have been recently presented by
\cite{gull14}. A more decisive indication that such links indeed
exist comes from the tracks computed by \cite{vigano13} by
coupling the magnetic field and period evolution (see
Fig.~\ref{ppdot-evol}). The main effect of magnetic field decay is to
produce a sharp bending of the track downwards after a time
$\approx 10^5\ \rm{yr}$ for strong initial fields. This implies
that the star's period does not increase indefinitely but freezes at
an asymptotic value which depends on the initial magnetic field,
the mass of the star and the crust resistivity \cite[see also][]{dallosso12}. A comparison
between the theoretical tracks and the positions in the $P$--$\dot
P$ plane of INSs of different types (see again
Fig.~\ref{ppdot-evol}) suggests that ``moderate'' magnetars ($B_0=\rm{a
\ few}\times 10^{14}\ \rm G$) evolve into XDINSs.

\section{Persistent emission}
\label{pers}

\subsection{Magnetospheric twist}\label{twist}

The current picture of a magnetar magnetosphere relies on the
notion that the star's external magnetic field differs from a
simple, potential dipole, which is usually assumed to be the case
for ``standard'' neutron stars. The reason for which the external field
is not dipolar is to be sought in the structure of the internal
magnetic field. Over the last decade, analytical and numerical
investigations have shown that any stable configuration for the
internal magnetic field of a star has necessarily to contain both
a poloidal and a toroidal component
\cite[e.g.][]{Tayler73,flowers77,brsp06,brnor06,br08,br09}. In
particular, \cite{br09} investigated stable, axisymmetric magnetic
equilibria and found that the ratio of the two components must be
such that $aE/U\lesssim E_p/E \lesssim 0.8$, where $E$ and $U$ are
the total magnetic and gravitational energies, $E_p$ is the energy
associated with the poloidal component and $a$ is a numerical
factor. Given that $E/U \lesssim 10^{-23}$ and $a\approx 10^3$ for
a neutron star, its internal magnetic field likely comprises a
toroidal component at least of the same order as, and possibly
much stronger than, the poloidal one. 
The instability of purely
poloidal or toroidal magnetic configurations was proven also by
Newtonian \cite[e.g.][]{Lander11,Lander11b} and
general-relativistic \cite[e.g.][]{Ciolfi11,Ciolfi12} numerical
simulations \cite[see also][for a recent overview]{ciolfi14}.
Although, earlier attempts with the twisted-torus model \cite[a
likely configuration for the internal stellar field,
e.g.][]{brnor06} pointed towards poloidal-dominated geometries,
which are themselves unstable \cite[][and references
therein]{ciolfi14}, more recent calculations indicate that large
toroidal fields (comprising up to 90\% of the total magnetic
energy) can indeed be achieved in this framework \cite[][see also
\citealp{reisen09,ak13} for magnetic configurations with $B_{tor}\gg
B_{pol}$]{Ciolfi13}. Due to the complexity of the problem, most of
those studies considered either the internal field structure
(given an assumption for the magnetosphere) or the external
magnetosphere (assuming an internal current distribution). The
first global models, recently presented by \cite{rui14, gla14},
and \cite{pili15} in both Newtonian gravity and GR, appear
promising, although a proper analysis of their stability has not
been carried out yet.

\begin{figure*}
\centering
\includegraphics[width=3in,angle=0]{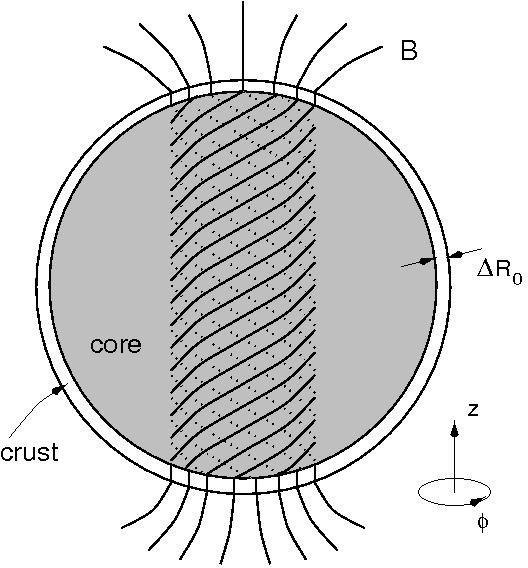}
\caption{\label{twist-gen} A schematic view of a magnetar internal field
\cite[][\copyright AAS. Reproduced with permission. A link to
the original article via DOI is available in the electronic
version]{Thompson01}.} \end{figure*}

In a magnetar, where the internal field can exceed $10^{15}\,
\mathrm G$, magnetic stresses can overcome the crustal tensile
strength \cite[][]{thdun95}. The easiest way in which the crust
reacts to the applied forces is through horizontal displacements,
parallel to the magnetic equipotential surfaces, i.e. a
magnetically-stressed crustal patch tends to rotate by an angle
$\Delta\phi$ \cite[][]{thom00}. This can be understood by
considering a flux tube in which the toroidal component is
non-zero in the crust and vanishes outside the star
\cite[][]{tlk02}. Because the conductivity is much higher in the
star's interior, the currents supporting the non-potential $B$-field
will close in a thin surface layer. The Lorentz force acting on
the current, and hence on the layer, is $\vec F_L=\vec j\times\vec
B/c=j\times(\vec B_p+\vec B_t)/c$, where $\vec j$ is the current
density. The part of $\vec F_L$ due to the toroidal field $\vec
B_t$ tries to produce a vertical displacement, which is unlikely
to occur due to the strong stratification \cite[][]{reisgold92},
while the part associated with the poloidal component $\vec B_p$
results in a slippage in the horizontal direction (see
Fig.~\ref{twist-gen}).

A direct effect of the magnetically-induced rotation of a surface
platelet is the twisting of the external field. Since the external
magnetic field lines are anchored to the crust, a torsional
displacement of the surface layers produces a transfer of magnetic
helicity from the interior to the exterior. If the external field
is initially dipolar, it will acquire a non-zero toroidal
component, a twist, confined to the field lines whose footpoints
are on the displaced layer. In a twisted magnetosphere, currents
necessarily flow also along the closed field lines to support the
non-potential field. This is at variance with what is usually
assumed to occur in ``normal'' radio-pulsars, where charges (the
Goldreich-Julian currents) move only along the open field lines
(again because the $B$-field is non-potential in that region). The
presence of large-scale currents in a magnetar magnetosphere has
major implications in shaping the emergent spectrum through
repeated resonant cyclotron scatterings, as will be discussed in
Sec.~\ref{spectra}. The gradual implant and subsequent decay of a
magnetospheric twist has been often invoked to explain the long
terms evolutions of some magnetars  \cite[][]{sandro05, cam07},
for example the behaviour observed before or after a series of
bursts or a giant flare. For instance, before the giant flare
emitted by SGR 1806-20, the source properties changed remarkably:
a study of the observed long-term variations indicated a clear
correlation among the increases in spectral hardening,  spin-down
rate, and bursting activity \cite[][]{tlk02,sandro05}. The
proposed scenario assumes the onset of a gradually increasing
twist: this, in fact,  results in an increasing optical depth for
resonant cyclotron scattering, and causes a progressive hardening
of the X-ray spectrum. At the same time, the spin-down rate is
expected to increase because, for a fixed dipole field, the
fraction of field lines that open out across the speed-of-light
cylinder grows. Since both the spectral hardening and the
spin-down rate increase with the twist, the model predicts that
they should be correlated in agreement with the observations.  

Although magnetospheric twists are expected to be localized,
meaning that they do not affect the entire magnetosphere
\cite[]{tlk02,belo09}, nearly all studies on the properties of the
persistent emission from magnetars rely on the ``globally twisted
magnetosphere'' first proposed by \cite{tlk02}. In this model it
is assumed that the external magnetic field is initially dipolar
and that, as a consequence of crustal displacements, a certain
amount of shear is added to the field. If one restricts to
magnetostatic equilibria in a low-density plasma, the momentum
equation reduces to\footnote{SGRs/AXPs are slow rotators and the
Coulomb force is negligible in the inner magnetosphere.} $\vec
j\times\vec B=0$, which, combined with the Amp\`ere-Maxwell
equation $\vec\nabla \times \vec B = (4\pi / c) \vec j$ gives the
force-free condition \begin{equation}
  \label{fff} (\vec\nabla \times \vec B) \times \vec B = 0. \end{equation}

\begin{figure*}
\centering
\includegraphics[width=4in,angle=0]{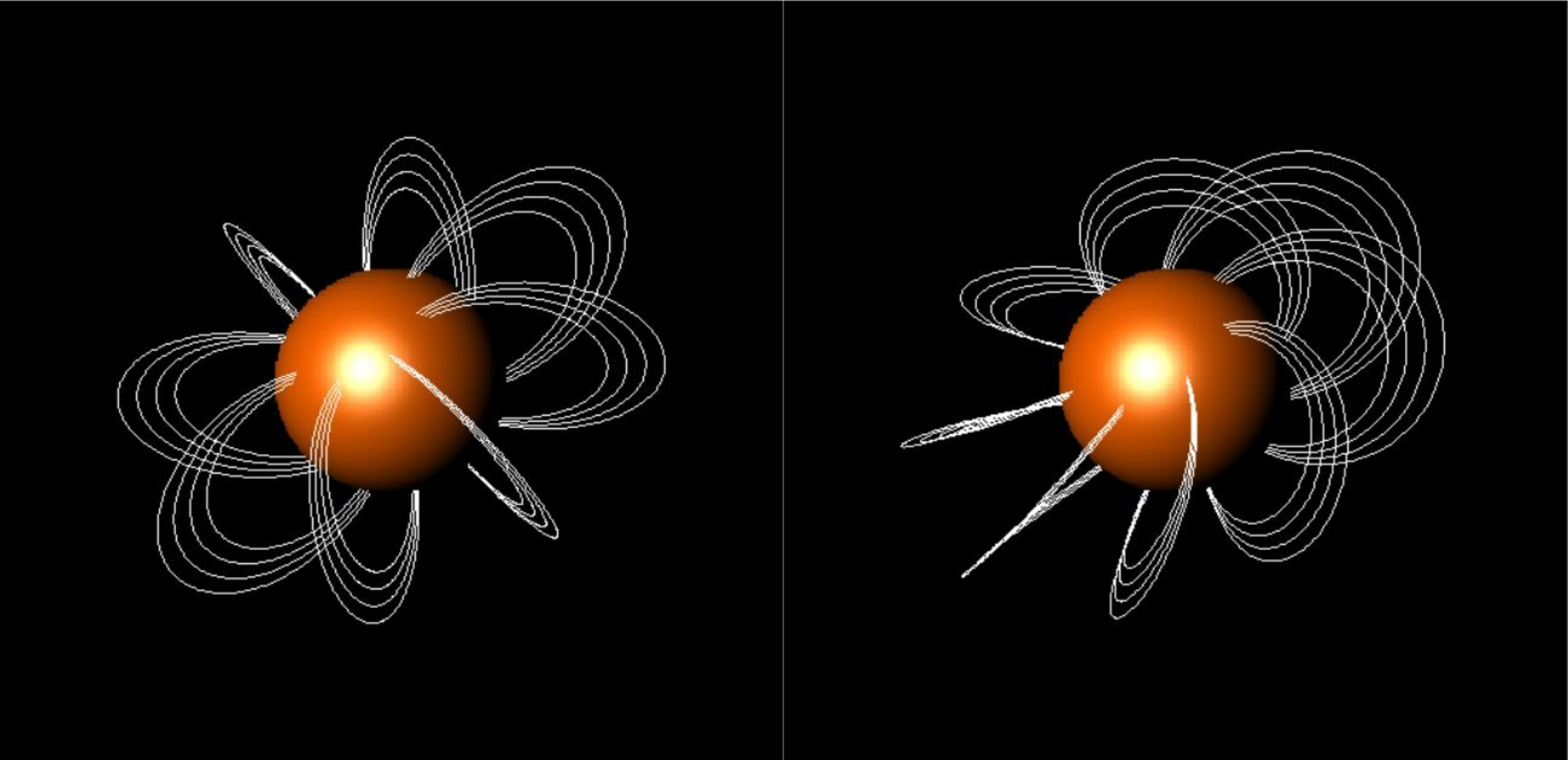}
\caption{\label{global-tw} A globally twisted dipolar field (right panel)
as compared with a pure dipole
magnetic configuration (left panel).}
\end{figure*}

By expressing the poloidal component through the flux function
$\PP$, an axisymmetric field has the most general form

\begin{equation}
  \label{axi-B} \vec B = \frac{\vec\nabla \PP(r,\theta) \times
  \vers{\phi}}{r \sin\theta}
  + B_\phi(r,\theta) \vers{\phi}\,, \end{equation} where $B_\phi$ is the
    toroidal component and $\vers{\phi}$ the unit vector in the $\phi$
    direction. By exploiting the force-free condition one can explicitly
    write the magnetic field as

\begin{equation}\label{bgen}
  \vec B= \frac{B_{p}}{2} \left(\frac{r}{R_{NS}}\right)^{-p-2} \left[-f',
  \frac{pf}{\sin\theta},
    \sqrt{\frac{C\ p}{p+1}} \frac{f^{1+1/p}}{\sin\theta}\right]
\end{equation} where a prime denotes a derivative with respect to
$\mu\equiv\cos\theta$, $B_p$ is the polar value of the magnetic
field, $R_{NS}$ is the star radius, $C$ is a constant and $0\leq
p\leq 1$ is the radial index. The function $f(\mu)$ satisfies the
Grad-Shafranov equation

\begin{equation}
  \label{ODE} (1-\mu^2)f'' + p\ (p+1)f + C f^{1+2/p}=0 \end{equation}
which is a second order ordinary differential equation for the
angular part of the flux function. Since equation (\ref{ODE}) must
be (numerically) solved subject to three boundary conditions
\cite[][]{tlk02, pav09}, the constant $C$ is an eigenvalue and is
completely specified once $p$ is fixed. The solution of equation
(\ref{ODE}) completely determines the external magnetic field and
provides a sequence of magnetostatic, globally-twisted dipole
fields by varying the index $p$. A picture illustrating a
globally-twisted dipole is shown in Fig.~\ref{global-tw}.

Besides controlling the radial decay, the value of $p$ also fixes the
amount of shear of the field. In fact, the twist angle, i.e. the
angle through which a field line has rotated when it comes back to
the stellar surface, is defined as

\begin{equation}\label{twistangle}
  \Delta \phi= \int_{{field\ line}} \frac{B_\phi}{(1-\mu^2)
  B_\theta}\,d\mu =\left[\frac{C}{p\ (1+p)}\right]^{1/2} \int_0^1
  \frac{f^{1/p}}{1-\mu^2} \,d\mu \,.
\end{equation} The effect of decreasing $p$ is to increase $B_\phi$ with
respect to the other components, and consequently to increase the
shear. The limiting values $p=0\,, 1$ correspond to a split
monopole and an untwisted dipole, respectively.

Primarily to assess the role played by the magnetic geometry on the
emergent spectra, the effects on the spectra of other sheared
magnetospheric configurations have
been investigated. In these models the helicity is not uniformly
distributed and, in a sense, they can be thought of as closer to the
realistic case in which the twist is localized. Globally-twisted
multipolar fields have been considered by \cite{pav09}, following
essentially the same approach adopted by \cite{tlk02} for the
dipole. More recently, \citet[][see also \citealt{dan11}]{dan12}
explored more general, non self-similar, force-free configurations
for the external $B$-field in which an arbitrary function is used
to control the spatial distribution of the twist. The implications
for spectral calculations will be discussed in Sec.~\ref{hard}

Once implanted by a starquake, the twist must necessarily to
decay. In a genuinely static twist ($\partial\vec B/\partial t =
0$), in fact, the electric and magnetic fields are orthogonal.
This implies that the voltage drop between the footpoints of a
field line vanishes since $E_\parallel = 0$, so that there is no
force that can extract particles from the surface and lift them
against gravity thus initiating the current required to sustain
the sheared field, $\vec j_B = c\vec\nabla \times \vec B/4\pi$. As
discussed by \citet[][see also \citealt{thom00}]{belotho07}, the
twist decays precisely to provide the potential drop required to
accelerate the charges. A non-vanishing $E_\parallel$ is
maintained by self-induction and the twist evolution is regulated
by the balance between the conduction current $j$ and $j_B$,
$\partial E_\parallel/\partial t = 4\pi(j_B-j)$. If $j<j_B$ the
magnetosphere becomes charge starved and $E_\parallel$ grows at
the expense of the magnetic field, injecting more charges into the
magnetosphere. On the other hand, when $j > j_B$ the field
decreases, reducing the current. The magnetosphere is then in a
dynamical (quasi-)equilibrium with $j\sim j_B$ over a time-scale
$<t_\mathrm{decay}$. The potential drop across a field line is
maintained close to the pair production threshold, $e\Phi\approx
1\, \mathrm{GeV}$, and the rate of magnetic energy dissipation is
$\dot E_{mag}\approx I\Phi$, where $I\approx j_B l^2$ is the
current and $l$ the linear size of the twisted region
\cite[][]{belotho07}. The magnetic energy stored in the twist is
$E_{mag}\approx I^2R_{NS}/c^2$, and the twist decay time
$t_{decay}\approx E_{mag}/\dot E_{mag}$ turns out to be $\approx
1\, \mathrm{yr}$ for typical parameter values. 
The detailed evolution of a twisted magnetosphere has been investigated by \cite{belo09}.

\subsection{Current distribution}\label{currents}

A twisted magnetosphere can be regarded as a force-free
configuration, threaded by currents that flow along the $B$-field
lines with $\vec j\sim \vec j_B$. Charges are extracted from the
star's surface and accelerated by the electric field parallel to
$\vec B$. In the simplest picture \cite[][]{tlk02}, the charge
flow consists of two counter-streaming currents: electrons and
ions moving in opposite directions, so that charge neutrality is
ensured. Using a simple, unidimensional circuit analogue, a
twisted flux tube is akin to a relativistic double layer
\cite[][]{belotho07,carl82}, in which electrons/ions leave the
anode/cathode and are accelerated by a potential drop, which, in
turn, depends on the current. \cite{belotho07} pointed out that
such a configuration cannot be realized in the magnetosphere of a
magnetar. In order to produce $j\sim j_B$, in fact, the Lorentz
factor of the electrons needs to be sufficiently high ($\gamma\approx
10^9$)
to make one-photon pair production in the strong magnetic field
through resonant cyclotron up-scattering unavoidable well before
$\gamma$ attains such large values. Currents are expected to be
carried mostly by pairs, the corona being in a state of
self-organized criticality with a voltage drop near the threshold
for the ignition of pair cascades.

The analysis by \cite{belotho07} revealed much of the (complex)
physics of a twisted magnetosphere. Still, being based on an
idealized circuital model, it was not particularly suited for
being used in spectral modelling. For this reason most
investigations in this direction have resorted to the simpler, albeit
less physically sound, picture of electron/ion currents. Under
this assumption and having specified the magnetic configuration,
the density of magnetospheric particles is automatically fixed
once the particle velocity is known. In particular, for a
force-free globally twisted dipolar field
\cite[][]{tlk02,ft07,ntz08a}, the charge density can be derived
from the condition $j=j_B$

\begin{equation}\label{magcurr} n_e ( \vec r, \beta ) = \frac{ p + 1 } {4
\pi e} \left ( \frac{B_\phi }{B_\theta} \right ) \frac {B }{r\vert
\langle \beta \rangle\vert } \, , \end{equation} where
$\langle\beta\rangle$ is the average charge velocity (in units of
$c$). The previous expression gives the co-rotation charge density
of the space charge-limited flow of ions and electrons from the NS
surface, that, due to the presence of closed loops in a twisted
field, is much larger than the Goldreich-Julian density, $n_{GJ}$.
In a general scenario, positive and negative charges (with
densities $n_\pm$) flow in opposite directions with velocities
$v_\pm$ and $j=j_B = e \left (v_+ n_+ - v_-n_- \right ) $, where
$v_+v_- < 0$. Electrons are assumed to flow from north to south
and conversely for ions. This breaks the symmetry between the star's two
hemispheres, and, for instance, implies that the observed
spectrum will be different when viewed from the north or the south
pole \cite[see][]{ntz08a}. The presence of ions introduces
negligible effects on the continuum spectra. Photons may still
scatter off ions, which are heavier and concentrated toward the
star's surface, but this is likely to give rise at most to a narrow
absorption feature at the ion cyclotron energy \cite[][and
discussion in Sec.~\ref{lowb}]{tlk02,ft07,tiengo13}. For this
reason, these models are often referred to as ``unidirectional
flows'', with reference to electrons only, while the term
``bidirectional flows'' is used when pairs are accounted for.

In the absence of any detailed modelling of the current flow (e.g.
particle acceleration, interaction of charges with radiation
traversing the magnetosphere), the velocity distribution is
assumed to be spatially independent so that the charge velocity is
a free parameter of the model. A major difference between the
various models \cite[][]{ft07,ntz08a} is in the adopted
description of the velocity distribution of the scattering
particles. In a strong magnetic field the electron distribution is
expected to be largely anisotropic: $e^-$ stream freely along the
field lines, while they are confined in a set of cylindrical
Landau levels in the plane perpendicular to $\vec B$. For this
reason, \cite{ntz08a} assumed a collective (bulk) electron motion
with velocity $v_{bulk}$ associated with the charge flow in the
magnetosphere, superimposed on a 1-D relativistic Maxwellian
distribution at a given temperature $T_e$ which simulates the
particle velocity spread (and the dissipation due to local
turbulence and possible instabilities).

The (invariant) distribution function is then

\begin{equation} \label{distribf} \displaystyle{\frac{d n_e}{d
(\gamma\beta) }} = \displaystyle{ \frac{n_e\,
\exp{(-\gamma'/\Theta_e )} }{2 \, K_1(1/\Theta_e)}} = n_e f_e(\vec
r , \gamma \beta) \end{equation} where $\gamma' =
\gamma\gamma_{bulk} (1-\beta\beta_{bulk} )$, $\Theta_e = kT_e/m_e
c^2$, $K_1 $ is the modified Bessel Function of the first order
and $f_e = \gamma^{-3} n_e^{-1} d n_e/d\beta$ is the momentum
distribution function.

In this model electrons are, then, assumed to move isothermally
along the field lines, whilst at the same time receiving the same boost
from the electric field. This is at least in qualitative agreement
with the results of the simplified bidirectional model by
\cite{belotho07}.  A different choice was made by \cite{ft07}, who
did not include the charge bulk motion in their models (despite
this being a necessary ingredient to reproduce the current flow) and
assessed the effects of other possible (local) distributions,
either thermal or not thermal, in a few representative
cases. In particular, they considered: \\
a) a mildly relativistic, 1-dimensional flow described by a
Boltzmann distribution at a temperature $k_{\rm B}T_0 =
(\gamma_0-1)m_ec^2$ \begin{equation}\label{eq:boltzone}
f(\beta\gamma) = {1\over
K_1(1/[\gamma_0-1])}\exp\left[-{\gamma\over {\gamma_0-1}}\right]
\, ; \end{equation} b) a mildly relativistic, one-dimensional gas
with the same Boltzmann distribution but extending over positive
and negative momenta, in order to simulate an electron-positron
flow; and \\ c) a broad power-law in momentum,
\begin{equation} \label{eq:broad_dist} f(\beta\gamma) \propto
(\beta\gamma)^{\alpha}, \end{equation} which mimics a warm
relativistic plasma.

As we discuss in the next section, charges must flow at mildly
relativistic speed ($\gamma\simeq 1$) in the twisted magnetosphere
for the model to successfully reproduce the observed X-ray
spectra. While this is not a problem for the (over) simplified
unidirectional flows discussed earlier on, where the velocity is a
tunable parameter, the question of what occurs in a more realistic
description which includes pairs is a crucial one. As discussed by
\cite{belotho07}, in a twisted magnetosphere electrons and ions,
lifted from the star's surface and accelerated by the
self-induction electric field, must efficiently produce $e^\pm$,
at least if the current circulating in the circuit is $\sim j_B$.
According to their analysis, $e^\pm$ flow with highly relativistic
speed ($\gamma\approx 10^3$) and a large velocity spread in the
inner magnetosphere ($r\sim R_{NS}$). This poses a problem for the
mildly-relativistic, counter-streaming model which is only valid
in the (unphysical) assumption that pair production is neglected
\cite[][]{belo13a}. On the other hand, in the presence of pairs,
the electric field along the $B$-lines, $E_\parallel$, is
incapable of counteracting the radiative pull outwards because, at
the same time, it acts as an accelerator for the charges of
opposite sign. The result is that charges are accelerated outwards
at relativistic velocity and no self-consistent solution yielding
mildly relativistic flows is possible.

Pair production in a twisted magnetosphere has been investigated
in several works. As discussed by \cite{ml07}, for an iron crust
and magnetic fields as high as $\sim 10^{15}$G, vacuum gaps may be
formed above the polar regions of SGRs/AXPs, with subsequent pair
creation. Near the stellar surface, where the magnetic field $B$
exceeds the quantum limit $B_Q \sim 4.4 \times 10^{13}$~G,
scattering between fast electrons and $\sim 1$ keV seed photons
generates high-energy gamma rays that immediately convert to
electron/positron pairs via one-photon pair production (e.g.
\citealt{Harding+Lai}). This idea, originally proposed by
\cite{belotho07}, has been more recently reconsidered in detail by
several teams \citep{nob11, currentsnoi}. The main point is that
single photon pair production requires photons with energy higher
than the threshold value, $\sim $~1 MeV. Therefore, in a region
dominated by resonant scattering, pair creation occurs in two
steps: (i) a seed photon with energy $\epsilon \sim$~1 keV is
up-scattered by a relativistic particle with
$\gamma=\gamma_{res}\sim (m_ec^2/\epsilon)(B/B_Q) \sim 1000$,
where $\gamma_{res}$ is the charge Lorentz factor at resonance,
gaining a considerable energy
$\epsilon'\sim\gamma_{res}^2\epsilon/(1+\gamma_{res}\epsilon/m_ec^2)$;
(ii) quite immediately, the high-energy photon converts to a
$e^\pm$ pair, via single photon pair production. As discussed by
\cite{currentsnoi}, the pair-dominated region is very thin and
located just above the star's surface where $B > 0.05 B_Q$. Here a
quasi-equilibrium configuration is reached with a pair
multiplicity $\sim L/\lambda_{acc,res}$ of a few, where $L$ is the
length of the field line and $\lambda_{acc,res}$ is the distance
travelled by a charge before reaching a Lorentz factor
$\gamma_{res}$. Screening of the electric field limits the
potential drop to $e\Phi_0/m_ec^2\approx\gamma_{res}\sim
500(B/B_Q)$ and the maximum $e^\pm$ Lorentz factor is
$\gamma_{res}$. Charges undergo only a few scatterings with
thermal photons, but they lose most of their kinetic energy in
each collision. In practice, the result is that a steady situation
is maintained against Compton losses because electrons and
positrons are re-accelerated by the electric field before they can
scatter again.
The newborn charges are accelerated by the huge electric field that permeates the magnetosphere up to a
limit value, so that a cascade of pairs is generated. This runaway
process limits the value of $\gamma$ to the threshold value for
pair production, $\sim$1000. Since pairs with $\gamma\sim
\gamma_{res}$ are injected from this inner region into the
external region, the circuit represented by the field lines
behaves quite differently from a double layer, allowing the
current to be conducted with only a small potential drop (see also
\citealt{belo11}).

\begin{figure*}
\centering
\includegraphics[width=3in,angle=0]{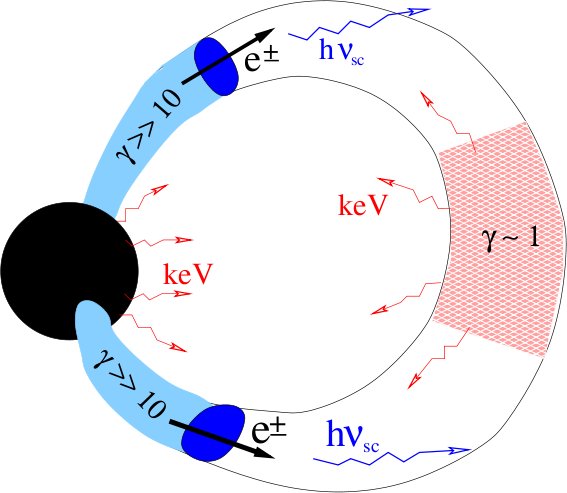}
\caption{\label{curr}
Sketch of an activated magnetic loop as proposed by \cite{belo13b}.
Relativistic particles are
injected near the star where $B>B_{QED}$.
Large $e^\pm$ multiplicity ($M \sim 100$)  develops in
the adiabatic zone $B>10^{13}$~G (shaded in blue).
The outer part of the loop is in the radiative zone; here the scattered
photons escape and form the
hard X-ray spectrum. The outflow decelerates
(and annihilates) at the top of the loop, shaded in pink;
here it becomes very opaque to the thermal keV photons flowing from the
star
\cite[from][
\copyright AAS. Reproduced with permission. A link to
the original article via DOI is available in the electronic
version]{belo13b}.}
\end{figure*}

A detailed investigation of charge distribution in a twisted
magnetosphere, based on analytical considerations and corroborated
by numerical tests has been recently presented by \cite{belo13a, belo13b}.
This work confirms the presence of an inner region with intense
pair production. This region consists of two parts. The innermost
one, where $B\gg B_Q$, is self-organized maintaining the near
critical condition of pair production with multiplicity $M \sim
1$, and here the circuit operates as a global discharge (i.e. the
accelerating voltage, which is screened by pairs, is distributed
smoothly along the field line). Field lines that extend to larger
distance from the star's surface enter an outer corona, which
extends until $B\sim B_Q$, where both scattering and pair
production are much more efficient and $M\sim 100$. Here, due to
efficient radiative coupling, plasma and radiation organize
themselves into a ``locked'' outflow with decreasing Lorentz
factor.  This leads to the formation of an extended equatorial
zone in the outer corona, where the flow is slowed down by the
combined effect of a large radiation drag and the onset of a
two-stream instability with consequent strong Langmuir
turbulence. The pair density is near annihilation balance and the
charges, decelerated down to mildly relativistic velocities,
creates an opaque layer which efficiently up-scatters the soft
X-ray photons by distorting the surface thermal spectrum. Outside
this equatorial region, and further away in the extended external
magnetosphere, charges flow at ultra-relativistic velocity and
scattering is relatively inefficient. The charge distribution is
illustrated in Fig.~\ref{curr}. Despite this being the most
complete study of magnetospheric currents presented so far, it
contains some drastic simplifying assumptions. The pair
multiplicity, for example, is a constant parameter assumed a priori. This
clearly affects many of the model results: local screening of the
electric field, velocity distribution of the two species,
development of the two-stream instability, efficiency of the
radiative drag, and ultimately the formation of a zone filled with
slowly moving charges. The challenging problem to find a fully
consistent solution of current dynamics including the interaction
with the radiation field remains so far unsolved.

\subsection{Soft and Hard X-ray spectral modelling}\label{spectra}

\subsubsection{Soft X-ray spectral modelling}\label{rcs}

The soft X-ray band ($0.3-10$~keV) is the energy range in which magnetar
spectra are best studied, thanks to a large amount of observations that
have been collected in more than two decades with X-ray satellites such as
{\it XMM-Newton}, {\it Chandra}, {\it Swift}. The observed spectra are
generally fitted by a double component model consisting of a thermal
component
(a blackbody, at about $\sim 0.5$~keV) and a steep power law (photon index
$\sim$
3-4) \citep{mere08}. In a few cases (a notable example being the AXP XTE
J1810-197) good spectral fits are also obtained with two blackbodies
\citep{hal05}. The thermal component, which often dominates in the lowest
energy band, most often has an inferred emission region (for the best
estimated distances) much smaller than the whole surface of the NS. These
spectral fits are of course phenomenological descriptions, but they
indicate that, although the emission is mostly thermal, it is more complex
than a blackbody at a single temperature. It is also interesting to note
that when one compares the average temperature (from the thermal
luminosity) of magnetars with those of other classes of isolated neutron
stars, there is a clear correlation between temperature and
magnetic field \cite[][]{aguil08} and the magnetars are systematically
more
luminous than rotation-powered neutron stars of comparable characteristic age \cite[see
a discussion in][]{sandro14}.
The morphology of the soft X-ray
pulse profiles of magnetars is varied. A few sources exhibit an (almost
symmetric) double-peaked light curve \cite[e.g. 1E 2259+586 and 4U
0142+0162;][]{pat01, wo04, rea07} with pulsed fraction in the range
$10-20\%$. For most other magnetars, instead, the pulsed component is
single peaked and often the pulsed fraction is high \cite[see e.g. 1E
1048.1-5937, XTE
J1810-197, 1E~1547.0-5408, SGR 0418+5729, and SGR
J1822.3-1606][]{tam08,bern09, bern11, hal11, dib12, rea13, rea12}. As
originally suggested by \cite{mawh01}, as a general rule sources with
larger spin-down rate have smaller photon index in the soft X-ray band.
This fact has been confirmed with more recent data and it appears to be
valid for both persistent and transient sources in outburst, but only for
rotational frequencies derivatives
$\dot \nu \gtrsim 10^{-14}$~s$^{-2}$, and with some exceptions
\cite[including the transients in outbursts and the recently discovered
low-B magnetars, see Sec.~\ref{lowb} and][]{sandro14}.
The long term evolution of the power law component and timing properties
of SGR~1806-20 indicates that the same correlation between spectral
hardness and average spin-down rate also holds for a single source
\cite[][]{sandro05}. On the wake of this, other correlations
between the spectral hardness and the timing properties have been
investigated. In particular, that with the dipole strength $B_{dip}
\propto (P \dot P)^{1/2}$ appears the most robust \cite[][]{kabo10}.

It has been widely suggested that the blackbody plus power law
spectral shape that is observed below $\sim10$~keV in magnetars'
spectra may be accounted for if the soft, thermal spectrum emitted
by the star's surface is distorted by resonant cyclotron scattering
(RCS) onto the magnetospheric charges. Since electrons permeate a
spatially extended region of the magnetosphere, where the magnetic
field varies, resonant scattering is not expected to give rise to
narrow spectral lines (corresponding to the successive cyclotron
harmonics), but instead to lead to the formation of a hard tail
superimposed on the seed thermal bump. This model is also in general
agreement with the hardness-$\dot P$ or hardness-magnetic field
correlation
mentioned above: stronger and more twisted fields yield a larger spin down
rate as well a higher magnetospheric charge density that in turn produces
a harder spectral tail.
In recent years, several teams have tested the resonant cyclotron
scattering model quantitatively against real data in the soft
X-ray range, using different approaches and under different
approximations. The first, seminal attempts in this direction were
presented by \citet{lg06} who developed a very simplified one
dimensional model. They assumed that seed photons are emitted by
the NS surface with a blackbody spectrum, and propagate backward
and forward in the radial direction. A thin, plane parallel
magnetospheric slab, permeated by a static, non-relativistic, warm
medium at constant electron density is assumed to exist above the
star's surface. Magnetic Thomson scattering occurs between photons
and the charges in the slab, and the process is computed by
neglecting all effects of electron recoil, as are those related to the
current's bulk motion. Despite being clearly over-simplified, this
model has the main advantage of being semi-analytical and, when
systematically applied to X-ray data, has proved successful in
capturing at least the gross characteristics of the observed soft
X-ray spectrum \citep{nanda1,nanda3,nanda2}.

The same model has been extended by \citet{guv07}, who relaxed the
blackbody approximation for the seed surface radiation and made an
attempt to include atmospheric effects, treating the star's surface
emission like that of a passive cooler, i.e. using an atmospheric
code akin to those originally developed for sources with purely
thermal emission \citep[e.g.][and references
therein]{zav96,sil01,fer03,pot14}. This is a quite drastic and
somewhat unphysical simplification for sources like magnetars, which
are characterized by strong magnetospheric activity leading to
particle back-bombardment, heat deposition and other similar
effects. Despite this, the model has been applied to real data in an
attempt to estimate the surface magnetic field through data
fitting of the soft X-ray continuum \citep{guv08, guv11, fer13}.
At present the problem appears to be still open: while it is
commonly recognized that thermal radiation from the star's surface
is likely to be different from a simple blackbody, either because
of local reprocessing by some sort of (non passive) atmosphere or
because the surface itself may be in a condensed state, a self
consistent inclusion of these effects in numerical models has not yet
been carried out.

\begin{figure*}
\centering
\includegraphics[height=.2\textheight]{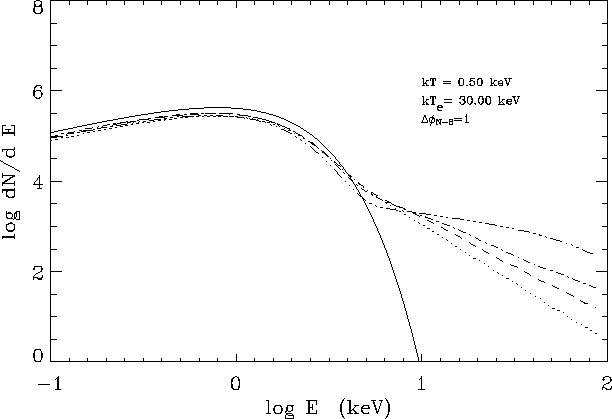}
\includegraphics[height=.21\textheight]{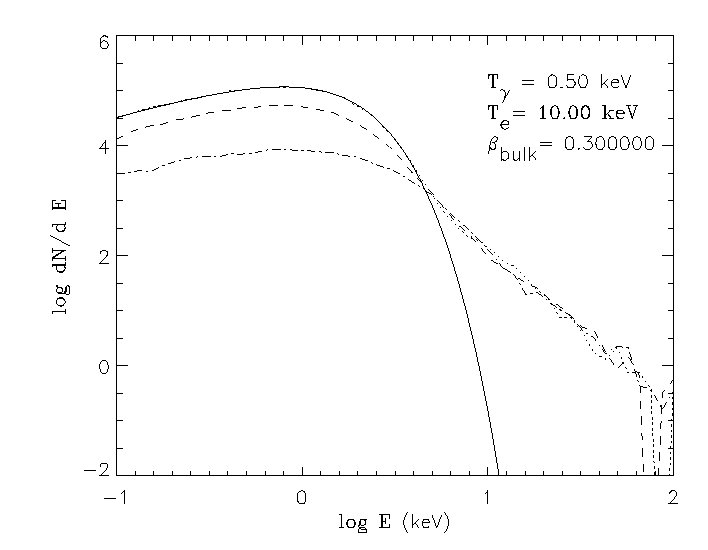}
\caption{\label{spect}
Left: Computed spectra from Monte Carlo simulations
for $B = 10^{14}$~G, $kT = 0.5$~keV, $kT_e = 30$~keV, $\Delta \phi = 1$
and different values of $\beta_{bulk}$: $0.3 $ (dotted), $0.5 $ (short
dashed), $0.7 $ (dash-dotted) and $0.9$ (dash-triple dotted). The solid
line represents the seed blackbody and spectra are computed at
a magnetic colatitude: $\Theta_s=64^\circ$. Right:
Spectrum from a single emitting patch on the star surface. The line of sight is at
$\Theta_s=90^\circ$ and
$\Phi_s=20^\circ$ (dotted line), $140^\circ$ (dashed line) and $220^\circ$
(dash-dotted line). These three values
correspond to having the emitting patch in full view (seen nearly face
on), partially in view and screened
by the star.
The solid line represents the seed
blackbody
\cite[readapted from][with OUP permission]{ntz08a}.}
\end{figure*}

In order to perform a more physical, 3-D treatment of the RCS
problem, the most suitable approach is to make use of a Monte
Carlo technique, which is quite easy to code, and, when dealing
with purely scattering media at moderate optical depths,
relatively fast. The Monte Carlo scheme allows one to follow
individually a large sample of photons, treating probabilistically
their interactions with charged particles. These simulations have
been developed by only a few teams \citep{ft07, ntz08a}. The numerical
codes that have been developed are completely general, inasmuch that in
principle they can handle different 3-D geometries (so highly
anisotropic thermal maps and magnetic fields) and different
radiative models of surface emission. On the other hand, since our
understanding of these ingredients is still limited, in order to
minimize the number of degrees of freedom, simulations were
computed by assuming, for simplicity, that i) the whole surface
emits isotropically as a blackbody at a single temperature, ii)
the magnetic field is a force-free, self-similar, twisted dipole
and iii) the electron velocity distribution is assumed a priori
(see Sec.~\ref{currents}). Besides, resonant scattering was treated
in the magnetic Thomson limit, which allows one to account for
polarization (under the two stream approximation) but it neglects
electron recoil, limiting the applicability of the results to
energies up to a few tens of keVs ($h \nu < mc^2/\gamma$~keV,
$B/B_{Q} < 10$). By comparing the results from the different
teams, one may conclude that, while the general effects induced by
magnetospheric RCS on primary thermal photons (i.e. the formation
of a ``thermal-plus-power-law'' spectrum) are not very sensitive
to the assumed particle velocity distribution, the details of the
spectral shape do, and, as we will discuss later on, this is
particularly critical for the model predictions in the hard X-ray
band. Nevertheless, in the soft X-ray band, for several
combinations of the parameters, the general shape of the continuum
is that of a thermal bump and a high-energy tail (see Fig.~\ref{spect}),
which is in agreement with what is observed in the {\it XMM-Netwon}  and
{\it Chandra}
spectra (below $\sim 10$~keV). The spectral index of the high
energy tail changes with the parameters and, in particular, harder
spectra are found for increasing twist angle. This was invoked as
a possible mechanism to explain the correlated flux-hardening long
term variations in some sources \cite[e.g.][]{tlk02, sandro05,
rea05, cam07, ntz08a}.

\begin{figure}
\centering
  \includegraphics[height=.3\textheight,angle=270]{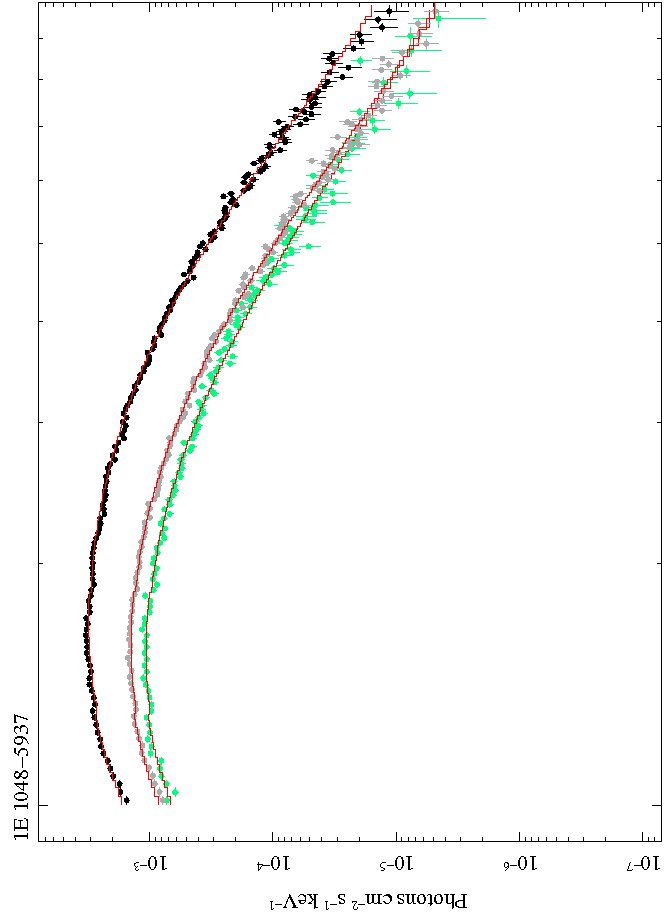}
  \includegraphics[height=.3\textheight,angle=270]{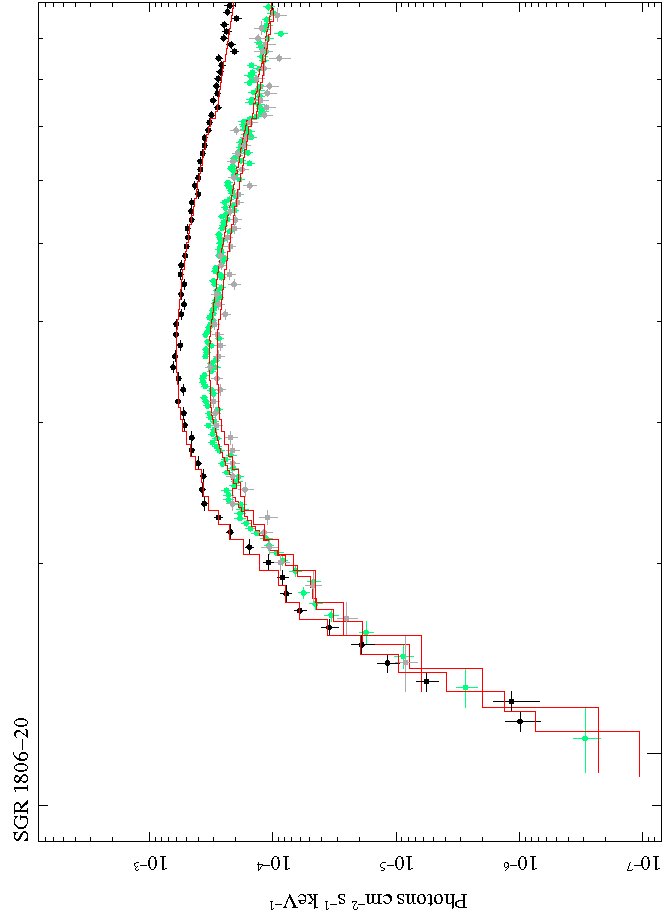}
\end{figure}

\begin{figure}
\centering
  \includegraphics[height=.3\textheight,angle=270]{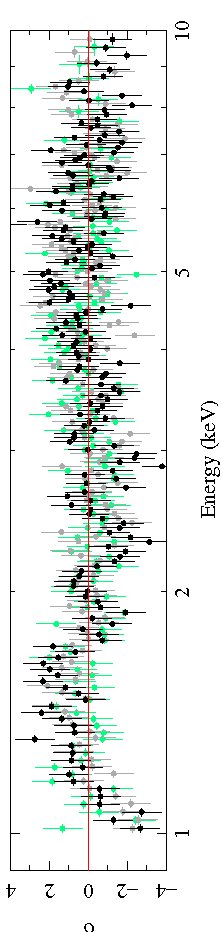}
  \includegraphics[height=.3\textheight,angle=270]{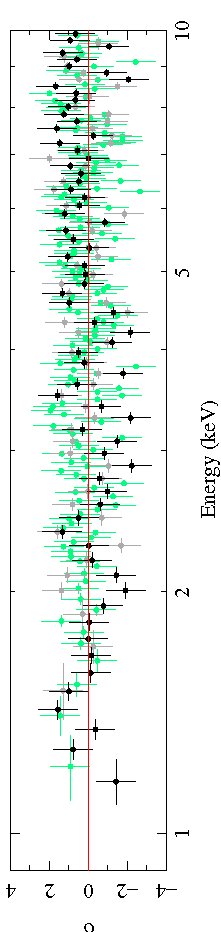}
  \caption{
Fit of the {\it XMM-Newton} spectra of the AXP
1E\,1048-5937 and SGR\,1806-20
with the NTZ model. The panels
show a joint
fit of spectra taken at three different epochs and the fitting
has been restricted to the $1-10$~keV range \cite[adapted from][with
permission]{zrtn09}.}
\label{figfit2}
\end{figure}

The numerical spectra computed by \cite{ntz08a} have been implemented in
XSPEC and successfully fit the XMM-Newton spectra of most of the magnetar
sources in quiescence \cite[NTZ model,][see also
Fig.~\ref{figfit2}]{zrtn09}. This allows one to derive of the gross
characteristics of the magnetosphere in such sources and to obtain a
better estimate of the thermal component.

Interestingly, in the case of two sources, \ea\ and \uu, it was
not possible to find a satisfactory fit with the NTZ XSPEC model
although these spectra were fitted by the simplified 1-D model
\citep{nanda1}. As suggested in \cite{zrtn09}, a plausible cause is that
the BB
peak appears to be less prominent to the observer because the
region that emits the soft seed photons is not completely in
view. By modelling the RCS spectra under the assumption that
photons are emitted by a single surface patch it is immediately clear
that the effects of the different viewing angle on the
spectrum are dramatic. When the emitting patch is in full view
both the primary, soft photons and those which undergo repeated
resonant scattering reach the observer, and the spectrum is
qualitatively similar to those presented earlier on, with a
thermal component and an extended power-law-like tail. If on the
other hand the emitting region is not directly visible, no
contribution from the primary blackbody photons is present (see
Fig.~\ref{spect}, right panel). The
spectrum, which is made up only by those photons which after
scattering propagate ``backwards'', is depressed and has a much
more distinct non-thermal shape, much more similar to the one
observed in \ea\ and \uu.

Unfortunately, in a realistic situation the thermal surface map is
expected to be complex and it cannot always be reconstructed by
fitting the X-ray spectrum alone. As discussed in \citet[][see
also \citealt{bern11}]{alba10}, a better strategy consists of
performing a simultaneous fitting of the energy-resolved
X-ray lightcurves, since they carry a much more defined imprint of
the surface thermal distribution (see Sec.~\ref{transient} and
Fig.~\ref{fig_1810} therein). While this is
not possible for
all sources, transient AXPs, for which a set of observations
spread over few years and at different flux levels are available,
provide a spectacular laboratory for this exercise. \cite{alba10}
were the first (and so far the only) team to present a
comprehensive study of the outburst decay of the two transient
AXPs (TAXPs) XTE J1810-197 and CXOU J164710.2-455216, reproducing
both the spectral and pulse profile evolution based on fits with
three-dimensional Monte Carlo simulations. This allowed them to prove
the presence of distinct temperatures zones (up to three) at the
star's surface, some of them possibly heated by the energy released
during the outburst, to model their evolution during the outburst
decay, and to constrain the viewing geometry of the sources,
i.e., the inclination of the line of sight and the magnetic axis
with respect to the rotation axis.

\begin{figure*}
\centering
\includegraphics[height=.3\textheight]{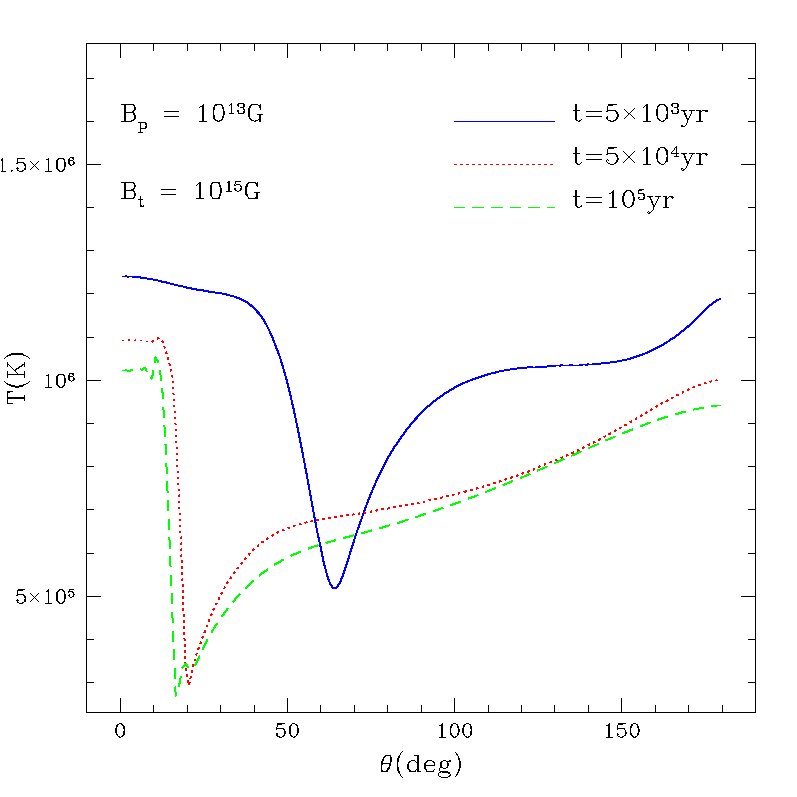}
\includegraphics[height=.3\textheight]{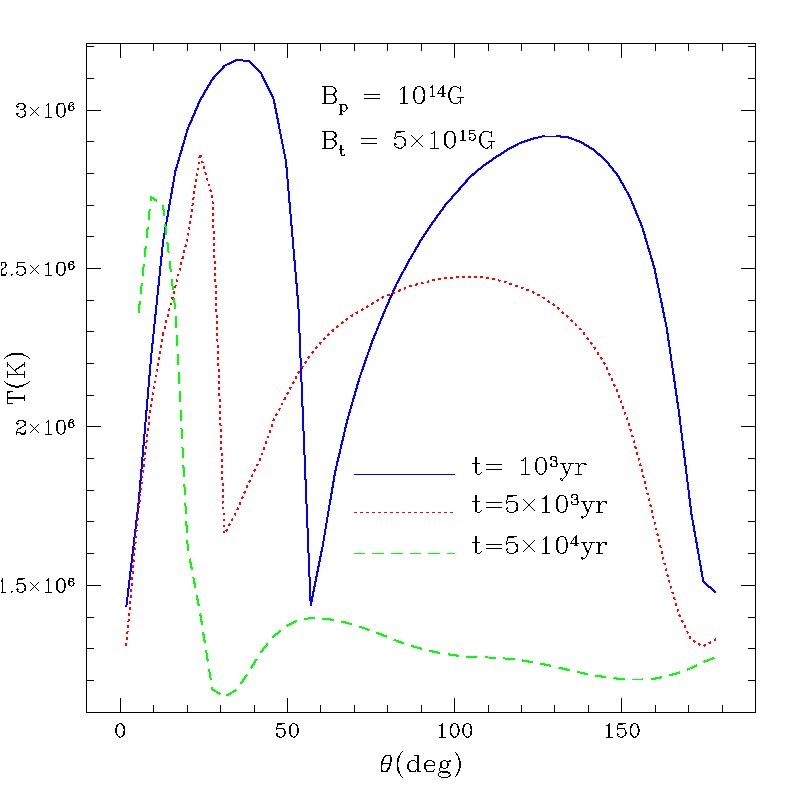}
\caption{\label{rostemp}
Temperature surface distribution for sources with a magnetic field at
birth which has both a poloidal and a toroidal component. For ages
typical of magnetars ($10^4-10^5$~yrs) the expected thermal map consists
of a hot polar spot and an extended colder region; the latter may be
undetected if the source is highly absorbed \cite[from][with
OUP permission]{perna13}.}
\end{figure*}

A similar conclusion concerning the need to investigate the pulse
profile behaviour when trying to reconstruct inhomogeneous surface
thermal maps was reached by \cite{perna13}. These authors
considered the case in which the temperature anisotropy is
observed also in quiescence, as is expected if complex magnetic
field components in the NS crust and interior make heat transport
from the core outward highly anisotropic. They used state-of-the
art numerical codes \citep{dan12,vigano13} for the coupled
magneto-thermal evolution of neutron stars and computed the
expected pulse profiles and spectra (under the assumption of
blackbody emission) for a range of magnetic configurations.
Particularly compelling is the finding that, while in presence of
purely dipolar fields the pulse profile is always double-peaked
and with a relatively low pulsed fraction, when strong toroidal
components are present the pulse fraction can exceed 50--60\% and
the pulse profile can be single peaked (as often observed in AXPs
and SGRs). Moreover, if the simulated spectra are fitted with a highly
absorbed BB model, only relatively concentrated hot peaks
are visible, so that the inferred BB radius turns out to be much
smaller than the NS radius (even as low as 1-2~km, see
Fig.~\ref{rostemp}). Strong
toroidal crustal B-field components, coupled with large absorption
column densities ($>10^{22}$~cm$^{-2}$, see e.g.
\citealt{paolo08}), can therefore explain the small caps very
often required by the spectral fits of AXPs and SGRs. Even smaller
(sub--km) hot spots are measured in certain sources (e.g.
CXO~J164710.2-455216, see e.g. \citealt{isra07}), but they look
more likely to be produced by particle bombardment and heat
deposition from currents highly concentrated in twisted magnetic
polar bundles \citep{belo09,tur11} which emerge from the crust,
rather than anisotropic internal heat transfer. In this respect it is
worth mentioning the work by
\cite[][]{buc15}, who presented a comprehensive and detailed parameter
study, in general relativity, of the role that the current distribution
investigating several equilibrium global field configurations derived
using a Grad-Shafranov approach \cite[][]{pili15}. These authours found
that the structure and strength of the magnetic field at the surface is
strongly influenced by the location and distribution of currents inside
the star, with the result that the surface field can easily be dominated
by higher multipoles than the dipole. This means that in some cases the
magnetic field at the equator can be even much higher or much smaller than
the value of the field at the pole and implies that signatures observed in
features originating at or near the surface might differ from the
expectations of a dipole dominated model, while observations of processes
related to the large scale field, as spin-down, will not (see
also the discussion in Sec.~\ref{lowb}).

\subsubsection{Hard X-ray spectral modelling}\label{hard}

Hard X-ray observations with the {\it INTEGRAL}, {\it RXTE} and {\it
Suzaku}  satellites have shown
that in some magnetar candidates \cite[namely 4U 0142+614,
1RXS J1708-4009, 1E 1841-045, 1E 2259+586,
SGR 1806-20, SGR 1900+14;][]{kui04, kui06,mere05,mol05,go06, eno11} a
large
fraction of
the total quiescent flux is emitted at energies well above $\sim 20$~keV.
These ``hard'' tails have a non-thermal (PL) character, and
extend up to a few hundreds of keV.
Pulsed phase spectroscopy has been performed for a few sources,
although with limited statistics due to the low number of counts,
revealing that the emission is likely characterized by different
components that emerge at different, and sometimes only in
limited, phase intervals \citep{denhar08a, denhar08b}.

This discovery came quite unexpectedly and suggests that a new magnetar
characteristics may be that a considerable fraction of their bolometric
luminosity is emitted in the hard, rather then in the soft X-rays. In
fact, limits on the non-detected sources are not deep enough to exclude
the presence of a similar hard tail. The first observations of
emission at $> 20$~keV revealed a difference between the (at that time
separated) AXPs and SGRs
\cite[][]{go06}: in the {\it Integral } and
{\it RXTE} data the AXPs' hard power law is considerably harder than that
observed below $10$~keV, while SGRs' hard spectra are considerably
steeper. More recent observations with {\it NuStar} indicate that the
division is
not as sharp and magnetar sources show a varied behaviour, which is
consistent with the fact that SGRs and AXPs are now considered a single
class \cite[][]{an14b}. {\it NuStar}, which has a sensitivity roughly two
orders of magnitude better than previous missions in this energy band and
a high angular resolution, is currently observing a selected sample of
magnetars as part of its priority A targets \cite[see][for a review and
references therein]{an14b}. One of the major goal is to
study the spectral location of the soft/hard X-ray turnover, which is
expected to correlate with $B_{dip}$ \cite[][]{kabo10}. Interestingly,
while
for some sources the new {\it NuStar} results are in agreement with the
{\it Integral} and {\it RXTE} ones, in some other cases (e.g. 1E~2259+586)
it appears
evident that the hard band spectrum for the total emission is not as hard
as the pulsed one. Despite that, the analysis of 1E~2259+586 data
confirmed that an additional
component, such as a power law, is needed to describe the emission in the
hard X-ray band \cite[][]{vogel14}. This suggests that at
least in some sources the non thermal mechanisms responsible for the
emission in the soft and hard X-rays are distinct, or that the charge
properties in the two emission regions are different.

\begin{figure}
\includegraphics[height=.4\textheight]{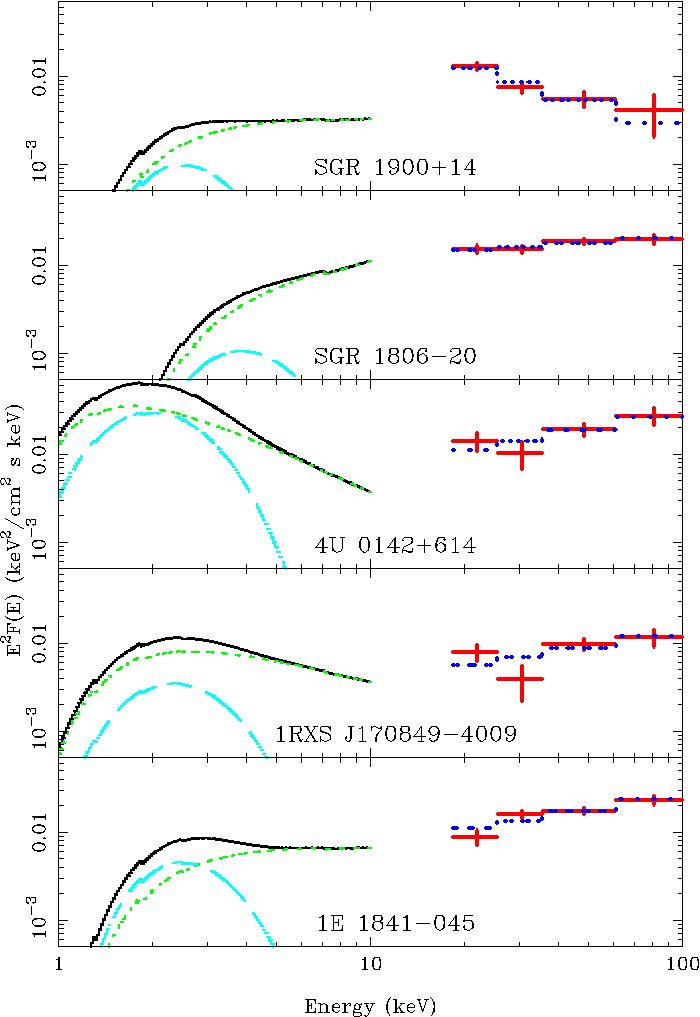}
\includegraphics[height=.4\textheight]{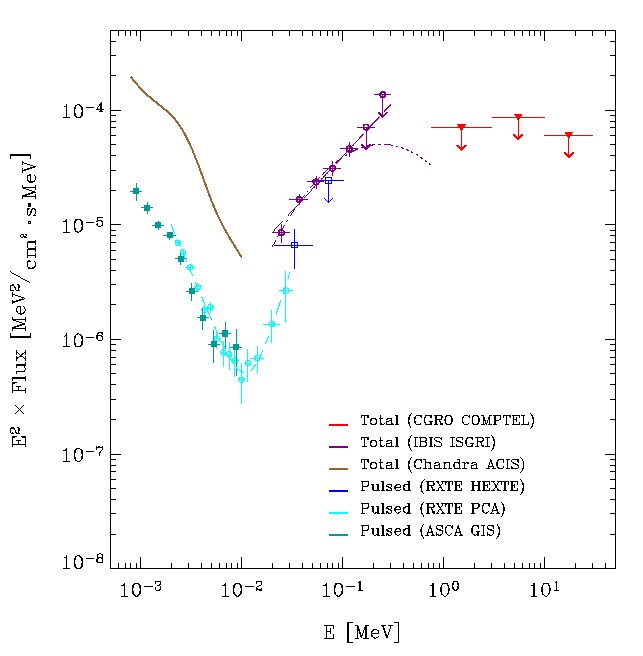}
\caption{Left: XMM-Newton and INTEGRAL spectra of magnetars
\cite[from][reproduced with permission \copyright ESO]
{go06}.
Note the different behavior of SGRs (two top
panels) and AXPs: in the latter sources the spectra turn upward
above 10 keV, while in the SGRs the spectra steepen.
Right:
Broad band spectral energy distribution of \uu  \cite[from][
\copyright AAS. Reproduced with permission. A link to
the original article via DOI is available in the electronic
version]
{kui06}. Both the total and the pulsed emission are
indicated.
Both figures are readapted from
\cite{mere08}, with kind permission
from Springer Science and Business Media.}
\label{fig-integral_spec}
\end{figure}

Observations at higher energy with Comptel and Fermi LAT failed to
detect magnetar emission, implying the presence of a spectral
break above a few hundred keV \citep{kui06,denhar06, sas10}. The
only exception reported so far is a possible Fermi Large Area
Telescope (LAT) detection of $\gamma$-ray pulsations above 200 MeV
from the AXP 1E 2259+586 \citep{wu13} which however still needs a
robust confirmation. Some example of few magnetar spectral energy
distributions in the soft/hard X--rays are shown in
Fig.~\ref{fig-integral_spec}.

\begin{figure*}
\includegraphics[width=8cm]{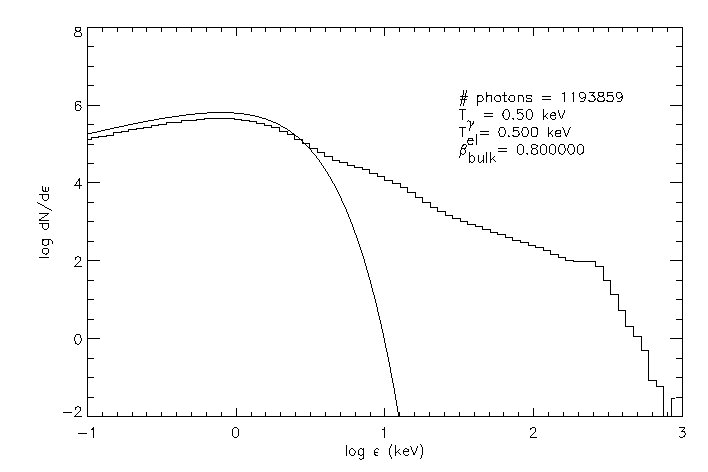}
\includegraphics[width=8cm]{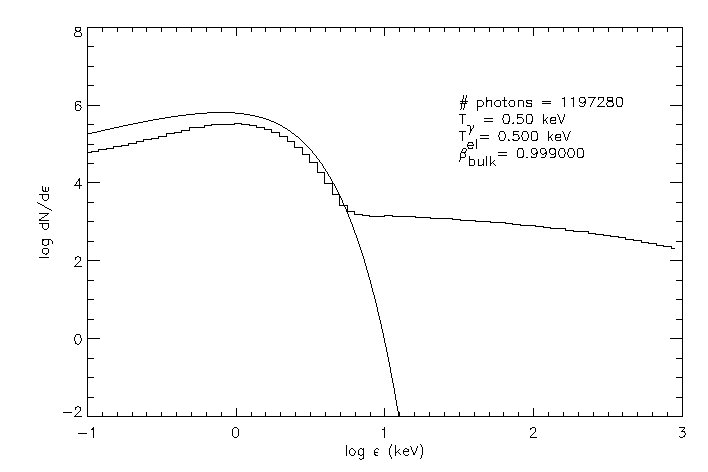}
\caption{\label{hardsp}
Monte Carlo spectra for $B = 10^{14}~G$ and
$\Delta \phi = 1$ computed by using the full
QED
expression of resonant scattering. Left: mildly
relativistic electrons ($\gamma = 1.7$). Right:
highly relativistic electrons
($\gamma = 22$) \cite[reprinted from][Copyright 2011, with
permission from Elsevier]{totnoi}.}
\end{figure*}

The mechanism responsible for the high energy emission is still
poorly understood, at least in its quantitative details. High
energy emission from currents moving in the highly magnetized
magnetosphere is expected for a number of reasons. \cite{thbel05}
originally suggested that hard X-rays may be produced either by
thermal bremsstrahlung in the surface layers heated by returning
currents, or by synchrotron emission from pairs created higher up
($\sim$ 100 km) in the magnetosphere. A further possibility,
according to which the soft $\gamma$-rays may originate from
resonant up-scattering of seed photons on a population of highly
relativistic electrons, has been proposed by \cite{bar05, bar08}.
As mentioned earlier, most of the RCS models computed with Monte
Carlo simulations are based on the assumption that scattering can be
treated in the Thomson approximation, which limits their validity up to a
few tens of keV. Instead, a proper investigation of the effects of
electron recoil and of multiple scatterings from high energy
photons demand the use of the full QED cross section. Relatively
simple expressions of the QED cross section at resonance, in a form
that is simple to include in Monte Carlo simulations, have been computed
by \cite{ntz08b} and a few examples of the emerging spectra, computed
under the assumption of self-similarly distributed twist and
constant electron velocity, have been discussed in \cite{totnoi}.
These simulations show that, if magnetospheric electrons are
mildly relativistic, when considering self-consistently electron
recoil and QED effects the spectrum exhibits a break at a few hundred of
keV. This is due to the fact that, in order to
populate the hard energy tail, soft seed photons need to experience
a series of successive scatterings, each characterized by a
limited energy gain because the Lorentz factor of electrons is
only $\gamma \sim$ a few.  In parallel, the efficiency of the QED
cross section decreases with increasing energy (or, since the
process is resonant, with increasing magnetic field) and the
combination of these two effects leads to the appearance of the
spectral break. The energy of the break depends on the effect of
the cumulative scatterings and is sensitive to the details of the
magnetic field topology and of the currents' distribution and
therefore cannot be predicted a priori nor estimated using a
simple expression. On the other hand, in the case in which
magnetospheric electrons are ultra-relativistic, the energy gain
per scattering is so large that the hard tail becomes efficiently
populated after just a few scatterings. In this case, now
independently of the details of the cross section and magnetic
topology, the spectral tail is predicted to be quite flat and
unbroken, even up to $>1000$~keV (see Fig.~\ref{hardsp}).

Although these studies are extremely useful in shedding light on
the basic behaviour of the QED scattering process, the assumptions
of constant charge velocity and self-similar magnetic twists are
clearly two major oversimplifications. This is crucial when trying
to mimic the hard X-ray emission, since the responsible emitting
region is likely to constitute quite a large portion of the whole
magnetosphere. We already mentioned that deep INTEGRAL
observations of two AXPs 1RXS J1708-4009 and 4U 0142+61 have
revealed several different pulse components (at least three) in
the hard X-rays, with genuinely different spectra
\cite[][]{denhar08a, denhar08b} and a quite spectacular
phase-dependence. The hard X-ray spectrum gradually changes with
phase from a soft to a hard power law, the latter being
significantly detected over a phase interval covering $\sim$1/3,
or more, of the spin period. This richness in the phenomenology
requires more complex magnetospheric topologies to be explained.
Unfortunately, despite many efforts having been devoted to the development
of techniques for solving the force-free equation, $\nabla \times B =
\alpha (x) B$, no general, affordable method has been presented so
far. \cite{pav09} developed a general method to generate twisted,
higher-order multipoles solving the Grad-Shafranov equation, and
analyzed in detail quadrupolar and octupolar fields.  The case of
an octupolar field has a special interest because it can be used
to mimic a twist localized in a region close to the magnetic
pole(s), and hence to investigate the consequence in the expected
spectra. Model (Monte Carlo) spectra and lightcurves have been
presented for the cases in which the twist is confined to one or
both polar regions (each region has semi-aperture of
$\sim 60\deg$), by assuming that only the polar lobes have a non-vanishing
shear while the equatorial belt is potential. Interestingly,
a configuration with a twist confined to a single lobe has been
found to be capable of qualitatively reproducing the main features
of the high-energy emission observed with INTEGRAL from the AXPs
1RXS J1708-4009 and 4U 0142+61, in particular the large variation
in the pulsed fraction at different energy bands, and a hard tail
which is quite pronounced at the peak of the pulse but
depressed by almost an order of magnitude at pulse phases close to
the minimum of the hard X-ray lightcurve.

\begin{figure*}
\centering
\includegraphics[width=.23\textwidth]{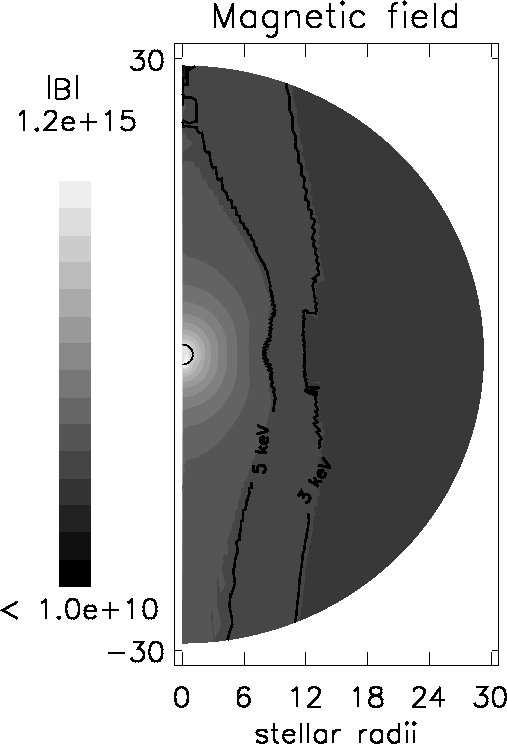}
\includegraphics[width=.23\textwidth]{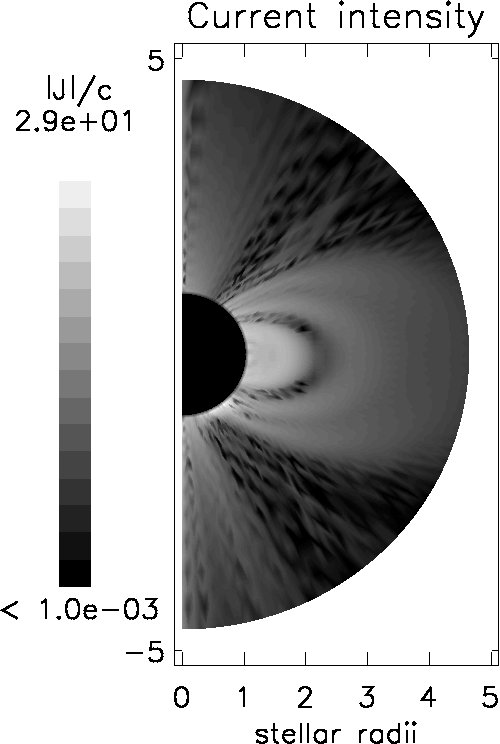}
\includegraphics[width=.47\textwidth]{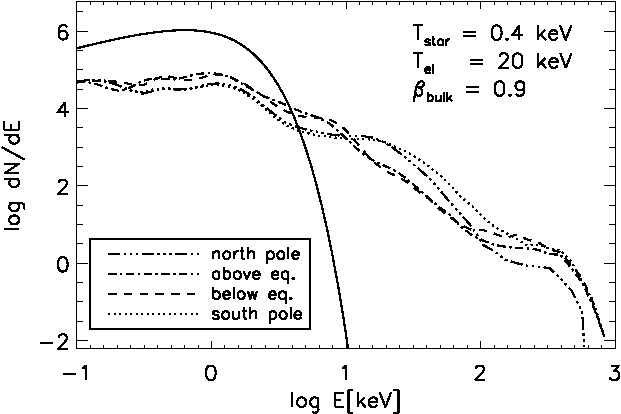}
\caption{\label{magspec} Left and center panel: magnetic field
and current distribution for a high helicity model with current and twist
concentrated in a closed bundle near the equatorial region and a
j-bundle near the southern semi-axis. This configuration is  likely more
realistic than the self-similar models, since currents are more
concentrated near the axis. The two panels show
$|\vec B|$ (grey logarithmic scale) with superimposed the
scattering surfaces
for photons of 1, 3 and $5~$keV and the current intensity $|\vec{J}|/c$
(gray linear scale)  in
units of $10^{14}~$G$/r_{NS}$.
Right: Corresponding synthetic spectra
computed with a Monte Carlo simulation. Different curves correspond
to four different viewing angles; the seed blackbody is shown for
comparison as a solid line.
\cite[readapted from][with permission; a link
via DOI to the original version is available in the
electronic version of this paper]{dan12}.}
\end{figure*}

More recently, an alternative to the (mathematically simple)
self-similar models, has been presented by \cite{dan12}, who
discussed the effects of more realistic magnetic field geometries
on the synthetic Monte Carlo spectra. They presented a numerical
method to build general force-free field magnetic configurations,
starting from an arbitrary, non-force-free poloidal plus toroidal
field and employing artificial dissipation to remove the
non-parallel currents. In particular, they considered configurations in
which the
currents are concentrated in a bundle along the polar axis, as
expected for a spatially-limited twist.
In this case the pulse fraction is larger with respect to that of
self-similar models, and the spectrum observed at different angles
varies in a much more irregular way (see Fig.~\ref{magspec}). Instead of a
simple PL, it
shows different spectral bumps the relative importance of which can vary
by one order of magnitude or more at different colatitudes. The
different spectral components inferred from the data may therefore
be due to these bumps. Even if a real fit has not been
attempted, we may speculate that this is qualitatively in  line
with the observations of the AXPs 1RXS~J1708-4009 and 4U 0142+61.

\begin{figure*}
\centering
\includegraphics[width=3.1in,angle=0]{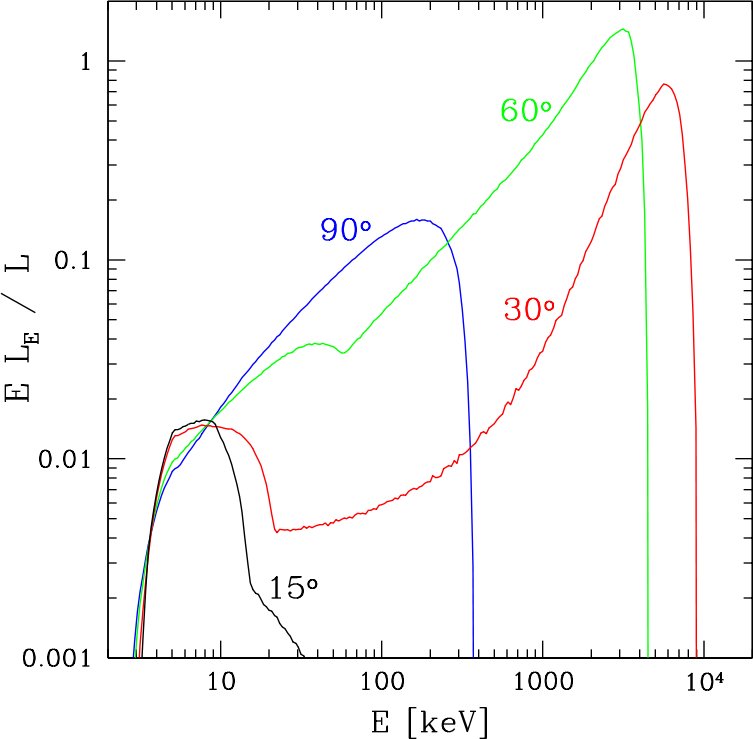}
\caption{\label{andsp}
Left: Spectrum emergent from scattering onto charges populating an
axisymmetric j-bundle. Different lines represent the spectrum as observed
at four
different angles with respect  to the magnetic axis: $15^{\rm o}$,
$30^{\rm o}$, $60^{\rm o}$, and $90^{\rm o}$.
The spectrum is normalized to the total luminosity of the j-bundle.
\cite[from][
\copyright AAS. Reproduced with permission. A link to
the original article via DOI is available in the electronic
version]{belo13b}.
}\end{figure*}

As mentioned earlier, a further poorly known ingredient of all
these simulations is the charge velocity distribution and probably
the most detailed solutions published have been presented by
\cite{belo13a, belo13b}. In this scenario, the electron-positron
flow decelerates as it propagates away from the neutron star
surface (due to Compton drag in the resonant scattering region),
then it reaches the top of the magnetic loop where it annihilates.
While computed with a Monte Carlo simulation, the corresponding
spectra show a distinct peak at $E>1$~MeV (so far unobserved),
the position of which is however strongly dependent on the
viewing angle
(or, equivalently, on the magnetic colatitude at which the
spectrum is emitted, see Fig.~\ref{andsp}).

A well defined change in the power law slope from soft to hard
X-rays is seen, which makes these spectral models incapable of
explaining observations of sources that have a similar slope below
and above $\sim10$~keV (historically these were referred as
SGR-like spectra), but instead in good qualitative agreement with
the observed spectra in which a marked turnover is present
(historically, AXP-like spectra). In an attempt to perform a
quantitative comparison without to resort to the time-consuming
computation of a complete table of Monte Carlo models,
\cite{hasc14} developed a simplified model by assuming the same
charge velocity distribution but by computing the spectra through
a simple calculation of the angle-dependent emissivity alone. The
magnetosphere is assumed to be axysymmetric and dipolar, apart
from the presence of a current-carrying region (the j-bundle),
which is filled with the electron-positron flow. These spectra
have been then applied to multi-year hard X--ray observations of
4U~0142+61, 1RXS J1708-4009 and 1E 1841-045, finding that they
successfully reproduce the emission observed above $\sim 10$~keV
in both the phase-average and phase-resolved spectra.
Unfortunately, the authors found that the model predictions cannot
be self consistently applied to explain also the data below 10~keV
(probably because of the lack of the particle back-bombardment
effects in the simulation). These predictions have therefore been
ignored in the fits and instead a blackbody component(s) was used
to account for the soft X-ray part of the spectrum. A concern
regarding these fits is the fact that the spectra used by
\cite{hasc14} do not include photon splitting, which is the main
factor responsible for the bump visible in the Monte Carlo spectra
at $\sim 1$~MeV \cite[][]{belo13b}. Although the simplified and
complete calculations are in good agreement below $\sim 400$~keV,
they are expected to be markedly different at higher energies.
Actually, there is no guarantee that, should the spectra be
computed self-consistently with the Monte Carlo code, the high
energy CGRO-Comptel upper limits are not violated (either always
or for the derived geometry). This casts some doubts on the
robustness of the study and on the constraints on the viewing
geometry reported by \cite{hasc14} fitting only the hard X-ray
part of the spectrum.

It is interesting to note that spectra presented by \cite{wad13},
using a new Sokolov and Ternov formulation of the QED Compton
scattering cross section in strong magnetic fields and accounting
for spin-dependent effects at resonance, also show a  spectral
cut-off whose energy is critically dependent on the observer
viewing angles and electron velocity, with substantial emission
expected up to 1~MeV except for very selected viewing angles.

At the present, unfortunately, what causes the high energy
emission and its richness in phenomenology is still an open issue:
models have been computed, but they depend dramatically on a large
number of degrees of freedom for the magnetospheric setting at
large scale. New and future missions such as {\it Astro-H}, {\it NuSTAR},
and possibly {\it LOFT} would provide the possibility of collecting
simultaneous soft and hard spectra, to study the correlation
between the variability in the two bands, to perform high
resolution pulsed phase spectroscopy and to reveal the detail of
the slope turn-over (when present) in the soft-to-hard emission.
In turn, this will provide a powerful tool to break the model
degeneracy and an unprecedented insight into the details of the
field and current distributions.

\subsection{X-ray polarization}\label{polar}

As discussed in Sec.~\ref{rcs}, comparison of RCS models with the
soft X-ray spectral data of magnetar candidates has proved quite
successful. However, spectroscopy alone cannot provide complete
information on the physical properties of the magnetosphere, due
to the inherent degeneracy in the RCS model parameters. Moreover,
computed spectra are rather insensitive to the source geometry,
although in principle they do depend on the angles that the line
of sight and the magnetic axis make with the star's rotation axis
\cite[][]{ntz08a,zrtn09}. Although a simultaneous fit of both the
(phase-averaged) spectrum and the pulse profile is effective in
this respect, at least for TAXPs
\cite[][]{pergot08,alba10,bern11}, polarization measurements at
X-ray energies would provide an entirely new approach to the
determination of the physical parameters in magnetar
magnetospheres.

X-ray radiation from a magnetar is expected to be polarized for
essentially three reasons: i) primary, thermal photons, coming
from the star's surface, can be intrinsically polarized, because
emission favors one of the modes with respect to the other; ii)
scattering can switch the photon polarization state; and iii) once
the scattering depth drops, the polarization vector changes as the
photons travel in the magnetosphere  \cite[the so called ``vacuum
polarization'',][see also \citealt{Harding+Lai}]{hs00,hs02}.

\begin{figure*}
\centering
\includegraphics[width=4in,angle=0]{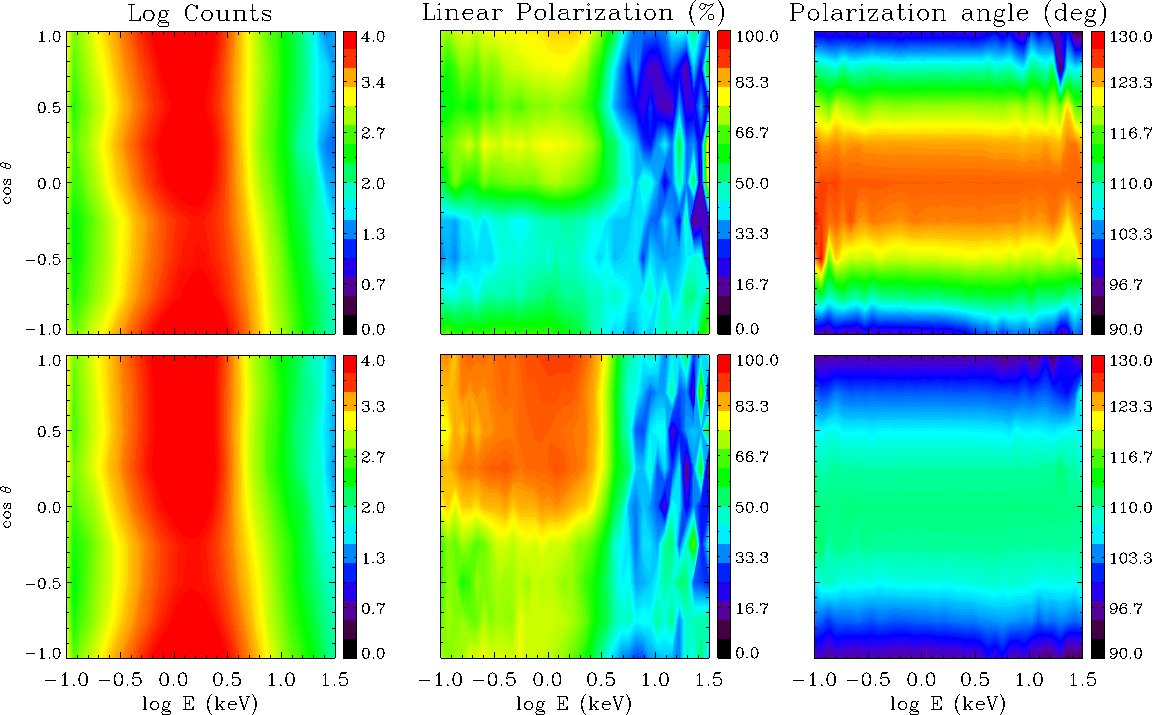}
\caption{\label{polfra} Contour plots for number of counts (in
arbitrary units; left column), polarization fraction (middle
column) and polarization angle (right column) as functions of the
photon energy and $\cos\theta$ for different values of the twist
angle and the electron bulk velocity:
$\Delta\phi=1.3$ rad, $\beta=0.3$ (top row) and
$\Delta\phi=0.7$ rad, $\beta=0.4$ (bottom row). In
both cases it is  $B_\mathrm p=5\times 10^{14}$ G
\cite[from][with OUP permission]{tav14}.}
\end{figure*}

Although a preliminary analysis was already contained in
\cite{ft07} and \cite{ntz08a}, a detailed study of the
polarization properties of magnetar radiation in the X-ray band
has been presented by \cite{Fernandez+Davis} and, more recently,
by \cite{tav14}, by means of Monte Carlo simulations.
Phase-averaged as well as phase-resolved results indicate that the
linear polarization fraction $\Pi_{\mathrm{L}}$ and the
polarization angle $\chi_{\mathrm{pol}}$, are very sensitive to
the magnetospheric twist angle $\Delta\phi$ and the
charge velocity $\beta$, and also to the geometric angles $\chi$
and $\xi$. This allows one to remove the
$\Delta\phi$-$\beta$ degeneracy which spectral
measures alone cannot disambiguate. An example is shown in Fig.~\ref{polfra} where
the photon spectrum, polarization fraction and
polarization angle are plotted as functions of viewing angle and energy
for two choices of the model parameters. While the photon
spectrum is practically the same, the pattern of $\Pi_L$ and
especially $\chi_{pol}$ is markedly different in the two cases.
According to the simulations by \cite{tav14}, polarimetric
measurements in bright magnetar candidates, like the AXP 1RXS
J1708-4009, are within reach of recently proposed
polarimeters which should hopefully fly in future missions.

\subsection{IR/optical emission}\label{IR}

As mentioned in Sec.~\ref{intro}, variable IR counterparts have now
been identified, or in some cases proposed, for a number of
magnetars \citep{mere11,gial14}. In addition, three magnetars have been
detected in the optical band, the two AXPs 4U 0142+61
\citep{hul00} and 1E 1048.1-5937 \citep{dur05}, and the SGR
0501+4516 \citep{fat08, dhi11}. The origin of this emission is
still under debate, and the main dispute is whether it is due to
the presence of a fossil disk \citep{per00} or if it has a
magnetospheric origin \citep{eic02, belotho07}.

AXP 4U 0142+61 is the only persistent magnetar for which the
optical/near IR (NIR) spectrum was measured, although the faintness of the
source required extensive observational data \citep{hul00}. In the
case of this source, an IR ``excess`` (or ``flattening``) was clearly
detected with respect to the extrapolation of the additional black
body used to account for the optical emission
\citep{isr04,isr05a}. This may suggest that the IR emission is due
to a distinct spectral component, a conclusion that might well
hold also for other AXPs, considering the similarity of
their IR magnitudes and $F_{\mathrm{X}} / F_{\mathrm{IR}}$ ratios \citep{dur11}.

The optical/NIR spectrum of 4U 0142+614 was found to be well
fitted by a  multi-temperature (700-1,200 K) thermal model,
leading to the suggestion it originates in an extended disk or
shell \citep{wang06}. If this detection of a fallback disk is
real, it would be the the first direct evidence for supernova
fallback in any context. On the other hand, disks may then be
ubiquitous while, despite intense campaigns, no similar direct
evidence has been found for other sources \cite[see also][for a
report on recent deep limits on fallback disks]{poss14}. The only indirect
evidence is one source, in which there is a detected correlation
between the NIR and X-ray fluxes \cite[][]{tam04}, which, as pointed
by \cite{wang06}, may indicate that the IR emission arises in an
X-ray-heated debris disk.

On the other hand, very deep optical and NIR observations of the
field of the low-B magnetar SGR 0418+5729, which is the nearest
and least extincted magnetar known, failed to detect the source
counterpart \citep{dur11}.  Shallower  observation of the field of SGR~0418+5729
with the new Gran Telescopio Canarias 10.4-m telescope and with the William
Herschel Telescope where also taken
closer to the onset of the outburst \cite[][]{esposito10, rea13} and only gave upper
limits
on the countarpart.
This negative result is in better
agreement with a magnetospheric origin interpretation, in which
case the IR/optical flux is expected to be fainter for lower field
strengths. Moreover, a magnetospheric scenario is more likely to
explain the fact that, when detected, the observed optical
emission is pulsed at the pulsar period, with pulsed fractions
$>50\%$, i.e., higher than in soft X-rays \citep{ker02, dhi09,
dhi11}.

\begin{figure*}
\centering
\includegraphics[width=3in,angle=0]{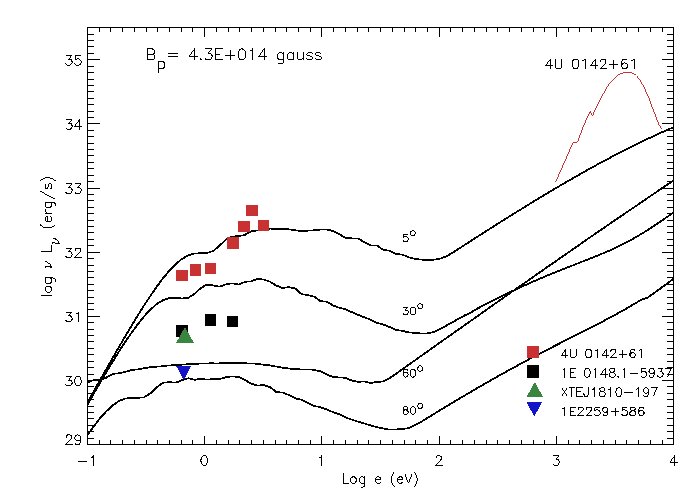}
\caption{\label{irspectra}
Model spectra for different values of the viewing angle and
$B=4.3\times 10^{14}$~G. The {\it XMM-Newton} X-ray spectrum of
1RXS\,J1708-4009 is from \citet[red solid line]{nanda2}. The
AXPs IR/optical data are from \citet[4U\,0142+614
and 1E\,1048-5937]{dur05}
and \citet[XTE\,J1810-197 and
1E\,2259+586]{mig07}. The adopted
distances for de-reddening are 5~kpc (4U\,0142+614, 1RXS\,J1708-4009),
3~kpc (1E\,1048-5937,
1E\,2259+586), 4~kpc (XTE\,J1810-197), 8.5~kpc (1E\,1841-045). Curvature
emission spectra have been
computed accounting for particle bunching \cite[Fig.~2
from][With kind permission
from Springer Science and Business Media]{currentsnoi}
.}
\end{figure*}

Magnetospheric emission in the IR/optical is expected
from the inner region of the magnetosphere \citep{belotho07,
currentsnoi}. As discussed in Sec.~\ref{currents}, this region is
expected to be pair-dominated, and the Lorentz factor of the
electrons and positrons is likely to be frozen at the threshold
for pair production ($\gamma\sim 500-1000$). In order to estimate
the amount of curvature radiation, \cite{currentsnoi} developed a
simple geometrical model accounting for misalignment between the
observer line of sight and the magnetic axis. Although the model
is too simple to allow for a proper spectral fitting, it
is detailed enough to make a prediction about the energetics and
the computation indicates that curvature radiation is sufficient
to explain the amount of observed IR/optical flux, at least if a
particle bunching mechanism is efficient (see also
\citealt{belotho07} and see Fig~\ref{irspectra}). This is not
unlikely, since many models have been suggested to explain the
origin of the interactions which push particles together and can
lead to the  formation of bunches of charged particles localized
in phase-space. The most promising explanation seems to be
connected with plasma instabilities, like the two-stream
(electron-positron/electron-ion) instability. We notice that this
mechanism does not affect the emission in other bands. In fact, in
order the process to be effective, it has to have $N=n_el_B^3>1$,
where $l_B$ is the size of the bunch and $n_e$ is the electron
density, which in turns means that only emission at frequencies
$\nu <\nu_{co}=n_e^{1/3}c$ is efficiently amplified. Also,
a low-energy cut-off is present because of the strong absorption
below the electron plasma frequency.

\begin{figure*}
\centering
\includegraphics[width=3in,angle=0]{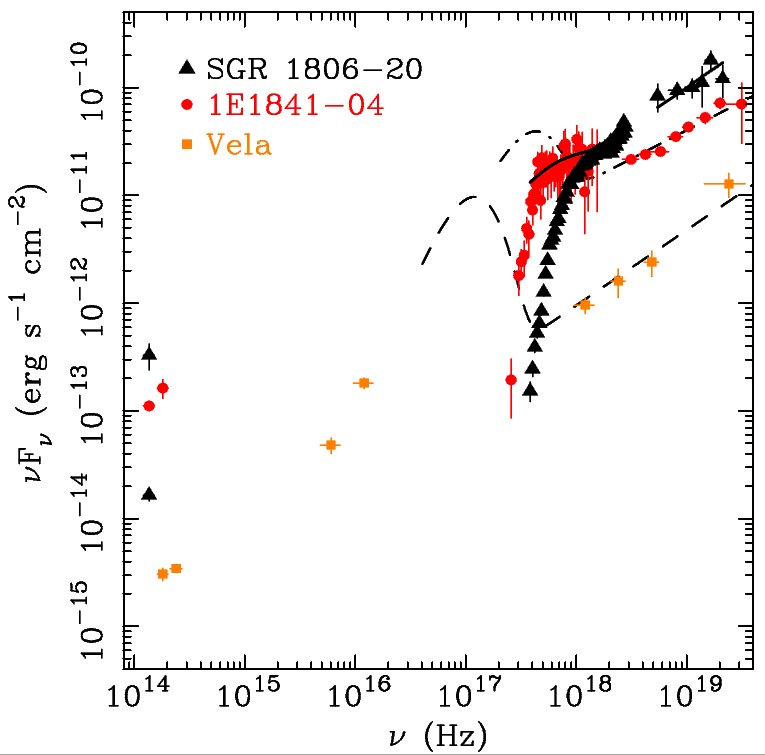}
\caption{\label{irhard}
Broad band energy spectrum of SGR 1806-20 (triangles), the AXP 1E 1841+045 (circles)
and the radio pulsar Vela (squares). In the case of SGR 1806-20 and 1E 1841+045
high energy data are taken from \cite{mere05,sandro05}. Absorbed and unabsorbed IR fluxes (${\rm A_V} = 29 \pm 2$,
5th October 2004 NACO observation) are shown in the case of SGR 1806-20,
unabsorbed $({\rm A_V} = 13 \pm 1 )$
IR fluxes are instead reported for the likely candidate of
1E 1841+045 \cite[circles;][]{testa08}. All the data
for Vela are taken from \cite{kaspi06}. Solid curves (continuous, stepped
and dot-stepped) are the
unabsorbed fluxes, for the blackbody plus power-law model used to fit the
high energy part of the spectra \cite[courtesy G.L. Israel,
from][reproduced with permission \copyright ESO]{isr05b}.}
\end{figure*}

\cite{isr05b} proposed a link between the IR and hard X-ray
spectrum of AXPs, and correspondingly between AXPs/SGRs and
radio-pulsars, based on the analysis and comparison of their
broad band energy spectra. These authors pointed out that,
similar to the case of a number of young radio pulsars such as the
Vela, it is possible to bridge the IR to $\gamma$-ray emission of
AXPs with a power-law with index of about 0.5-0.6 (see Fig.~\ref{irhard}).
This peculiar
similarity, for classes of neutron stars with otherwise quite
different emission properties, may naturally explain the IR
excess or flattening in the spectra of AXPs and points toward a
similar origin for the less energetic bands and the hard X-ray
emission. This scenario may be unambiguously proven through
simultaneous studies of the correlated variability in the two
bands, which have unfortunately so far been hampered by the lack of
gamma-ray observatories sensitive enough to these faint sources.

\newpage

\section{Transient Magnetars}
\label{transient}

The persistent X-ray emission of a number of SGRs/AXPs has been
known to be variable all along, with typical flux variations
of a factor of a few over a timescale of days to months,
often in coincidence with periods of enhanced bursting activity
\cite[see Sec.~\ref{intro};][]{reaesp11}.
The first evidence that the luminosity of SGRs/AXPs can change much more
dramatically came from observations of SGR 1627-41
\cite[][]{Woods99a} and
AX J1845.0-0300 \cite[][although the latter source is only a
candidate
magnetar with no detection of $\dot P$]{Torii98, Vasisht00}.

It was not until 2002, however, that the existence of a new class
of magnetar sources with much more extreme variability was realized,
thanks to the discovery of the first transient AXP, XTE
J1810-197 \cite[][]{ibretal04}. At present eleven transients are
known and, remarkably, they almost make up all of the new magnetars
observed in the last 10 yrs \cite[the exception being CXOU
J171405.7-381031,][]{hal10}. The main properties of transient
magnetars are listed in table \ref{table_trans}\footnote{Only
sources for which the peak luminosity is $>10$ times the quiescent
one have been included.}.

\begin{table} \caption{\label{table_trans}Transient magnetar
sources\footnotemark[123]}
\footnotesize\rm \begin{tabular*}{\textwidth}{lcccc}
\br Source&$P$ &$\dot P$ &$B$ &$D$ \\ &(s)&$(10^{-11}$
s/s)&$(10^{14}$ G)&(kpc)\\ \mr CXOU J1647-4552&10.61& $<0.04$&
$<0.7$&3.9\\ XTE J1810-197&5.54& 0.77& 2.1& 3.5\\ SGR 0501+4516&5.76
&0.59 &1.9 &52.0\\ SGR 0418+5729&9.08 &0.0004 &0.06& 2.0\\ SGR
1833-0832&7.56 &0.35 &1.6 &--\\ PSR 1622-4950&4.33 &1.7 &2.2
&9.0\\ 1E 1547-5408& 2.07& 4.77& 3.2&4.5\\ Swift J1822.3-1606&8.4
& 0.02& 0.14&1.6\\
SGR 1627-41 &2.59 &1.9 &2.2&11.0\\
Swift J1834.9-0846&2.48 & 0.80& 1.4&4.2\\
SGR J1745-2900&3.76&1.38&2.3&8.3\\ \br
\end{tabular*} \end{table}

\footnotetext[123]{Data from the McGill magnetar catalogue
\cite[][]{ol14}.}

\begin{figure} \centering
\includegraphics[width=4in,angle=0]{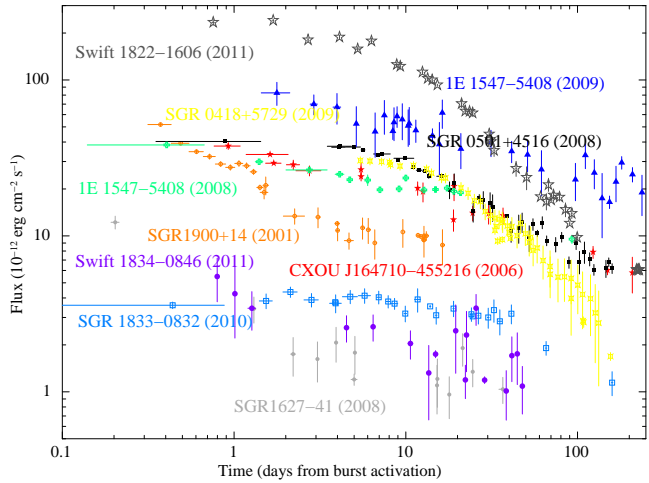}
\caption{\label{decay} Flux evolution over the first $\sim$200 days of all
magnetar outbursts (only if observed with imaging instruments, and for
which this period span is well monitored). Fluxes are reported in the 1-10
keV energy range, and the reported times are calculated in days from the
detection of the first burst in each source. See \cite{reaesp11} for the
reference of each reported outburst. \cite[From] [Copyright ? 2014
WILEY-VCH Verlag GmbH \& Co. KGaA, Weinheim. Reproduced with
permission.]{rea14c} } \end{figure}

These sources are characterized by a sudden ($\approx$ hrs)
increase of the X-ray flux, by a factor $\approx 10$--1000 over
the quiescent level, accompanied by the emission of short bursts.
This active phase, commonly referred to as an outburst, typically
lasts $\approx 1$ yr, during which the flux declines, the spectrum
softens and the pulse profile simplifies. Fig.~\ref{decay} shows
the decay of the X-ray flux for several transient magnetars. Some
objects have undergone repeated outbursts (SGR 1627-41, 1E
1547-5408) and the decay pattern is often different from source to
source (and even between outbursts from the same source). Outbursts,
besides revealing new magnetars which in quiescence are too faint
to be detectable, or which passed unnoticed among the host of
unclassified, weak X-ray sources, have also occurred in persistent
sources like 1E 2259+586 or 1E 1048.1-5937, although drawing a
precise line between outbursts and less extreme variability is somewhat
haphazard. The group of transient sources also harbours the two
peculiar ``low-field`` magnetars, SGR 0418+572 and Swift
J1822.3-1606 (see Sec.~\ref{lowb}), and the recently discovered
source in the Galactic Centre (see Sec.~\ref{radiomag}).

\subsection{Outburst Models}\label{outburst_model}

\begin{figure*}
\centering
\begin{minipage}{.5\textwidth}
  \centering
  \includegraphics[width=1.\linewidth]{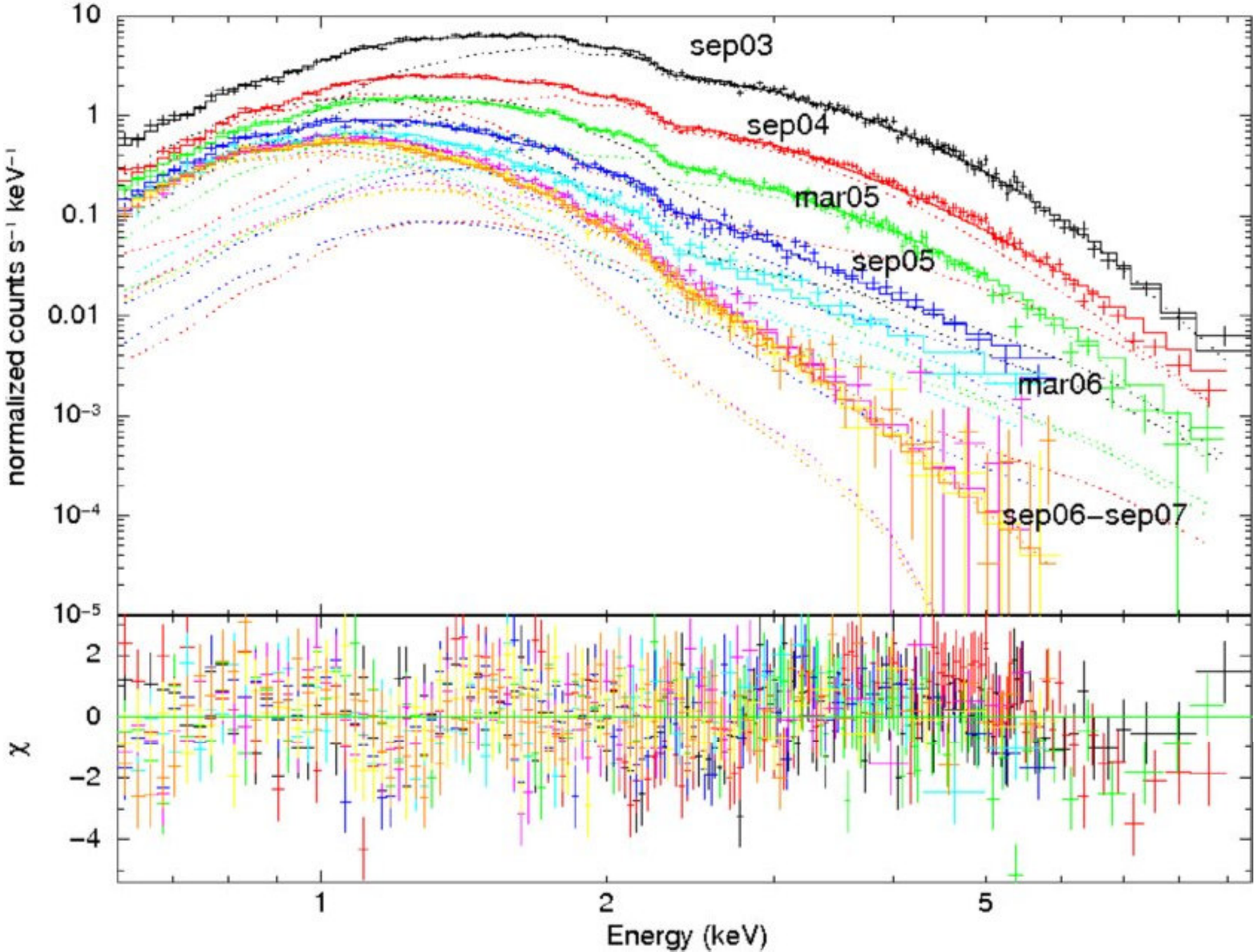}
\end{minipage}%
\begin{minipage}{.5\textwidth}
  \centering
  \includegraphics[width=1.\linewidth]{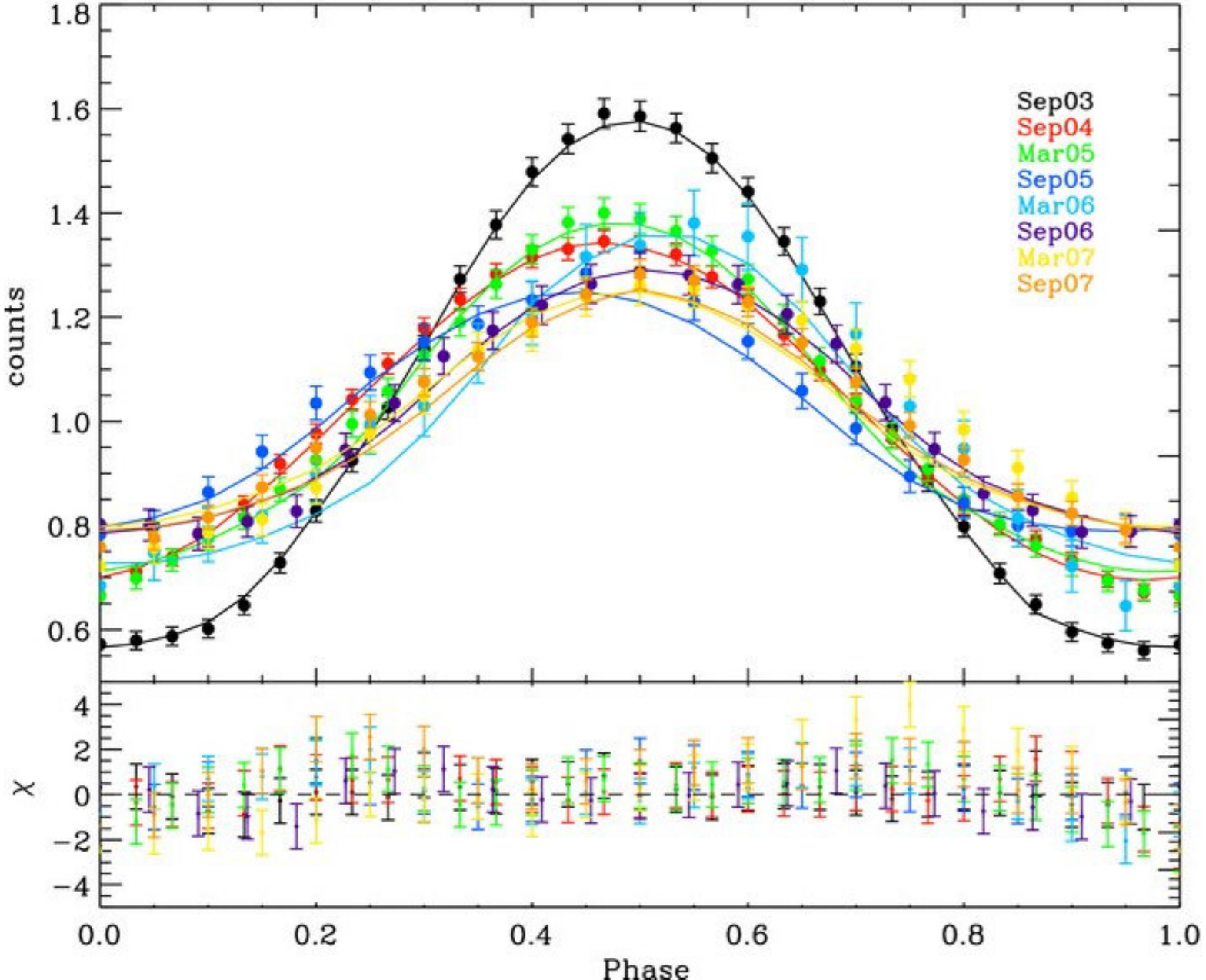}
\end{minipage}
\begin{minipage}{0.5\textwidth}
  \centering
  \includegraphics[width=1.\linewidth]{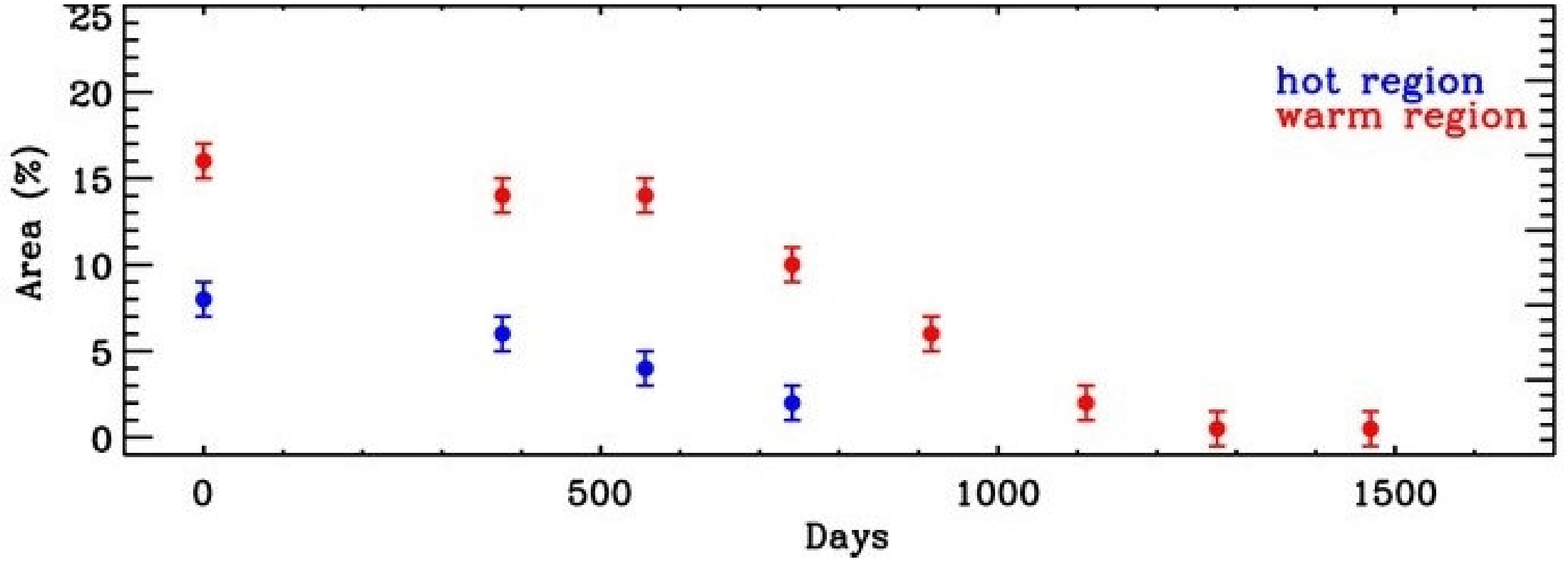}
\end{minipage}
\caption{The evolution of the
spectrum, pulse profile (top panels) and size of the emitting regions
(expressed as fraction of the
entire surface, bottom panel) during the
outburst of XTE J1810-179. Results are for the three temperature model (hot
cap, warm corona and cool surface) of \cite{alba10}; the cold temperature
is fixed at 0.15 keV. The adopted spectral model is the superposition of
two NTZ's (see Sec.~\ref{rcs})
at the hot and warm temperature \cite[adapted
from][
\copyright AAS. Reproduced with permission. A link to
the original article via DOI is available in the electronic
version]{alba10}.}
\label{fig_1810}
\end{figure*}

A common feature of all observed outbursts is the presence in the
X-ray spectrum of one (or two) thermal component(s) at higher
temperature ($\sim 0.3$--0.9 keV) with respect to that associated
with the cooling star surface during quiescence ($\sim 0.1$--0.2
keV). The (radiation) radius of these hotter regions is fairly
small ($\lesssim 1$ km) and usually decreases in time as the
outburst subsides, when the temperature also declines (e.g.
\citealt{alba10}, \citealt{rodrig14} for XTE J1810-197, CXOU
J1647-4552; \citealt{rea09}, \citealt{cam14} for SGR 0501+4516;
\citealt{rea13} for SGR 0418+572; \citealt{rea12} for Swift
J1822.3-1606; \citealt{isretal10} for 1E1547.0-5408, and
references therein). The variation in time of the spectrum, pulse
profile and size of the emitting regions for the AXP XTE J1810-197
during the the outburst decay is shown in Fig.~\ref{fig_1810}.

This has been interpreted as due to some form of heat deposition
in a limited region of the star surface which then cools and
shrinks. Until now, however, the heating mechanism has not been
unambiguously identified. One possibility is that energy is injected
deep in the crust, e.g. because of magnetic dissipation, and then
flows to the surface, as first suggested by \cite{let02}.

\cite{pons12} developed a quantitative model for the outburst
evolution by simulating the thermal relaxation of the neutron star
in response to an impulsive energy injection in the star crust.
They found that most of the energy is released in the form of
neutrinos, unless injection occurs in the outer crust ($\rho
\lesssim 3\times 10^{11}\, \mathrm{g\,cm}^{-3}$). The successive
evolution depends mostly on the energy input in the outer crust,
$E_{OC}$. However it has to be $E_{OC}\gtrsim 10^{40}\,
\mathrm{erg\,s}^{-1}$ to produce any visible effect on the surface
and, at the same time, $E_{OC}$ is bounded from above at $\sim
10^{43}\, \mathrm{erg\,s}^{-1}$ because any excess energy is
efficiently radiated away by neutrinos produced in the heated
crust. This limits the surface temperature to $\sim 0.5$ keV.
Because heat transport occurs mostly along the field, which is
predominantly radial in the outer layers, the size of the hot spot on
the star surface remains nearly constant in time (which
may be
problematic in explaining the observed shrinking of the heated
region).

Results were successfully applied by \cite{rea12} to fit the
outburst decay in Swift\,J1822.3-1606 over the entire period
covered by their observations, $\sim 250$~d after the first burst
that led to the discovery of the source. The case of
SGR\,0418+5729, for which a much longer time coverage is available
($\sim 1200$~d), is, however, much less conclusive in this respect
\cite[][]{rea13}. The calculated flux in the 0.5--10 keV band
systematically overestimates the observed one at later times
($\gtrsim 400$~d), when the luminosity suddenly drops and the
hotter blackbody (initially at $kT\sim 0.9$~keV) disappears
leaving only a cooler component at $\sim 0.3$~keV. A similar
effect has been recently reported in the decay curve of CXOU
J1647-4552 around 1000 d \cite[][]{rodrig14} and in Swift J1834.9-0846
\cite[][]{Esposito13}.

Alternatively, heating of the surface layers may be produced by
currents flowing in a twisted magnetosphere as they hit the star.
As discussed in \S\ref{twist}, once implanted by crustal
displacements, a twist must necessarily decay in order to supply the
potential drop required to accelerate the conduction current
\cite[][]{belotho07}. The evolution of an untwisting magnetosphere
proceeds through the expansion of a potential region, where
$\vec\nabla\times\vec B=0$, which progressively confines the twist
to a limited bundle of current-carrying field lines (the
j-bundle), until the twist is completely erased \cite[][]{belo09}.
Fractures, or plastic deformations, most likely affect only a
limited area of the star crust, so the twist is expected to
involve only the bundle of field lines whose footpoints are
anchored in the displaced region. As the magnetosphere untwists,
the area covered by the j-bundle shrinks and the luminosity
decreases. The rate of Ohmic dissipation, which depends on the initial
twist configuration and on how the twist angle $\Delta\phi$
evolves in time ($\Delta\phi(t)$, is not necessary monotonic, the
twist may first decrease and then increase to a maximum value
$\approx 1$ rad). The simplest model gives

\begin{equation} L\approx 10^{36}\left(\frac{B}{10^{14}\ \mathrm G}\right)
\left(\frac{\cal V}{10^{9}\ \mathrm V}\right)
\left(\frac{R}{10^{6}\ \mathrm{cm}}\right)^2\Delta\phi
\sin^4\theta_{j-b}\ \mathrm{erg/s}
\end{equation} where $\cal V$ is the discharge voltage and $\theta_{j-b}$
is the angular extension of the j-bundle on the star's surface
\cite[this is valid for a small polar bundle with maximal twist,
see][]{belo09}. Since heat is unlikely to leak outside the cap at
the base of the j-bundle, the area of the X-ray emitting region is
$\sim \pi R^2\sin^2\theta_{j-b}$. A quite definite prediction of
the model is, then, that the luminosity is proportional to the
square of the emitting area throughout outburst evolution (this
still holds, at least approximately, also for more elaborate
versions). A spatially-limited twist can also explain the nearly
thermal spectra observed in most transients. If currents fill only
a tiny fraction of the magnetosphere, thermal photons produced in
the hot surface regions have only a small chance of undergoing resonant
scattering to populate a high-energy tail.

\cite{belo09} found that the model can satisfactorily reproduce the
observed properties of the outburst of XTE J1810-197 for a small
twisted region ($\sin^2\theta_{j-b}\sim 0.03$) and large twist
angle ($\Delta\phi\sim 1\ \mathrm{rad}$). The application to other
transient sources is, however, not without difficulties. The main
problem is that the small size of the thermally emitting spot, and
hence the limited spatial extent of the twist, can make the
luminosity released by Ohmic dissipation too low (especially if
the magnetic field is $\lesssim 10^{14}\ \mathrm G$) to explain
the observed flux, like in SGR 0501+4516 \citep{rea09} or
SGR\,0418+5729, \citep{rea13}. Moreover, the relation $L\propto
A^2$ does not seem to be met in other sources \cite[][]{rodrig14}.
Different geometries of the j-bundle, not necessarily involving the
polar region, may however ease the energetic requirement.

\subsection{Low-field Magnetars}\label{lowb}

Recently, the commonly accepted picture in according to which the
activity in SGRs and AXPs is necessarily related to a super strong-field
(the `supercritical $B$' paradigm) has been challenged by the discovery of
two fully-fledged magnetars, SGR\,0418+5729 and Swift\,J1822.3--1606,
\cite[][]{rea10,tur11,rea12,livingstone11,rea13,scholz12} with a dipole
magnetic field $\lesssim 10^{13}\, \mathrm G$, well within the range of
ordinary radio pulsars. A third candididate, XMM J185246.6+003317, has
been reported and awaits further monitoring \cite[][]{rea14}.

SGR\,0418+5729 was discovered on 2009 June 5, thanks to the
detection of a couple of SGR-like bursts \cite[][]{vdh10}.
Follow-up observations revealed a previously unknown bright X-ray
source at a flux level of a few times \mbox{$10^{-11}$ erg
cm$^{-2}$ s$^{-1}$}, pulsating at a period of $9.1$ s
\cite[][]{vdh10,esposito10}. Based of \emph{ROSAT} All-Sky Survey
data, the upper limit on the source flux in quiescence is of the
order of $10^{-12}$ erg cm$^{-2}$ s$^{-1}$. Despite the dense
monitoring, no spin-down was detected during the first $\sim 5$
months of observations following the outburst onset. The estimated
upper limit on the period derivative, $1.1\times10^{-13}$ s
s$^{-1}$, translates into an upper limit on the surface dipole
magnetic field strength of $3\times10^{13}$ G
\cite[][]{esposito10}.

This made SGR\,0418+5729 the magnetar with the lowest dipole magnetic
field ever discovered (for comparison, the previous record holder, the AXP
1E\,2259+586, has a spin-down magnetic field about twice as strong,
$6\times10^{13}$ G). It took nearly another 3 years of monitoring to
finally pinpoint the source spin-down rate from a coherent timing analysis
of all the X-ray data spanning $\sim 1200$ days:
\mbox{$(4\pm1)\times10^{-15}$ s s$^{-1}$}, corresponding to $B\sim
6.1\times10^{12}$ G and to a characteristic age
$\tau_c=P/(2\dot{P})\simeq36$ Myrs \cite[][]{esposito10,rea10,rea13}.

Swift\,J1822.3-1606 was detected on 2011 July 14, when it emitted
several magnetar-like bursts \cite[][and references
therein]{livingstone11,rea12}. A few days after the outburst
onset, a new, persistent X-ray source was discovered at a flux
level of $\sim 2\times10^{-10}$ erg cm$^{-2}$ s$^{-1}$, pulsating
with a period of $\sim 8.4$ s. Contrary to the case of
SGR\,0418+5729, the source has been already detected in X-rays at
a flux level of $\sim 4\times10^{-14}$ erg cm$^{-2}$ s$^{-1}$,
although its presence in two \emph{ROSAT} X-ray catalogues passed
unnoticed \cite[][]{rea12,scholz12}. Swift\,J1822.3--1606 was
intensely monitored between 2011 July and 2012 August with
different X-ray satellites
\cite[][]{livingstone11,rea12,scholz12}. Phase coherent timing
analyses yield spin-down rates between \mbox{$\sim
0.7\times10^{-13}$ s s$^{-1}$} and $\sim 3.1\times10^{-13}$ s
s$^{-1}$, and a dipole magnetic field between $2.4\times10^{13}$ G
and $5.1\times10^{13}$ G \cite[][]{rea12, scholz12}. Despite a
precise measurement of $\dot P$ is not available as yet, any of
the values proposed so far makes Swift\,J1822.3--1606 the magnetar
with the second lowest dipole magnetic field after SGR\,0418+5729.
Given the low value of the period derivative, the characteristic
age of Swift\,J1822.3--1606 is quite long, $\tau_c\sim 0.8$ Myr,
although the source appears not as old as SGR\,0418+5729.

\begin{figure*}
\centering
\includegraphics[width=6in,angle=0]{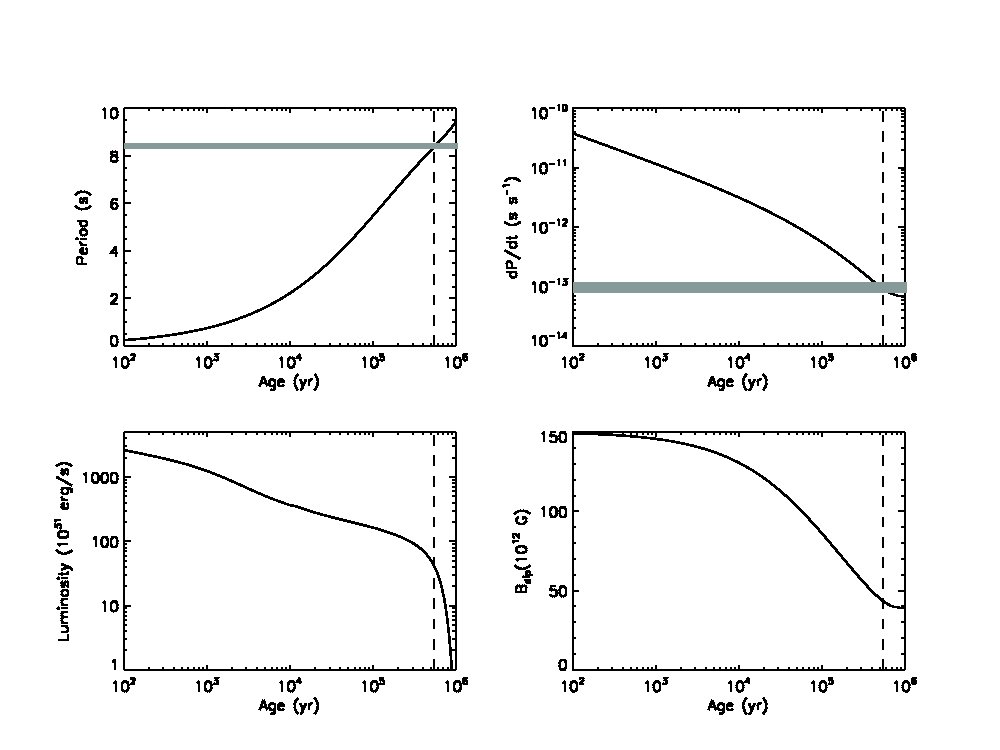}
\caption{\label{fig1822_evol}
From top left to bottom right: time evolution of $P$, $\dot P$,
$L_{\mathrm{X}}$ and $B_p$ for Swift\,J1822.03--1606. The dashed vertical
line marks the estimated age of the source; the gray strips in the first
two panels show the observed values of $P$ and $\dot P$ with their
uncertainties. The model is for $B_{tor}(t=0)=5\times 10^{15}$~G
\cite[from][ \copyright AAS. Reproduced with permission. A link to the
original article via DOI is available in the electronic version
]{rea12}.} \end{figure*}

The large characteristic age, the small number of detected bursts
with comparatively low energetics and the low persistent
luminosity in quiescence have been taken as suggestive that these
are ``old magnetars'' approaching the end of their active life, in
which the magnetic field has experienced substantial decay
\cite[][]{esposito10,rea10,tur11}. A key question is if, and to
what extent, the present (internal) magnetic field is still strong
enough to stress the crust and produce bursts/outbursts. Indeed,
\cite{tur11} and \cite{rea12} have shown that the magneto-thermal
evolution \cite[][see \S\ref{magevol}]{pons09,vigano13} of an
initially ultra-magnetized neutron star, $B_p(t=0)\sim 2\times
10^{14}$ G, can reproduce the observed $P$, $\dot P$, $B_p$ and
$L_{\mathrm{X}}$ in SGR\,0418+5729 and Swift\,J1822.3--1606, for
an age $\sim 1$ Myr and $\sim 0.5$ Myr, respectively, provided
that the initial internal toroidal field $B_{tor}(t=0)$ is high
enough\footnote{More recent calculations of magneto-thermal
evolution including the Hall drift actually show that a strong
toroidal component develops regardless of the initial topology the
the magnetic field \cite[][]{vigano13}.}. The evolution of the
period, period derivative, dipole $B$-field and luminosity for
Swift\,J1822.03--1606 is shown in Fig.~\ref{fig1822_evol}. The
fact that the characteristic age (assuming a constant $B$) in
these two sources is largely in excess of that derived from
magneto-thermal evolution reflects a quite general property of
magnetar sources, for which characteristic ages are longer than
those derived using other estimators. 

According to the calculations by \cite{perna11}, who modeled the evolution
of the internal magnetic stresses in a magnetar, the occurrence of crustal
fractures (and hence of bursts/outbursts) can extend to late phases
($\approx 10^5$--$10^6$ yr). Both the energetics and the recurrence time
of the events evolve as the star ages and depend on the initial field. The
models which successfully reproduce the properties of the two low-field
sources imply that the two low-$B$ magnetars could become
burst-active despite their age, with an expected (current) event rate of
$\approx 0.01$--$0.1\, \mathrm{yr}^{-1}$.

\begin{figure*}
\centering
\includegraphics[width=5in,angle=0]{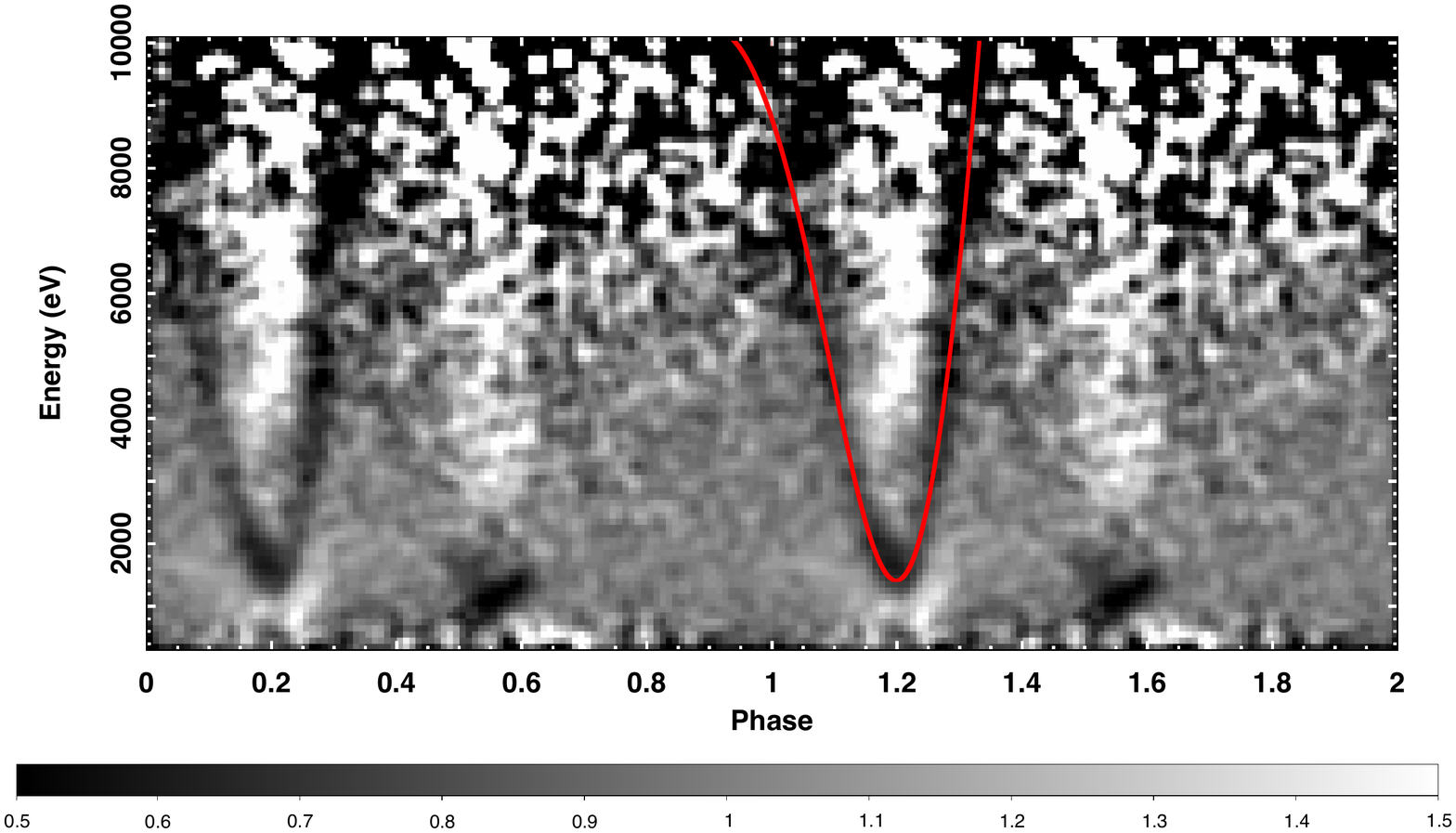}
\caption{The phase-dependent spectral feature in the EPIC data of
SGR 0418+5729. Normalized energy versus phase image obtained by
binning the EPIC source counts into 100 phase bins and 100-eV-wide
energy channels and dividing these values first by the average
number of counts in the same energy bin (corresponding to the
phase-averaged energy spectrum) and then by the relative 0.310
keV count rate in the same phase interval (corresponding to the
pulse profile normalized to the average count rate). The red line
shows (for only one of the two displayed cycles) the results of a
simple proton cyclotron model consisting of a baryon-loaded plasma
loop emerging from the surface of a magnetar and intercepting the
X-ray radiation from a small hot spot \cite[from][
the authors acknowledge Nature Publishing Group for
reproduction]{tiengo13}.}
\label{line-0418}
\end{figure*}

Quite recently, \cite{tiengo13} reported the discovery of a phase-variable
absorption feature in the X-ray spectrum of SGR\,0418+5729. The feature is
best detected in a 67 ks XMM observation performed on 2009 August 12, when
the source flux was still high ($5\times
10^{-12}\,\mathrm{erg\,cm}^{-2}\,\mathrm s^{-1}$ in the 2--10 keV band)
but is also visible in the RXTE and Swift data collected in the first two
months after the outburst onset. The line energy is in the range $\sim
1$--5 keV, in XMM data, and changes sharply with rotational phase, by a
factor of $\sim 5$ in one-tenth of a cycle (see Fig.~\ref{line-0418}).
These seem to favour an interpretation in terms of a cyclotron line. If
protons, e.g. contained in a rising flux tube close to the surface, are
responsible for the line, the local value of the magnetic field within the
baryon-loaded structure is close to $10^{15}$ G. SGR\,0418+5729 would then
be, at the same time, the magnetar with the lowest dipole field and the
neutron star with the largest (small-scale) field ever measured.

\subsection{Transient Radio Emission}\label{radiomag}

For a long time SGRs and AXPs were thought to be with no exceptions radio
quiet, to
the point that the lack of (pulsed) radio emission was often quoted as one of their
defining properties. In fact, an expanding radio nebula was detected
in the
aftermath of the Giant Flares from SGR 1806-20 \cite[][]{gaens05,camer05} and SGR
1900+14 \cite[][]{Frail99}, but this originated in the shocked material around the
neutron star and not from the star magnetosphere.

The first detection of pulsed radio emission from a magnetar came,
rather unexpectedly, from the archetypal transient XTE J1810-197
\cite[][]{camilo06}, opening a new window in the study of
magnetars. For many months, XTE J1810-197 was the brightest radio
pulsar in the Galaxy at frequencies above 20 GHz, exhibiting a
strong variability in both the radio flux and the pulse shape on
different timescales. The radio emission likely began about a year
after the onset of the X-ray outburst and lasted a few years
\cite[][]{camilo06,camilo07a,lazar08,seryl09}. Pulsed radio
emission was then discovered from the AXP 1E 1547-5408
\cite[][]{camilo07b}. This source has shown three X-ray outbursts in
the past 5 years. Radio emission was observed in the interval
between the last two events, during which it declined, to rise
again in coincidence with the last outburst, although there was a
delay of a few days between the onset of the X-ray outburst and
the radio activity \cite[][]{camilo09,burgay09}. Another possibility is that the
radio emission  was blinking, in a way uncorrelated with the start of the X-ray
outburst.
PSR 1622-4950
\cite[][]{levin10} was the first (and so far the only one) magnetar
which was discovered thanks to observations in the radio
band. New and archival X-ray observations have shown that the
source is likely a transient magnetar which underwent an outburst
in 2007 (before the first available Chandra observation). The
X-ray flux is still declining as the source approaches quiescence
\cite[][]{ander12}. Finally, pulsed radio emission has been detected
from the recently discovered magnetar in the Galactic Centre, SGR
J1745-2900 \cite[][]{reagc13,eat13,shajoh13,kaspi14}. The source
entered an outburst phase with the emission of a single burst on
April 24 2013. The radio activity switched on 4-5 days after the
outburst onset with similar properties to those of the other two
magnetars detected at radio wavelengths. Despite intensive
searches, pulsed radio emission was not found in other magnetar
sources \cite[][]{burgay06,crawf07,laz11}.

Although the sample is quite limited, a number of common features
in the radio emission from magnetars have started to emerge: i)
association with X-ray outbursts, ii) the radio flux decays
together with the X-ray flux but its onset is delayed, iii) marked
variability in the radio band, and iv) flat radio spectrum. All
these properties are much at variance with those of ordinary radio
pulsars (including the high-B PSRs). \cite{rearadio12} have shown
that radio-loud magnetars are characterized by $L_X<\dot E$ in
quiescence, much as ordinary radio pulsars, while the X-ray
luminosity always exceeds the rotational energy loss rate in
radio-silent magnetars. In fact, as indicated by \cite{ho13}, this
property seems to be characteristic also of other classes of
neutron stars, although the inverse is not true: not all sources with $L_X<\dot E$
emits in radio. SGR J1745-2900, which was
not included in the
original sample considered by \cite{rearadio12}, further confirms
this picture, having $\dot E\sim 5\times 10^{33}\
\mathrm{erg\,s}^{-1}$ and $L_X\lesssim 10^{32}\
\mathrm{erg\,s}^{-1}$ in quiescence \cite[][]{reagc13}. This seems
to point to a common origin for the radio emission in PSRs and in
transient magnetars. In magnetars with $L_X/\dot E<1$ particle
acceleration and the subsequent ignition of the cascade process
could proceed as in normal pulsars, and their radio emission might
basically follow the same rules, with rotational energy driving
pair creation through a cascade. The largely different radio
properties between the two groups might result from the presence
of a substantial toroidal component in the magnetosphere of the
magnetars, contrary to the nearly dipolar field of PSRs. The
influence of the large charge density required to support the
non-potential field may also act in quenching the radio emission
in the brightest magnetars with $L_X/\dot E>1$
\cite[][]{thomp08a,thomp08b}, although this latter interpretation is hard
to be reconciled with the fact that, in at least few sources
(SGR~J1745-2900, XTE~J1810-197 and possibly also in 1E~1547-5408 during
the 2009 outburst), radio emission
was seen while the source was bright with  $L_X>\dot E$.

\section{Magnetar bursts}
\label{bursts}

\subsection{Burst phenomenology}
\label{burstobs}

\begin{figure}
\centering
  \includegraphics[width=3.in]{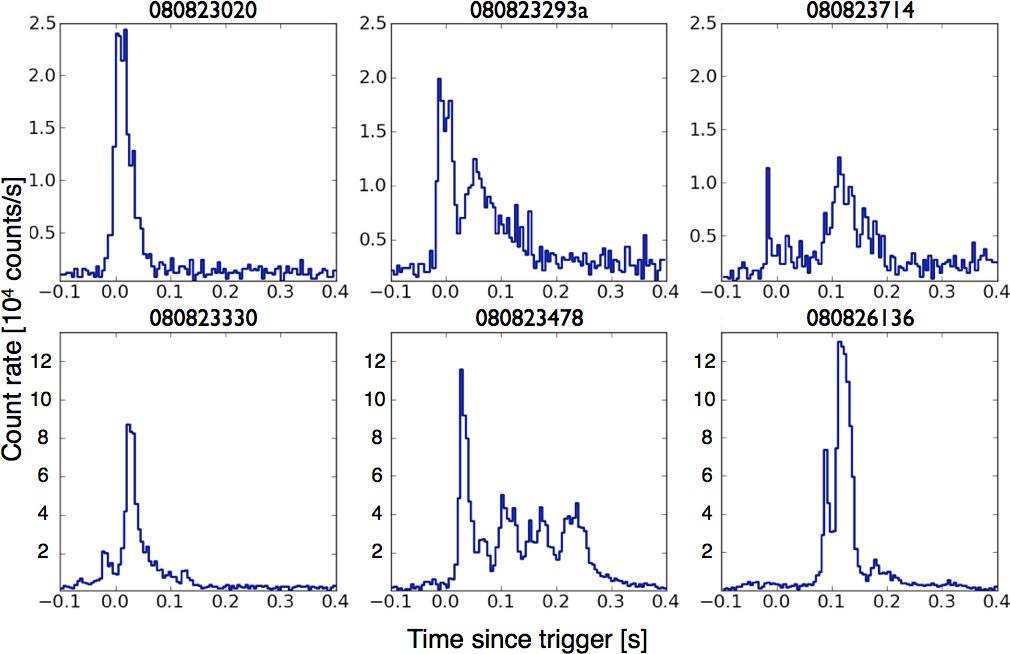}
  \includegraphics[width=2.5in]{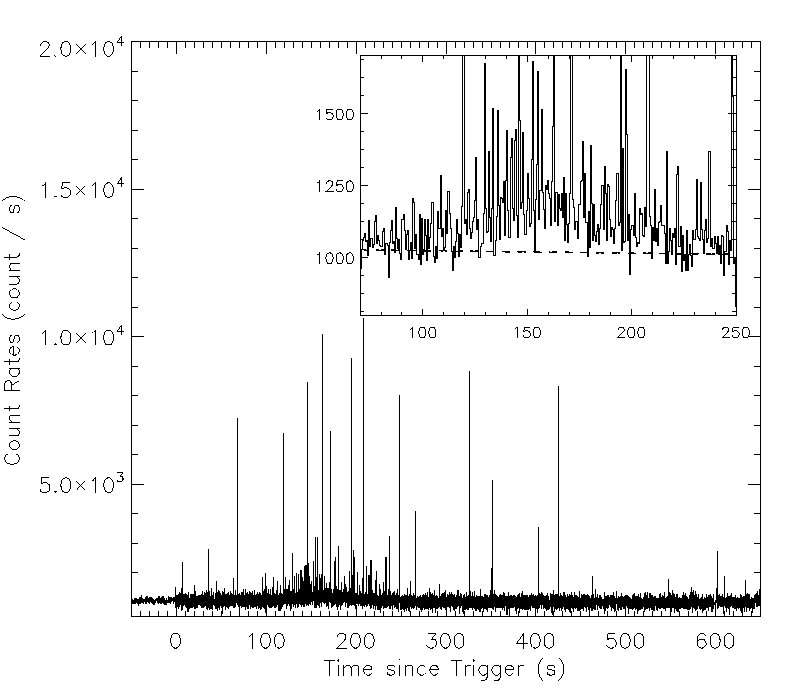}
  \includegraphics[width=3.in]{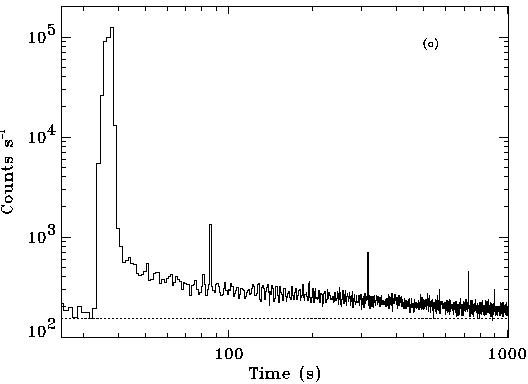}
  \includegraphics[width=3.in]{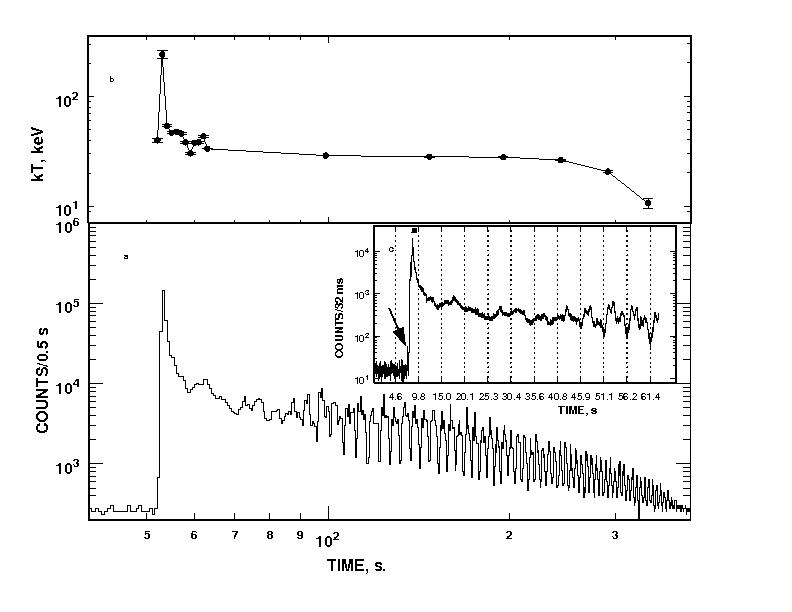}
\caption{Examples of different types of magnetar bursts.  Top left:
 Short bursts from SGR 0501+4516 recorded by Fermi/GBM
\cite[from][]{huppenkothen13}.  Top right:
  Burst storm from 1E 1547.0-5408 recorded by
Fermi/GBM, with inset showing the overall enhancement in emission during this
  event\cite[from][]
{Kaneko10}. Lower left:  Intermediate burst from SGR
1900+14 recorded by RXTE \cite[from][]{Ibrahim01}. Lower right: Giant
  flare from SGR 1900+14 recorded by Ulysses, upper panel showing OTTB
spectral temperature \cite[from][]{Hurley99}. The first three panels are
\copyright AAS, reproduced with permission. A link to
the original articles via DOI is available in the electronic
version. The last panel is reprinted by permission from Macmillan
Publishers Ltd: Nature, \copyright 1999. The authors
acknowledge Nature Publishing Group for allowing reproduction.
}
\label{burstlc}
\end{figure}

One of the hallmarks of magnetars is the emission of repeated soft
gamma-ray bursts, and this feature played a key role in their
discovery \cite[for a nice review of the history of the field,
see][]{woodth06}. Bursts have now been observed from 18 sources
whose spin has also been measured (confirming that they are
indeed neutron stars)
\footnote{See the Amsterdam Magnetar
Burst Library, http://staff.fnwi.uva.nl/a.l.watts/magnetar/mb.html}.
Magnetars emit bursts in the few keV to few
hundred keV energy band (hard X-ray/soft gamma-ray). As mentioned in the
Introduction, magnetar bursts are typically grouped into three
classes defined primarily by energy released and their duration:
short bursts (with energies up to $10^{41}$ erg), intermediate
flares (energies in the range $10^{41}-10^{43}$ erg), and giant
flares (energies in the range $10^{44}-10^{46}$ erg, see
Fig.~\ref{burstlc} for example lightcurves from the different types of
burst).

Bursting activity is highly variable.  Sources can experience long
periods of apparent quiescence (when bursting, if it occurs, is
in the form of low luminosity bursts below the detection threshold
of our current generation of space telescopes - note also that there have
been gaps in gamma--ray telescope coverage since the discovery of
magnetars, due to the lack of suitable telescopes, and that sky coverage
even now is never 100\%), sporadic and
highly occasional bursting, or periods of high burst activity. At their
most dramatic, these can climax in burst storms (when
several hundreds of bursts can be emitted over just a few hours), or
rare giant flares (Fig.~\ref{burstlc}).
During such outburst periods there may also be changes in the
overall emission (luminosity, pulsed flux and spectrum, see
\citealt{reaesp11} for an observational review, and
\citealt{pons12} for a recent theoretical study) and the timing
behaviour (spin-down rate and glitching, see for example
\citealt{Woods01,Woods02,Woods03,Woods07,Dib14}). There is no
clear dependence of short/intermediate bursting activity on the
dipolar magnetic field strength of the source.  The most energetic
giant flares come from three sources whose inferred dipole fields
are amongst the highest (a few times $10^{14}$ G up to $\sim
10^{15}$ G).  However bursts have also been seen from sources with
fields apparently below the quantum critical limit, such as Swift
J1822.3-1606 with a field $\approx 1.35\times 10^{13}$ G
\citep{Scholz14} and SGR 0418+5729 with field $6\times 10^{12}$ G
\citep{rea10}\footnote{Although note that the dipole field
strength as estimated from spin down provides only a lower limit
on the magnetic field strength, and there are no constraints on
the strength of either higher order poloidal components or
toroidal components.}.

\subsubsection{Short bursts}
\label{shortbursts}

Short bursts are the most common, with fluences in the range
$10^{-11} - 10^{-5}$ erg/cm$^2$/s, which for the assumed distances
implies isotropic energies in the range $10^{36} - 10^{41}$ erg
\citep{Gogus99,Gogus00,Woods99a,Gavriil04,Kumar10,Scholz11,Lin13}.
Peak luminosities are in the range $10^{36} - 10^{42}$ erg/s,
extending to well above the standard Eddington luminosity of
$\approx 2 \times 10^{38}$ erg/s for a non-magnetic neutron star.
The lowest luminosity bursts recorded are at the sensitivity limit
of our current generation of detectors.  For sources that have
shown sufficient numbers of bursts for meaningful statistical
analysis (SGR 1806-20, SGR 1900+14, SGR 1627-41, 1E 2259+286, SGR
0501+5416 and 1E 1547.0-5408), burst fluences are distributed as a
power law,  dN = E$^{-\gamma}$ dE with $\gamma \sim 1.4-2.0$
\citep{Cheng96, Aptekar01, Woods99a, Gogus99, Gogus00,
Gogus01,Gavriil04, Kumar10, Savchenko10,Lin11,
Scholz11,Prieskorn12,vanderhorst12}.
Burst durations are $\sim 0.01-1$s and are distributed
lognormally  with a peak at $\sim 0.1$s, less than the
rotational period of the star. Duration is correlated with
fluence. Bursts from the sources that have shown insufficient
events for full statistical analysis are consistent with this picture.
Lightcurves are
extremely variable in shape: the rise is in general faster than
the decay \citep{Gogus01,Gavriil04}, but bursts can be
multi-peaked, and no simple phenomenological model has yet been
found that would fit the morphologies of the different burst
lightcurves. Some short bursts (classified according to their
fluence) from five sources have extended faint tails of $\sim 100-1000$ s in
duration, leading to an overall energy release that can exceed
that of the original spike
\citep{Woods05,Gavriil04,Gavriil02,Gavriil06,Dib09,An14,Gavriil11,
Mereghetti09,Savchenko10,Scholz11}.
The tails appear to be pulsed at the rotational frequency.

From the sources with large samples, wait times (for bursts above
the detection threshold, in periods of continuous telescope
coverage\footnote[1]{The recent detection of short bursts from 1E
1547.0-5408, in the VLF radio band due to the ionospheric
disturbance that occurs as the incident gamma-rays ionize the
Earth's atmosphere opens up the possibility of using the VLF band
to obtain a more complete coverage of the waiting time
distribution with being dependent on sky coverage of space
telescopes \citep{Tanaka10}.}) form a lognormal distribution with
a peak at $\sim 100$s.  During the active period of SGR 1900+14 in
1998, for example, \citet{Gogus99} found wait times ranging from
less than 1s to more than 1000s\footnote{The longest wait times in this
study were set by the length of the observing window.}, and far longer
wait
times are clearly possible: SGR 1900+14, for example, has also had
quiescent periods $\sim$ years in duration. The shortest wait
times observed are comparable to the durations of individual
bursts, such that the distinction between single bursts and
multi-peaked events is not clear. There appears to be no
correlation between burst intensity and wait time to the following
burst \citep{Laros87,Gogus00,Gavriil04,Savchenko10}. Of
particular note in the discussion of wait times are
burst storms, periods of unusually high short burst activity (which may
include some intermediate flares), in which tens to hundreds of bursts
occur over only a few hours on top of an overall rise in emission
that can be strongly pulsed at the rotational phase \cite[see for
example][for the cases SGR~1900+14, 1E 2259+286, and 1E 1547.0-5408]{
Hurley99a,Israel08,Gavriil04,Mereghetti09,Kaneko10,Savchenko10}.
There have also been
efforts to determine whether the occurrence of bursts correlates
with rotational phase: here the evidence is mixed. For 1E
1048.1-5937, 1E 2259+286 and XTE J1810-197 bursts do seem to occur
preferentially at rotational pulse maxima
\citep{Gavriil02,Gavriil04,Woods05}. However for SGR 1806-20, SGR
1900+14, SGR 1627-41, 4U 0142+61 and 1E 1547.0-5408 no such
correlation is found \citep{Palmer99,Palmer02, Woods99a,
Gavriil11, Savchenko10,Scholz11,Lin12a}.

\subsubsection{Intermediate flares}
\label{IFs}

Intermediate flares, with (isotropic) energies in the
range $\sim 10^{41} - 10^{43}$ erg, and peak luminosities that
exceed the non-magnetic Eddington limit, have been seen
from SGR 1627-41, SGR 1900+14, SGR 1806-20, and 1E 1547.0-5408.
The primary bursts appear to be brighter and slightly longer (durations
$\sim 0.5$s up to a few s) versions of the short bursts. Morphologies,
however, are varied. Some have a
clear decay and an abrupt end \citep{Mazets99a, Mazets99b,Olive04,
Israel08}. In others the initial burst is followed by a
extended decaying tail that can last for up to several thousand
seconds, but contains less than $\sim 2$ \% of the energy
released in the initial peak \citep{Ibrahim01,Lenters03,Esposito07, Gogus11}. The
tails are pulsed at the rotational period of the star:  in some
cases the pulsed amplitude rises dramatically
\citep{Ibrahim01,Lenters03}, as seen during some burst storms
\citep{Kaneko10}; whilst in others no change is seen
\citep{Gogus11}.   The pulsations appear for the most part to be
phase-aligned with the pre-burst pulsations, but there are some
occasional exceptions \citep{Guidorzi04,Gogus11}.   There has also
been one burst with a decaying pulsed tail, where the sharp
initial burst peak appears to be absent, leading to a rather slow
rise time $\sim 10$s \citep{Kouveliotou01,Guidorzi04}.
Intermediate flares sometimes have short precursors
\citep{Ibrahim01,Gogus11}, and during some events short bursts are
seen during the extended tail \citep{Ibrahim01,Lenters03,
Gogus11}. 

\subsubsection{Giant flares}
\label{GFs}

The most energetic bursts, the giant flares, are extremely rare.
Only three have ever been seen, in 1979, 1998 and 2004, each from
a different magnetar (SGR 0526-66, SGR 1900+14, and SGR 1806-20).
The total energy released, if the emission is isotropic and
assuming reliable estimates of distance, is in the range
$10^{44}-10^{47}$ erg \citep{Fenimore96,Feroci01,pal05}.  The
overall properties of the three giant flares are, despite the
differences in energy, very similar (in marked contrast with the
heterogeneity of the intermediate flares).  They have a very
bright initial peak, followed by an extended decaying tail with a
duration of several hundred seconds that is strongly pulsed at the
rotational frequency of the star\footnote[1]{Note that none of the
giant flares were caught during pointed observations: they are so
rare but so bright that most are seen off-axis.  This means that
the sensitivity to late time weak emission is much less than for
some of the intermediate flares, which were observed during
pointed observations. One should bear this in mind when comparing
the apparent durations.} \cite[see e.g.][]{maz79b,Hurley99,hur05}.

The initial peaks, which can have rise times as short as $\sim 1$
ms, last $\sim 0.1-1$ s and are very hard.  Luminosities reach up
to $10^{47}$ erg/s \citep{hur05}, which causes substantial dead
time and pile up effects in space telescopes.  This renders
reliable spectral modelling very difficult, however the spectrum is
very hard, with emission being detected up to 2 MeV
\citep{maz79b,Hurley99}.   The initial peaks are strongly
variable, on timescales as short as a few ms
\citep{Barat79,Hurley99,Terasawa05,Schwartz05}.  Both the SGR
1900+14 and SGR 1806-20 giant flares were observed to have
precursors\footnote[1]{A precursor with the properties observed for
SGR 1900+14 giant flare would not have been detectable for the SGR
0526-66 giant flare given the instrumentation at the time
\citep{Gill10}.}.  For SGR 1900+14 the precursor resembled a
normal short burst, and occurred $ < 1$ s before the giant flare
\citep{Mazets99c}.  For SGR 1806-20 the precursor was flat-topped,
with an energy release that puts it in the intermediate flare
class, and occurred 142 s prior to the giant flare \citep{hur05}.  A discussion of whether the apparent precursors are in fact
genuinely causally connected to the giant flares was presented by
\cite{Gill10}.

The energies emitted in the tails of the three giant flares have been similar ($\sim 10^{44}$ erg). This is 
 $1-2$\% of the energy released in the initial peak of the 2004 giant flare: for the two earlier giant flares the energies 
 released in initial peak and tail were comparable.  The overall envelope of the tails (averaged over rotational
phase) decays smoothly as a power law, coming to an abrupt end
after several hundred seconds.  The pulsations in the
SGR 0526-66 giant flare tail were seen immediately after the
initial peak; for the other two giant flares they appeared only a
few tens of seconds later.  Pulse profiles can evolve during the
tails, in the case of SGR 1900+14 simplifying quite dramatically
\citep{Feroci01}, with evolution being much more minimal for SGR
1806-20 \citep{pal05,Mereghetti05,Boggs07,Xing11}.  The giant
flares are so strong that they have a detectable effect on the
Earth's electromagnetic field \cite[][]{man06} and
ionosphere \citep{Inan99,Inan07,Tanaka08},
with even the rotational pulsations being
clearly visible, and this has been
used to put lower limits on the strength of low energy ($< 10$
keV) emission from the burst (unaffected by satellite dead time
issues).

Radio afterglows were detected after both the SGR 1900+14 and SGR
1806-20 giant flares
\citep{Frail99,camer05,gaens05,Gelfand05,Taylor05,Fender06}. The
amount of energy in the radio afterglow is much less than that
emitted in the gamma-rays. This is different from what is observed in gamma
ray bursts, where the ratio between the two types of emission is of order unity, such that the lower energy and longer duration emission is associated
with re-processing of the gamma-ray energy by the surrounding material. The radio afterglow from the magnetar giant flares
is linearly polarized,
implying that it is
caused by electron synchrotron emission, and is observed to expand over
time.  Its generation requires an ejection of relativistic
particles and magnetic fields (in the form of a ``plasmon'', which
expands and cools) by the burst process (see Section \ref{emission}).

\subsection{Burst trigger mechanisms}
\label{trigger}

Rapid magnetic field reconfiguration is assumed to be
an integral part of the bursts: as we will
discuss in Sec.~\ref{emission}, the gamma-ray emission is
assumed to come from particles accelerated by rapid field change.
Thus slow magnetic evolution builds up stresses in the
system, some of which are released catastrophically in bursts,
which must either be driven by or result in rapid magnetic
field reconfiguration.  However the precise trigger mechanism, and
the role of the magnetic field within it (if any), remains
unclear.   Three main locations (and associated families of
instabilities) have been considered for the trigger mechanism:
below we review each in turn.

The first option, as suggested by \citet{thdun95}, is that the
magnetic field evolves into an unstable configuration within the
liquid core of the star \citep{Markey73,Wright73,Tayler73,flowers77}, which is then
susceptible to a large-scale magnetohydrodynamical instability
\citep{Lander11,Ciolfi11,Kiuchi11,Ciolfi12}.  This would develop
on the Alfv\'en crossing time of the core, which is $\sim 0.1$s
for a $10^{15}$ G interior field ($\tau \sim R/v_A$, where the Alfv\'en speed $v_A$ is given in Equation \ref{va} later in this paper) and hence broadly compatible with the durations of both the normal bursts and the initial peaks of the giant flares.

Whether such an unstable
state could develop is open to question, since if the core is
superconducting, the high conductivity should facilitate swift
reconfiguration and hence prevent the formation of such an
unstable state.  However the configuration could be stabilized by
currents in the crust, or superconductivity may be suppressed if
the core field exceeds $\sim 10^{16}$G.  Such rapid internal
magnetic reconfiguration would inject an Alfv\'en pulse into the
magnetosphere that would then generate the observed burst
emission.  The energy available to power the burst in this
scenario, since it originates in the core, is more than sufficient
to power even a giant flare. Some internal heat release may also
be expected due to dissipation in core and crust as the
instability proceeds.

The second option is that the decay of the core field places
magnetic stresses on the solid crust of the star (an ionic lattice
to which the field in the crust is locked).  The crust can deform
elastically to accommodate this up to a certain point, then
ruptures catastrophically once its breaking strain is exceeded
\citep{thdun95,Thompson01}.  The stored energy released in this
scenario would come from the crust and possibly also the core:
although initially it was thought that the crust alone could not
store enough elastic energy to power the giant flares
\citep{thdun95}, the latest molecular dynamics simulations predict
a higher breaking strain, indicating that this may be feasible
after all \citep{Horowitz09, Hoff12}.  Studies are now underway that aim
to determine how and where stresses would build up in the crust as
a result of core field evolution, the goal being to determine how
often, and where (location including depth) the crust is most
likely to fail \citep{perna11,Pons11,Beloborodov14,Lander14}.   Recent calculations by \citet{Lander14} of the strain induced in a crust by a changing magnetic field configuration find a characteristic burst energy 

\begin{equation}
\frac{E}{10^{45} \mathrm{erg}} \approx 0.25 
\left(\frac{\sigma_\mathrm{max}}{0.001}\right) \left(\frac{d}{R_c}\right)^2 \left(\frac{l}{2\pi R}\right)
\end{equation}
where $\sigma_\mathrm{max}$ is the breaking strain of the crust \citep[which could be as high as 0.1;][]{Horowitz09}, $d$ is the depth at which the crust ruptures, $R_c$ the crust thickness, $R$ is the star radius, and $l$ the rupture length.  The characteristic local field strength related to crust breaking (from the same study) is given by

\begin{equation}
B_\mathrm{break} = 2.4\times 10^{14} \mathrm{G}  \left(\frac{\sigma_\mathrm{max}}{0.001}\right)^{1/2} \, . 
\end{equation}

When the crust does rupture it must do so by rapid plastic deformation,
not via brittle fracture, due to the impossibility of opening up
voids under the conditions of extreme pressure that pervade in
neutron star crusts \citep{Jones03}.  The role of the magnetic
field during crust rupture however is not clear: \cite{Levin12}
have argued that under some circumstances the field deformation
induced by an incipient rupture may act as a brake on its
propagation. Shear wave timescales (which control the timescale on
which the crust ruptures, $\tau \sim \pi R/v_s$, where the shear speed $v_s$ is given by Equation \ref{vs} later in this paper) are compatible with those observed in
bursts and flares \citep[see for example][]{Schwartz05}, injecting
an Alfv\'en pulse into the magnetosphere.  However the transfer of
energy into the external magnetosphere may be slowed by a large
impedance mismatch at the crust-magnetosphere boundary
\citep{Link14}. Crustal rupture would most likely lead to local
heating as well.

The final possibility is that the core and crust evolve smoothly,
and that stress builds up instead in the magnetosphere.  Stress
release is then envisaged as taking place via a plasma instability
involving spontaneous magnetic reconnection
\cite[see][for a review]{Uzdensky11}.  A number of studies have looked
at how the external magnetosphere might respond to the expulsion
of magnetic helicity due to the decaying core field, and have
found that the development of unstable configurations with strong
magnetic shear, that might be prone to reconnection instabilities,
is feasible \citep{tlk02,belo09,Parfrey12,Parfrey13}. The
resulting explosive reconnection event would progress on the
Alfv\'en crossing time in the magnetosphere, which is $\lesssim 0.01$
s (since in the magnetosphere $v_A \sim c$), and the energy that can be released is
more than sufficient to power even the giant flares.  Specific
instabilities that have been considered primarily for the giant
flares (driven by earlier concerns about the ability of crust
ruptures to release enough energy) are the relativistic tearing
mode \citep{Lyutikov03,Lyutikov06,Komissarov07} and collisionless
Hall reconnection mediated by emission from precursor bursts
\citep{Gill10}, the precursors presumably being triggered in this
scenario by another mechanism.   Similarities between the giant
flares and reconnection driven coronal mass and flux tube ejection
events in solar physics have also been explored in some depth
\citep{Masada10,Yu11,Yu12,Yu13,Huang14a,Huang14b,Meng14}. Instabilities
may also arise from the interaction of MHD waves with the vacuum
in fields above the quantum critical limit \cite[where QED effects are
important,][]{Heyl05}. All of these mechanisms would lead
directly to particle acceleration and radiation in the
magnetosphere, with possible crustal heating via particles
impacting the surface.

At present it is by no means clear which of the various mechanisms
are in operation, and given the diversity of burst properties
(Sec.~\ref{burstobs}) more than one may be in operation.
Timescales, as discussed above, appear to be roughly
compatible with either burst rise times or durations (assuming that
the emission process timescales reflect
those of the trigger mechanism). Moreover both starquakes and
magnetospheric reconnection could in principle
explain the power law distribution of fluences
\cite[which is often taken as evidence for Self-Organised
Criticality, see for example][]{Aschwanden14}.
Serious efforts are now being made to simulate the build up of
stress and the development of instabilities, but in order to allow
meaningful tests of the data, consideration must be given to how
these various triggers connect to the emission that we see.   We
discuss this in the next section (Sec.~\ref{emission}).

\subsection{Burst emission processes}
\label{emission}

\subsubsection{Sources of emission}

Magnetar bursts are complex, with varied spectra and morphologies.
The emission process for all bursts is generally assumed to be
started by rapid rearrangement of the magnetic field (resulting
from one of the trigger mechanisms discussed in Sec.~\ref{trigger}),
possibly involving either induced or spontaneous
magnetic reconnection.   This accelerates charged particles with ensuing
gamma-ray emission, since the rapid
acceleration of electrons in a strong curved field leads to a
cascade of pair creation and gamma-rays \citep{Sturrock89}.  To
obtain the hardest emission, there must be very low contamination
by baryonic material (since scattering would lead to softening).
Fully self-consistent models of the emission resulting from
the various proposed trigger mechanisms do not yet exist.
Despite this, the rise timescale of the gamma-ray emission has frequently
been used as a key piece of evidence to argue for a particular trigger
mechanism, as discussed in the previous section.
However it is not clear that this is warranted: details of the gamma-ray emission
process may in fact completely obscure the timescales associated with the original
trigger mechanism \cite[see for example][]{Hoshino12}.

To obtain the radio afterglow seen in the giant flares, it is
necessary to postulate the ejection of a plasmoid of magnetic
fields and trapped shocked plasma, that gradually cools
(Sec.~\ref{burstobs}). Such plasmoid ejection is a natural and expected
consequence of a large-scale reconnection event in the
magnetosphere \citep[see Sec.~\ref{trigger} and for example][]{Lyutikov06}.

The initial spike of magnetar flares may also lead to radio emission \citep[explored for example in][]{Lyutikov02}.  More recently it has been suggested that the interaction between strongly magnetized relativistic ejecta (expelled by the initial spike of giant flares) and the surrounding wind nebula might be responsible for extragalactic Fast Radio Bursts \citep{Popov07,Popov13,Thornton13,Lyubarsky14}. At present this remains speculative. 

If the local energy generation rate is high enough, as explained
above, it will lead to copious production of electron-positron
pairs and gamma-rays. If this occurs in a closed field line
region, where the charged pairs cannot cross magnetic field lines,
they become trapped.  As density increases, so does optical
thickness, trapping the photons as well and leading to rapid
thermalization \citep{thdun95}.   The field necessary to confine the plasma can be estimated by requiring that magnetic pressure exceed the pressure of the radiation and the pairs at the outer boundary of the fireball, yielding

\begin{equation}
B_\mathrm{dipole} > 2 \times 10^{14} \left(\frac{E_\mathrm{fireball}}{10^{44} \mathrm{~erg}}\right)^{1/2} \left(\frac{\Delta R}{\mathrm{10~km}}\right)^{-3/2} 
\left(\frac{1+ \Delta R/R}{2}\right)^3 ~\mathrm{G}
\end{equation}
\citep{Thompson01}, where $\Delta R$ is the characteristic size of the fireball and $R$ the neutron star radius.  

Such a trapped pair plasma
fireball would then cool and contract due to radiative diffusion
from a thin surface layer, with the bulk of the radiation leakage
occurring close to the stellar surface since this is where the
field is strongest and scattering the most suppressed. The opacity
at the surface will be dominated by the electron-ion plasma ablated
from the neutron star surface (especially if the emergent flux is
close to the magnetic Eddington limit) which form the photosphere.
This trapping would prolong the emission from the burst by acting
as a reservoir for the energy.  Fireball formation is not
exclusively linked to any one trigger mechanism
\citep{thdun95,Heyl05}.

\begin{figure}
\centering
\includegraphics[width=3in]{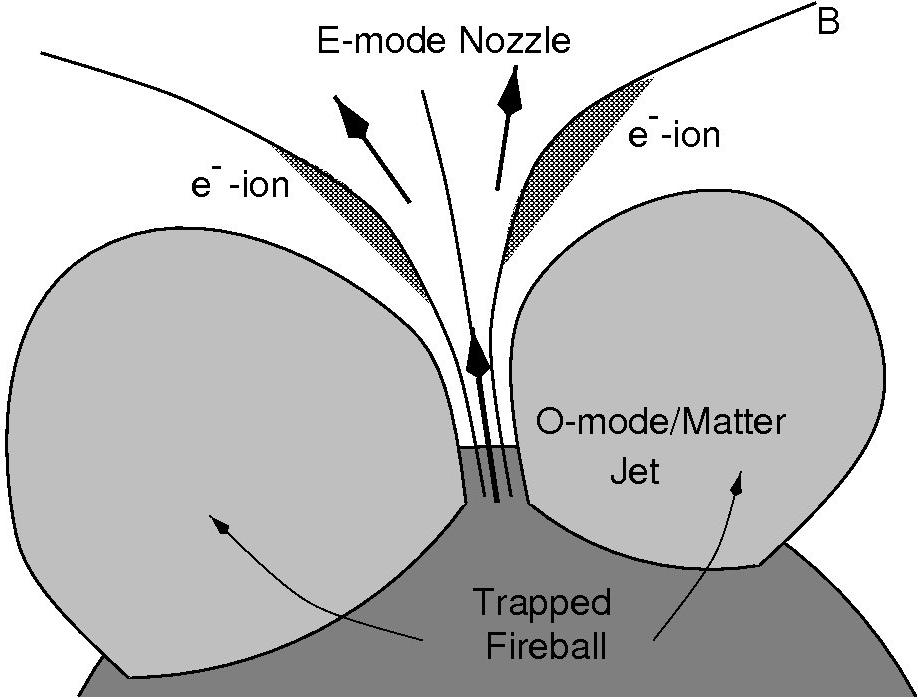}
\includegraphics[width=3in]{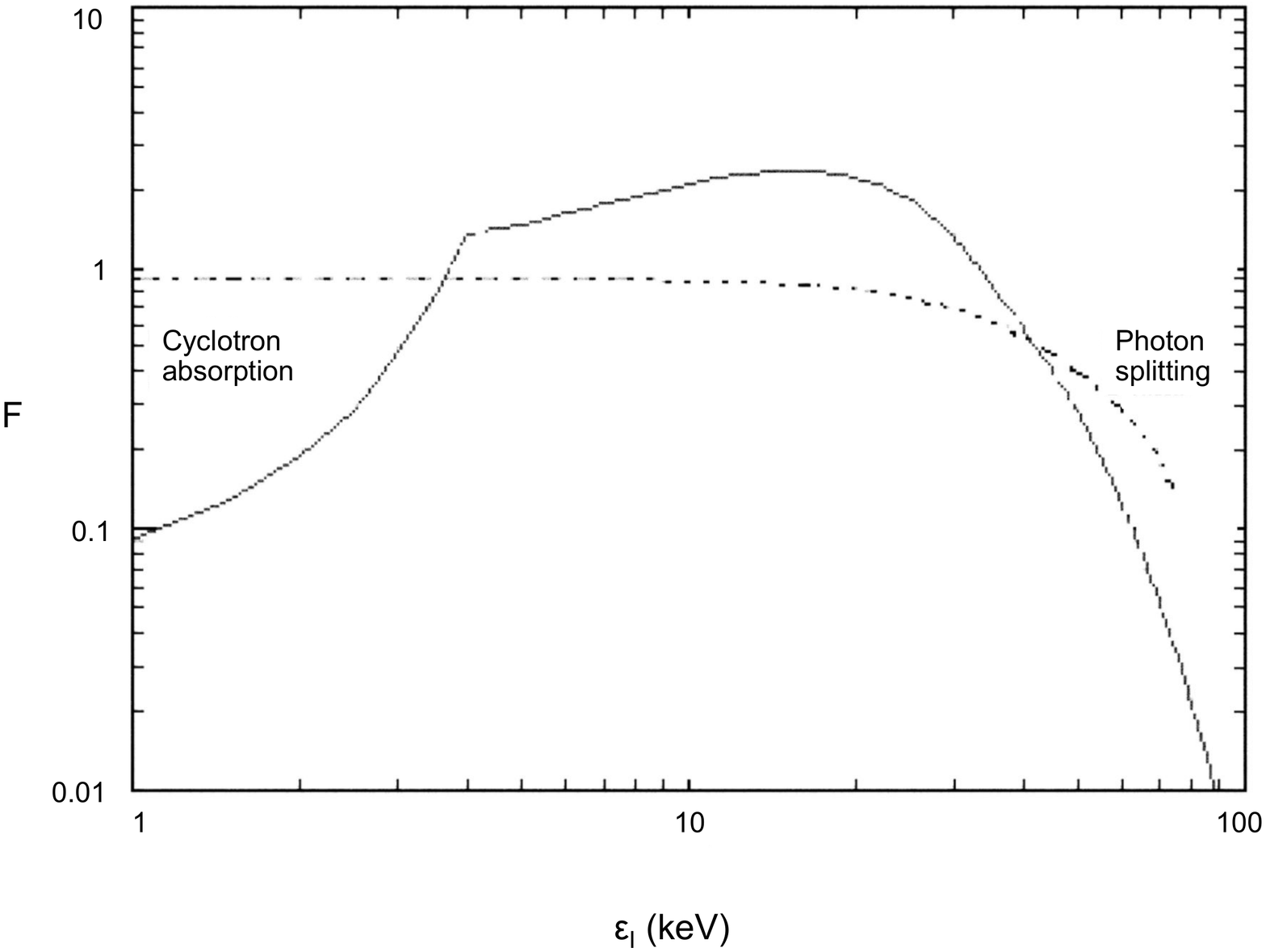}
\caption{Left: Figure from \citet{Thompson01},
to illustrate how radiation escapes
from a trapped photon-pair plasma fireball.  Scattering opacities are strongly
polarization dependent, with most radiation escaping from the fireball in the
E-mode.  This radiation is then collimated along open magnetic field lines (due to
the magnetic field dependence of the scattering opacties), forming a beam that is
then modulated by the star's rotation to give rise to a rotational pulse
\cite[from][\copyright AAS. Reproduced with permission. A link to
the original article via DOI is available in the electronic
version]{Thompson01}.
Right:
Figure from \citet{Lyubarsky02a} showing how the photospheric spectrum from a
fireball with temperature 15keV (dashed line) would be modified by various high
field radiation processes: cyclotron absorption, for a $6\times 10^{14}$ G
field, and photon splitting \cite[from][with OUP
permission]{Lyubarsky02a}.
}
\label{burstrad}
\end{figure}

The fireball model has been very successful at explaining the
later decaying tail phase of giant flares \citep{Thompson01}.  A cooling fireball trapped on closed field lines should have a luminosity $L$ whose time-dependence is described by the following function 

\begin{equation}
L (t) = L(0)\left[1 - \frac{t}{\tau_\mathrm{evap}}\right]^{a/(1-a)}
\end{equation}
where the cooling luminosity is assumed to vary as a power of the remaining fireball energy, $L \propto E^a$ 
\citep{Thompson01}.  This proves to be a good fit to giant flare tail data, with $\tau_\mathrm{evap}$ of order a few hundred seconds, 
and a value of $a$ that is close to that expected for a spherical fireball of uniform temperature \citep{Feroci01}. 
Spectral fitting indicates that the emitting area falls while the
temperature of the radiation remains roughly constant at the level
expected for the photosphere of a trapped fireball in a magnetic
field in excess of the quantum critical field
\citep{thdun95,Feroci01}. The observed photospheric temperature ($\sim
20-30$ keV) is lower than the inferred temperature in the core of
the fireball, which is $\sim 100$ keV \citep{thdun95,Thompson01}.  \citet{thdun95} argue that 
this is an intrinsic property of the way various processes act to preserve thermal equilibrium in the fireball, 
and the way that radiation gradually escapes. However other processes such as photon splitting as the emitted radiation 
propagates through the strong magnetic field may also be important \citep{Baring95}. The beaming of radiation as it
leaks from the base of the fireball and streams along field lines
provides a simple explanation for the strong rotational pulses
seen in the giant flare tails (Fig.~\ref{burstrad}). Whether fireballs
form in the
smaller bursts is still not clear (energy release may not occur at
a fast enough rate \citep{Gogus01}, although the similarity of the
spectra of the short bursts to the spectra in the tails of the
giant flares suggests that there may be a link.

Some of the energy released during the burst is likely to excite
vibrations, either of the star (crust/core, see Sec.~\ref{seis}) or in the
form of Alfv\'en waves in the magnetosphere.
This too can act as a store for energy that is then radiated on
longer timescales.  If the vibration rate is slow enough, this can act as a source of ongoing excitation that forms an extended pair corona (that obscures and scatters the radiation from the trapped fireball) from which the radiation emerges isotropically \citep{Feroci01,Thompson01}.  The presence of such an extended pair corona has
been invoked to explain the smooth emission immediately after the
peak of the giant flares, which then clears over 30-40 s to reveal
the strongly beamed rotational pulse emanating from the trapped
fireball beneath \citep{Feroci01}.

A burst may also have a thermal component of emission that is
produced by residual heat of crust rupturing
\citep{let02,Kouveliotou03}, extreme heating and possible melting
of the crust immediately underneath a trapped fireball
\citep{thdun95}, or bombardment of the stellar surface by
magnetospheric particles \citep{Lyutikov06,belo09}.  Such thermal
components, particularly deep crustal heating, are one possible
explanation for the presence of both the additional pulsed
components seen after some bursts (short and intermediate) - as a
localised hotspot - and the must longer decaying afterglows seen
after burst active periods, intermediate flares and giant flares
\citep{Kouveliotou03,Feroci03}.

The location of the emitting regions is further complicated by the fact that the
luminosity of the bright bursts may exceed the relevant Eddington limit, leading to
photospheric expansion and ejection of material as radiation pressure overwhelms the
gravitational force \cite[see for example][]{thdun95,Watts10}.  Magnetar bursts can
easily exceed the non-magnetic Eddington limit.  However strong magnetic fields
suppress scattering opacities, increasing the Eddington luminosity even before
magnetic confinement effects - which can increase the limit still further - are
taken into account \citep{paczynski92,thdun95,Miller95,vanputten13}.

\subsubsection{Radiative transfer processes}

Short burst spectra, in an era of improved broadband coverage, are typically well
fit as either two blackbodies (2BB, with temperatures $\sim 5$ keV and $\sim 15$
keV) or using a Comptonization (power law with a high energy cutoff) model
\citep{Feroci04,Olive04,Nakagawa07,Israel08,paolo08,Lin11,Scholz11,vanderhorst12,Lin12a}.
There is a sharp correlation between radius and temperature of the blackbodies in
the 2BB fits, with the softer BB component saturating before the harder one as burst
fluence increases.  For some sources bursts tend to harden with increasing fluence,
whereas for others they soften
\citep{Gogus01,Gavriil04,Gotz04,Kumar10,Savchenko10,Scholz11,vanderhorst12}.
Earlier burst papers tend to use an Optically Thin Thermal Bremsstrahlung
(OTTB) model
\citep{Gogus99,Woods99a,Gogus00,Aptekar01}, which fit the data
well above 15 keV although they overpredict the flux of photons at
low energies \citep{Fenimore94,Feroci04}.
Despite this, OTTB fits are often included in more recent analysis, to
allow comparison with earlier studies. OTTB temperatures are typically in the range 20-40 keV,
and in general no emission is seen from short bursts above 150-200
keV.  There have however been a handful of events from SGR 1900+14
with a much harder spectrum, and emission extending up to 500 keV
\citep{Woods99c}.   There have also now been studies exploring the
softest part of the burst spectrum, using data from {\it
XMM-Newton}, where the burst spectra appear to be well fit with a
more physically-motivated model comprising a modified blackbody
plus resonant cyclotron scattering \citet[][and see
Sec.~\ref{emission}]{Lin12,Lin13}.  There have also been strong efforts to
search for spectral lines in magnetar bursts.  For a long time the
only reported detections, from 5~keV to 13-14~keV, in bursts from  SGR
1806-20, XTE
J1810-197, 4U 0142+61, and 1E1048.1-5937 came from {\it RXTE}
data \citep{Ibrahim02, Woods05,Gavriil02,Gavriil06,Gavriil11}.  However
recently a similar feature has been detected in bursts from
1E1048.1-5937 observed by {\it NuSTAR} \citep{An14}, increasing
confidence that they are indeed intrinsic to the bursts.  The line
energy is close to that expected for the proton cyclotron line
given the inferred magnetic field strength.   In addition to
time-integrated spectra, data quality are now sufficiently good
that it is possible to do time-resolved spectroscopy.  These
studies indicate that although the best fit spectral model remains
the same during individual bursts, the  parameters can evolve
\citep{Israel08,Lin11,Younes14}.  However between bursting
episodes, the best fit model for individual sources may change
\citep{vonkienlin12}.

The spectra of the initial spikes of intermediate flares are similar to the short
bursts \citep{Mazets99a, Mazets99b,Olive04, Israel08}. The spectrum of the extended
decaying tails is however different (in contrast with the tails seen after some
short bursts, where the peak and tail have similar spectra).  Tail spectra for
intermediate flares are well fit by a BB, possibly with an additional power law
component, and the emission softens during the tail
\citep{Ibrahim01,Lenters03,Esposito07, Gogus11}.

For the giant flares, reliable spectral modelling in the initial spike is
complicated enormously by dead time and pile up \citep{Fenimore96,Mazets99c}.
However OTTB models or quasi-BB models yield spectral temperatures in the range
200-300 keV, and emission has been detected up to 2 MeV \citep{maz79b,Hurley99,hur05}.
The spectrum of the emission in the tail is very similar to that of the short
bursts, with OTTB temperatures $\sim 10-30$ keV \citep{Fenimore81,Hurley99} that
vary with rotational phase.  Other spectral models such as BB or 2BB, possibly with
a power law, also provide a good fit to the data.  A significant hard ($> 1$ MeV)
component was seen during the tail and subsequent afterglow from the SGR 1806-20
giant flare \citep{Mereghetti05,Frederiks07,Boggs07}.

Although the spectral models described above provide a reasonable fit for the data,
they are not based on physical models that take into account all of the scattering
and resonant processes known to be important in such strong magnetic fields.  Any
thermal emission, for example, as might be expected from a trapped fireball
\citep{thdun95,Thompson01}, would be strongly modified, see
Fig.~\ref{burstrad}.
Lower energy photons scatter less and can hence escape from deeper, hotter parts of
the atmosphere. The radiation at low energies should thus exceed that expected for
simple blackbody emission \citep{Ulmer94, Lyubarsky02a}.  Photon splitting and
merging will also be important in modifying the spectrum \citep{Miller95,
Baring95,Thompson01} at energies above around 30 keV, and resonant cyclotron
scattering (RCS, see Sec.~\ref{rcs}) will also be important.  Efforts to
fit
burst spectra using more physical models, or to intepret the phenomenological models
in terms of physical parameters, are however very rare.  \citet{Israel08} (also
\citet{Kumar10}) suggested that the two blackbodies in the 2BB model fits might be
the photospheres associated with the different polarization modes, although
theoretical calculations of the properties of the two photospheres do not match
those inferred from the fits \citep{vanputten13}. \citet{Lin11} attempted to
interpret the parameters of the Comptonization and 2BB model fits in terms of a
population of coronal electrons scattering surface emission (for example from a
fireball), see also \citet{vanderhorst12} and \citet{Younes14}.  More recently,
\citet{Lin12,Lin13} made an effort to fit soft burst emission using the modified
blackbody model developed by \citet{Lyubarsky02a}, augmented to include effects of
RCS. The fact that physical model interpretations are still so scarce, however,
emphasizes the huge uncertainty in terms of the location of the emission mechanisms,
how they form, and how the released energy is partitioned between them.

\subsection{Burst seismology}
\label{seis}

\begin{figure}
\centering
  \includegraphics[width=3in]{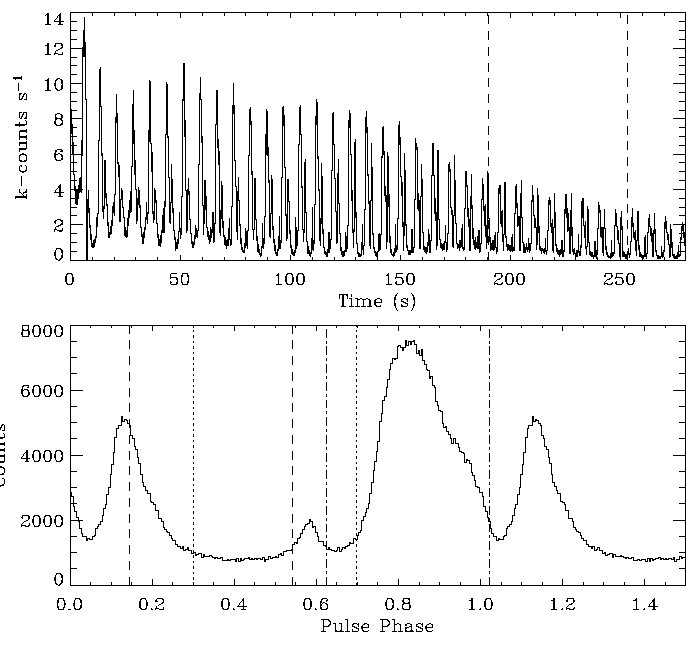}
  \includegraphics[width=3in]{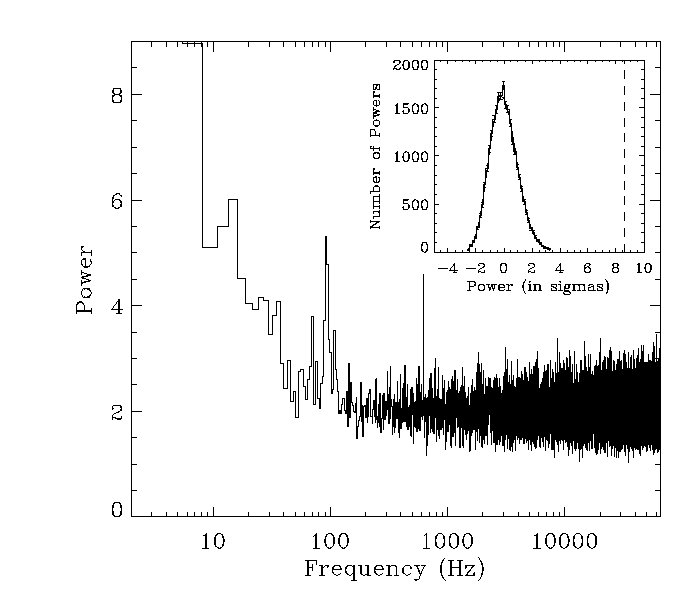}
\caption{Figure from \citet{strohmayer06} showing the two strongest QPOs
detected in the tail of the SGR 1806-20 giant flare.  Top left:  RXTE
lightcurve of the event.  Lower left: Rotational pulse during this time:
the power spectra shown are computed using the segments enclosed by the
dashed lines.  Right:  Power spectrum made by averaging nine 3 s segments
from the time interval marked by dashed lines in the top left panel. The
92 Hz and 625 Hz QPOs are clearly visible, and the inset illustrates the
significance of the 625 Hz feature \cite[from][\copyright AAS. Reproduced
with permission. A link to the original article via DOI is available in
the electronic version]{strohmayer06}
}
\label{qpos}
\end{figure}

Asteroseismology is a precision technique for the study of stellar
interiors, and it is magnetars that have opened up this field for
neutron stars.  This began when Quasi-Periodic Oscillations (QPOs)
in the hard X-ray emission were found in the tails of the giant
flares from the magnetars SGR 1806-20
\citep{israel05,watts06,strohmayer06} and SGR 1900+14
\citep{strohmayer05}.  In the tail of the SGR 1806-20  giant flare
(Fig.~\ref{qpos})
there were several QPOs in the range 18-150 Hz, and two isolated
higher frequency signals at 625 Hz and 1840 Hz.  The QPOs detected in
the tail of the giant flare from SGR 1900+14 had frequencies in
the range 28-155 Hz.   Widths (FWHM) were in the range 1-20 Hz,
with fractional amplitudes up to $\sim 20$ \% rms that are
strongly rotational phase-dependent.

The idea that giant flares might excite global seismic vibrations was first predicted by \citet{duncan98}, and this is the most plausible explanation that has yet been advanced to explain the QPOs \citep{israel05}.  If
this interpretation is correct, such vibrations offer an
unprecedented opportunity to constrain the interior field strength
and geometry (something that is very hard to measure directly),
and also perhaps the dense matter equation of state
\citep{samuelsson07,watts07}.    In order to do this, however, the
modes must be correctly identified.

The QPOs were initially tentatively identified with torsional
shear modes of the neutron star crust and torsional Alfv\'en modes
of the highly magnetized fluid core.   These identifications were
based on the expected mode frequencies, which are set by both the
size of the resonant volume and the relevant wave speed.  For
crustal shear modes, the appropriate speed is the shear speed $v_s
= (\mu_s/\rho)^{1/2}$ where $\mu_s$ is the shear modulus and
$\rho$ the density. The shear modulus is of the order of the
Coulomb potential energy $\sim Z^2e^2/r$ per unit volume $r^3$,
where $r\sim(\rho/Am_p)^{-1/3}$ is the inter-ion spacing, while
$Z$ and $A$ are the effective atomic number and mass number,
respectively, of the ions in the crust. Using the shear modulus
computed by \citet{Strohmayer91} and scaling by typical values for
the inner crust \citep{Douchin01}, the shear velocity as shown by
\citet{Piro2005} is:

\begin{equation}
v_s      =  1.1 \times 10^8 \mathrm{cm/s}
\left(\frac{\rho}{10^{14} \mathrm{g/cm}^3}\right)^{1/6}
\left(\frac{Z}{38}\right)  \left(\frac{302}{A}\right)^{2/3}
\left(\frac{1-X_n}{0.25}\right)^{2/3}
\label{vs}
\end{equation}
where $X_n$ is the fraction of neutrons. This yields a rough
estimate for the frequency for the fundamental crustal shear mode
of $\nu \sim v_s/2\pi R$ = 18 (10 km/$R$) Hz.  Full mode
calculations find similar values, but with additional dependencies
on the mass and radius of the star due to relativistic effects
(see for example \citealt{samuelsson07}), and it is this dependence
that makes the modes potentially powerful diagnostics of the dense
matter equation of state \citep{Lattimer07}.  Many of the lower
QPO frequencies could be explained as angular harmonics with no
radial nodes, whilst the two highest frequencies in the SGR
1806-20 giant flare were identified as radial overtones of these
crustal modes.

For torsional Alfv\'en modes of the core, the appropriate wave
speed is the Alfv\'en speed $v_A = B/\sqrt{4\pi\rho}$ where $B$ is
the magnetic field strength, giving

\begin{equation}
v_A = 10^8 \mathrm{cm/s}  \left(\frac{B}{10^{16}
\mathrm{G}}\right) \left(\frac{10^{15}
\mathrm{g/cm}^3}{\rho}\right)^{1/2} \, . 
\label{va}
\end{equation}
This yields a very rough estimate for the frequency of the
fundamental torsional Alfv\'en mode of $\nu \sim v_A/4R =$ 25 (10
km/$R$) Hz \citep{Thompson01}.  Note however that the value of the
field strength $B$ in magnetar cores is highly uncertain, as is
the appropriate value of the density $\rho$. In principle only the
charged component ($\sim$ 5-10\% of the core mass) should
participate in Alfv\'en oscillations, reducing $\rho$, however
there are mechanisms associated with superfluidity and
superconductivity that can couple the charged and neutral
components, leading to additional mass-loading. As above, full
mode calculations that take into account relativistic effects lead
to additional dependencies on neutron star mass and radius (see
for example \citealt{sotani08}).  It should also be noted that the
Alfv\'en modes constitute continua rather than a set of discrete
frequencies, since the field lines within the core have a
continuum of lengths.  The observed QPOs would then be associated
with turning points of the Alfv\'en continuum, since these tend to
dominate the oscillatory properties when one computes the time
evolution of systems with continua \citep{levin07,sotani08}.

In fact, for a star with a magnetar strength field, crustal
vibrations and core vibrations should couple together on very
short timescales \citep{levin06,levin07}.  Considering them in
isolation, as described above, is therefore not appropriate.  The
current viewpoint, based on more detailed modelling that takes
into account the magnetic coupling between crust and core, is that
the QPOs are in fact associated with global magneto-elastic
axial (torsional) oscillations of the star
\citep{glampedakis06,lee08,andersson09,steiner09,vanhoven11,vanhoven12,
colaiuda11,colaiuda12,gabler12,gabler13,passamonti13,passamonti14,asai14,glampedakis14}.
However since magneto-elastic oscillations depend on the same
physics described above, albeit now in a coupled system, they have
frequencies in the same broad range as the simple estimates given
above.  Current magneto-elastic torsional oscillation models can
thus in principle explain the presence of oscillations at
frequencies of 155 Hz and below.

\begin{figure}
\centering
\includegraphics[width=3in]{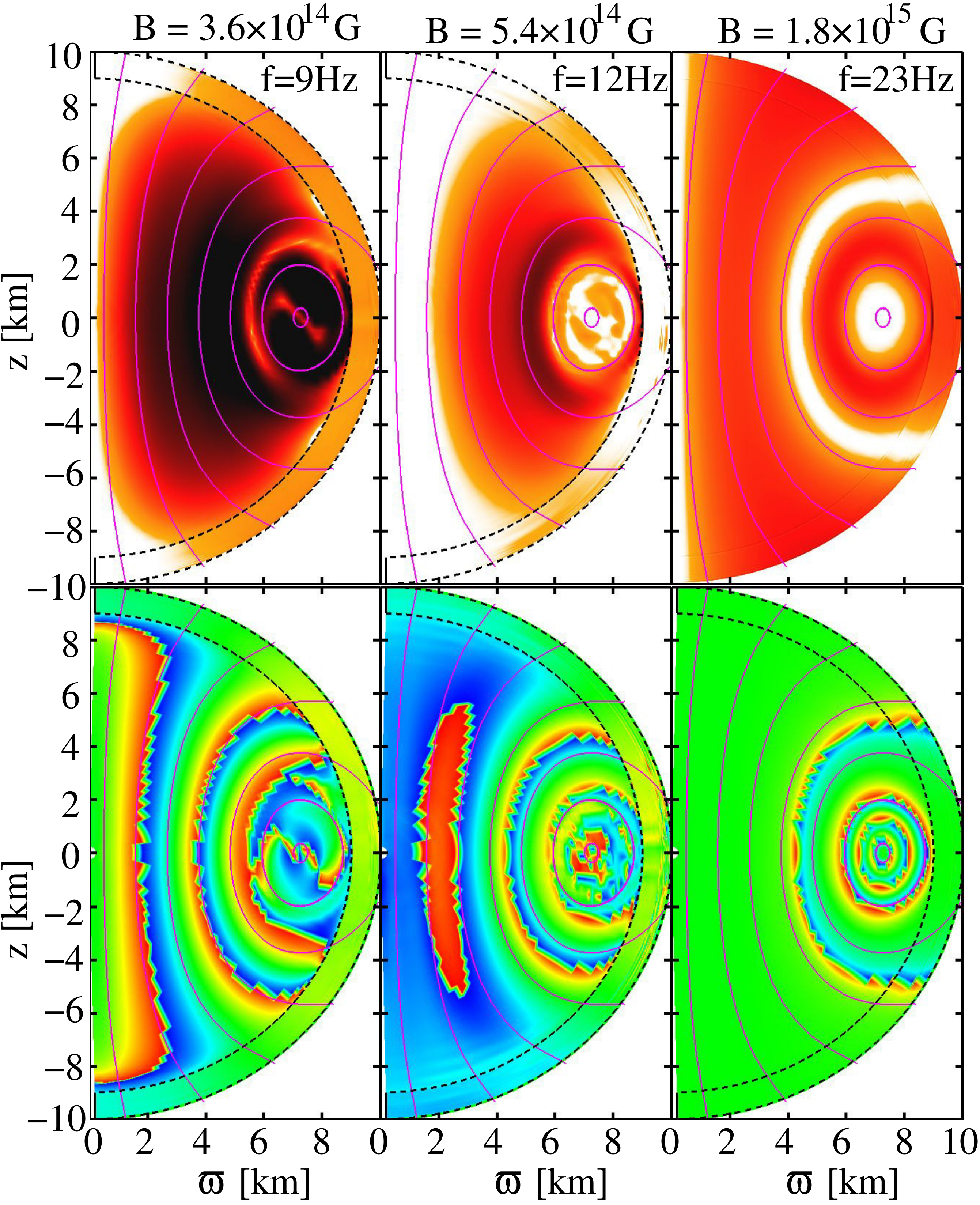}
\caption{Simulations of QPOs of a magnetized neutron star with a solid crust and
superfluid core, in General Relativity, from \citet{gabler13b}.
The upper panels
show the Fourier amplitude and the lower panels the phase. Both frequency
and mode structure change as the field strength varies.  The color scale
ranges from white-blue (minimum) to orange-red (maximum) in the top panel,
and from $\theta=\pi$/2 (blue) to $\theta=-\pi$/2 (orange- red),
respectively. The crust is indicated by the dashed black line, and
magnetic field lines by the solid magenta lines \cite[from][\copyright
2013 American Physical Society, reproduced with permission]{gabler13b}.}
\label{qposims}
\end{figure}

Until very recently, however, it appeared that there was a
significant problem with the higher frequency QPOs.  This is
because although there are crust shear modes in this frequency
range, they should overlap with the various Alfv\'en continua
(there are no gaps between the harmonics of the continua as the
frequency increases).  As a result, the coupled oscillation should
damp very rapidly, on timescales of less than a second
\citep{vanhoven12,gabler12}.  The data analysis, however,
indicated that the oscillations persisted for up to $\sim 100$s
\citep{watts06,strohmayer06}.  Various solutions to this problem
have been explored, including coupling to polar modes
\citep{lander10, Lander11, colaiuda12}, and resonances between
crust and core that might develop as a result of superfluid
effects \citep{gabler13b, passamonti14}.   It is clear from these
studies that superfluidity in particular can have a large effect
on the characteristics of the mode spectrum: and since
superfluidity is certainly present in neutron stars, mode models
must start to take this into account before we can make firm mode
identifications (Fig.~\ref{qposims}).

However the debate over this issue also exposed the fact that the
initial data analysis  did not actually test whether the signal
could also be there in much shorter data segments, more consistent
with the theoretical predictions.  \citet{huppenkothen14b} have
since re-analysed data for the 625 Hz QPO in the SGR 1806-20 giant
flare and found that the data are in fact consistent with a
short-duration signal that damps and is re-excited several times
(rather than a long-lasting low-amplitude QPO).   What might cause
late time excitation and re-excitation remains an open question,
and is relevant to the lower frequency QPOs as well since several
seem to appear only late in the tails of the giant flares.
Aftershocks may play an important role in exciting and re-exciting
the QPOs that we see, and there may also be intrinsic delays in
the process whereby vibrations are excited by the flare due to
impedance mismatching between the different components of the star
\citep{Link14}.

Since giant flares are very rare, there have also been efforts to
search for seismic vibrations in the much more frequent lower
energy bursts\footnote{There have also been a number of searches
for gravitational waves associated with magnetar flares and any
associated starquakes or global seismic oscillations
\citep{Abbott07,Abbott08,Abbott09,Abadie11}.  So far only upper
limits have been reported, but new analysis techniques are being
developed for the next generation of detectors \citep{Murphy13}.
}.  As discussed above, it is not yet entirely clear whether these
bursts are caused by the same mechanism as the giant flares.
However if they are, it is quite possible that they might excite
seismic vibrations at frequencies similar to those seen in the
giant flares, particularly if we are genuinely seeing global modes
of vibration of the star\footnote{The search for global seismic
vibrations in small and intermediate magnetar flares is a core
science driver for future hard X-ray and gamma-ray missions such
as the proposed {\it Large Observatory for X-ray Timing}
\citep{Feroci14}.}.  Searching for QPOs in the smaller bursts is
however complicated by the short, transient nature of the burst
lightcurves themselves, and this has required the development of
specially tailored statistical methods
\citep{huppenkothen13,huppenkothen14c}.

So far these techniques have been applied to several data sets.  A
sample of 27 bursts from the magnetar SGR 0501+4516, using Fermi
GBM data, made one candidate detection, but its significance was
weak \citep{huppenkothen13}.  A search of a larger sample of 286
Fermi GBM bursts from SGR J1550-5418, however, found significant
QPOs at 93 Hz and 127 Hz after averaging together multiple bursts
from highly active episodes \citep{huppenkothen14}.   Similar
analysis using RXTE data from the most burst-active magnetars SGR
1806-20 and SGR 1900+14 (the two sources for which QPOs have been
observed in the giant flares\footnote{Very few bursts from these
sources have been observed in the period since Fermi GBM has been
flying, and no giant flares have yet been observed in the Fermi
era.}) led to the detection of a QPO at 57 Hz after averaging
together multiple bursts from SGR 1806-20 \citet{huppenkothen14c}.
These frequencies are in the range found in the giant flares, and
the QPO widths are also comparable.  It therefore seems plausible
that they are instances of the same phenomenon. If these
frequencies do indeed represent global magneto-elastic
oscillations the implication is that such vibrations are excited
not only by giant flares, but also by trains of shorter bursts.
This is important information when we start to consider how modes
are excited by the trigger mechanism.

The analysis of SGR J1550-5418, however, also revealed a QPO in a
single burst, at a much higher frequency of 260 Hz.  In addition
to being in a different frequency band, this QPO was much broader
than those seen in the giant flares and had very high fractional
amplitude.  If this is a magneto-elastic oscillation mode, then it
is of interest since models predict that modes in this frequency
range should die out on timescales comparable to the duration of
short bursts.  This could explain the observed low coherence,
since broad width is a natural consequence of a rapidly
exponentially decaying signal.  This signal could, however be
something quite different, such as a plasma instability associated
with magnetic reconnection \citep{kliem00} or a local oscillation
in a smaller, temporarily decoupled, cavity
\citep{huppenkothen14}.  In this case it may be a fingerprint of
the burst trigger process.  Variability in the impulsive phase of
the giant flares has previously been suggested, but dead time and
saturation effects strongly distort timing analysis for the very
brightest events \citep{barat83,terasawa06}.

Another open question is how magneto-elastic oscillations couple to the
magnetosphere and hence modulate the emission from the star. An important concern is
that the fractional amplitude of the QPOs is in some cases quite high, and certainly
much higher than the likely amplitude of any oscillations of the neutron star's
crust.  Emission in the tails of the giant flares is dominated by radiation leaking
from the trapped pair-plasma fireball.  This emission is strongly beamed (giving
rise to the strong rotational pulse), and thus in principle could act to amplify
small surface vibrations, however analysis of the beams from the giant flares
indicates that although the effect is real it is unlikely to be strong enough to
explain the highest observed fractional amplitudes \citep{dangelo12}. This suggests
that there is some additional effect modulating the intensity of the emission:
something that takes on added importance in the light of QPOs detected in the
smaller bursts, where it is not clear that fireballs even form.  A likely mechanism
is a modulation of the optical depth to Resonant Cyclotron Scattering
(see Sec.~\ref{rcs}), via changes in particle number density
\citep{timokhin08} and/or
magnetic field geometry \citep{gabler14}.  However the details of this process, and
the interaction with the fireball, remain to be worked out fully.

\section{Summary and Conclusions}
\label{conc}

There is now a general agreement that the key observational
phenomena that make soft gamma repeaters and anomalous X-ray
pulsars so unique are well explained by the presence of a
magnetar, a neutron star with ultra-strong magnetic field. In the
absence of accretion from a binary companions and of large enough
rotational energy losses,  magnetic energy appears the only
reasonable option to power both the persistent and bursting
emission at the observed levels. Recently, thanks to the discovery
of the so-called ``low-B'' sources, it has become increasingly
evident that to make an ``active'' magnetar what matters is not
(or not only) a large ($\gtrsim 10^{14}$ G) dipole field, but a
strong, residual poloidal component of the internal magnetic
field. In this section we summarize the status of theoretical
modelling, within the magnetar scenario, in the attempt to
highlight what are, in our opinion, the issues which are basically
settled and those which are still open.

A clear picture has now emerged of where magnetars stand in
relation to the population of isolated neutron stars, at least in
terms of their observational properties. However, borders among
the different classes are somehow blurry and the possible
(evolutionary ?) links remain still to be fully understood. The
existence of ``low-B'' and transient magnetars, implying that the
number of highly magnetized neutron stars may be much larger then
previously thought, makes it apparent that there may be a
``birthrate'' problem, unless there is an overlap between classes,
meaning that objects that we see as observationally diverse are
just neutron stars at different evolutionary stages, or magnetars
can form through channels different from standard supernova
events.

A better understanding of magnetar's formation path may help to
bring clarity here. One of the biggest open questions is how
magnetars acquire their super-strong magnetic field, as compared
to those of the, apparently much more abundant, radio pulsars.
Despite much effort, no definite conclusion has been reached as
yet. Nonetheless, the very recent discovery that the magnetar in
the Westerlund 1 cluster may have originated  in a binary system
points quite strongly towards a particular formation route
involving massive binaries. Much theoretical attention has also
been given to the idea that newborn magnetars are the central
engines for gamma-ray bursts and newly developed models have had a
great deal of success in explaining the ``plateau'' phase of the
observed lightcurve. A conclusive proof for this  seems now to
rest with the observers, and may come in the next few years if a
GRB is definitely linked to the gravitational wave detection of a
compact binary interaction.

A topic of the greatest relevance is the magnetic field
configuration in newborn magnetars. In particular how the field
structure varies across core, crust, and magnetosphere; the
balance between toroidal and poloidal components; and the
small-scale structure of the external field. The state of the
magnetic field at birth, and its subsequent evolution, are
critical input for many aspects of magnetar physics. This is an
active area of study that has also been given new impetus by
the discovery of magnetars with low surface dipole field
strengths. There has been a good deal of progress in this area in
recent years, although the various physical processes likely to
affect the evolution of the magnetic field in the crust, where
coupling of magnetic and thermal effects becomes important, remain
challenging to model.

There has been great progress in understanding the emission
processes of magnetars. The existence of a twisted magnetosphere
seems now to be generally accepted. Resonant cyclotron scattering onto
pairs flowing in a twisted
magnetophere provides spectra which are in good quantitative
agreement with the soft X-ray data, and also a natural explanation
for the emission in the other wavebands (hard X-ray, optical/IR).
However, a complete solution of the non-linear charge acceleration
problem, including the various QED effects leading to photon
splitting, positronium dissociation, and the fact that the
twisted magnetosphere is expected to be strongly dynamic, has not
yet been found. Another of the major open issues is the lack of a
credible model of the interplay between crustal, atmospheric
and magnetospheric emission, capable of explaining the broadband
spectral energy distribution of magnetar sources.

For the bursts, some aspects do seem clear. It is widely accepted, in the
absence of a better model, that reconnection is likely to be required to
explain the observed gamma-ray emission. For the giant flares, plasma
ejection must take place to explain the radio afterglows, and a
magnetically trapped pair plasma fireball seems the only viable hypothesis
for the pulsed tails of the giant flares. The trigger mechanism for all
bursts, however, remains unknown, as do the emission processes in the
smaller and intermediate bursts. It will also be important to ensure that
the bursts and persistent emission are considered as a whole: constraints
on the magnetospheric structure and radiative transfer environment
obtained from study of the persistent emission should be applied
consistently, for example, when considering the radiation propagating in
the aftermath of a burst. The detection of quasi-periodic oscillations in
bursts has also provided new insight. The idea that these are caused by
global seismic vibrations, excited by the burst process, is certainly the
most plausible model put forward to date, although details of how the
stars oscillate remain to be worked out. Far more work is required,
however, is to examine self-consistently the conditions for excitation,
decay, and modulation of the emission.

Some theoretical predictions must await more advanced
observational capabilities in order to be tested fully. {\it
NuStar} (and possibly in the future ASTRO-H) is now offering the first
opportunity to provide simultaneous data on the hard X-rays/soft
X-rays turn over (at few tens of keV). {\it ATHENA} (the second
large mission that will be developed by ESA, with launch in 2028)
is the most important X-ray mission on the horizon and will have
an unprecedented capability for compact objects and collapsar physics. A
large
area X-ray timing mission, such as the Large Observatory
for X-ray Timing ({\it LOFT}, studied as a candidate for an ESA M3 mission), would
also enable a large increase in capability to detect seismic
vibrations in magnetar bursts and resolve both any mode splitting
and their evolution on short timescales, two things that would
both help to distinguish theoretical models. This may open the
possibility of performing, systematically, asteroseismology studies
in neutron stars. The role of gravitational wave observatories in
pinning down the mechanism behind short gamma-ray bursts, and the
possible role of millisecond magnetars in that process, has
already been described above. In fact, magnetars are wonderful
candidates for detection by ground-based, long-baseline,
interferometric gravitational wave detectors such as LIGO and
Virgo. Magnetars are also powerful probes of the Galactic structure
and the interstellar medium: pulsar timing arrays are also
starting to be used to hunt background stochastic gravitational
waves.

Magnetar radiation is expected to be strongly polarized, and
the polarization observables may also  probe the so-called
``vacuum polarization'' effect, which is predicted by nonlinear
QED, but has not yet been verified experimentally. Future X-ray
polarimetry experiments, currently under consideration for several
small and medium missions (e.g. {\it IXPE}, a NASA SMEX
candidate, and {\it XIPE}, an ESA M4 candidate) may therefore open
a completely new window on our understanding of the radiation
processes around magnetars and on the physics of matter and
radiation in superstrong fields.

\ack

This work benefitted from discussions with a number of
colleaugues. In particular, we would like to thank Paolo Esposito,
Sandro Mereghetti, Yuri Lyubarsky, Sergei Popov, Luigi Stella for
a careful reading of the manuscript and for their useful comments.
ALW would also to thank Thijs van Putten, Chris Elenbaas, and
Daniela Huppenkothen for comments on an early draft of
Sec.~\ref{bursts}. The work of RT is partially supported by INAF
through a PRIN grant. ALW acknowledges support for her work on
magnetars from an NWO Vidi Grant, and from the Nederlandse
Onderzoekschool voor Astronomie NOVA's Network 3 programme.


\end{document}